\documentclass[twocolumn]{aastex702}
\usepackage{xspace}
\usepackage[hyphens]{url}
\usepackage{comment}
\pdfoutput=1 %for arXiv submission
\usepackage[figure,figure*]{hypcap}
\usepackage{footnote}
\usepackage{hyperref}
\usepackage{subfigure}
\usepackage{amsmath}
\usepackage{siunitx}
\usepackage{epsfig}
\usepackage{natbib}
\usepackage{xcolor} 
\usepackage{tablefootnote}
\usepackage{longtable, threeparttablex}

\usepackage[version=4]{mhchem}

\begin{document}

\def\Msun{\hbox{M$_{\odot}$}}
\def\Lsun{\hbox{L$_{\odot}$}}
\newcommand{\kms}{\ensuremath{km\ s^{-1}\xspace}}
\def\n2hp{N$_{2}$H$^{+}$}
\def\micron{$\mu$m\xspace}
\def\13CO{$^{13}$CO}
\def\etamb{$\eta_{\rm mb}$}
\def\Inu{I$_{\nu}$}
\def\kapnu{$\kappa _{\nu}$}
\def\ffore{f$_{\rm{fore}}$}
\def\tastar{T$_{A}^{*}$}
\def\nh3{NH$_{3}$}
\def\deg{$^{\circ}$\xspace}
\def\arcsec{$^{\prime\prime}$\xspace}
\def\arcmin{$^{\prime}$\xspace}
\def\Vlsr{\hbox{V$_{LSR}$}\xspace}
\def\CO{\ce{CO} $2-1$\xspace}
\def\HII{\mbox{H \small{II}\normalsize}}
\def\sgra{SgrA$^{\star}$\xspace}

\newcommand\HNCO{HNCO\;\mbox{4(0,4)--3(0,3)}\xspace}
\newcommand\SiO{\mbox{SiO\;(2--1)}\xspace}
\newcommand\CS{\mbox{CS\;(2--1)}\xspace}
\newcommand\SO{\mbox{SO\;3(2)--2(1)}}
\newcommand\HCtrN{\mbox{HC$_{3}$N\;(11--10)}\xspace}
\newcommand\HthCOp{\mbox{H$^{13}$CO$^{+}$\;(1--0)}}
\newcommand\HthCN{\mbox{H$^{13}$CN\;(1--0)}}
\newcommand\HNthC{HN$^{13}$C\;(1--0)}
\newcommand{\hcop}{HCO$^{+}$\xspace}
\newcommand{\ha}{H40$\alpha$\xspace}
\newcommand{\chtcho}{CH$_3$CHO $5_{(1,4)}-4_{(1,3)}$\xspace}
\newcommand{\HCFN}{\mbox{HC$^{15}$N}(1--0)}

\newcommand\paperI{\textit{Paper I}}
\newcommand\paperII{\textit{Paper II}}
\newcommand\paperIII{\textit{Paper III}}
\newcommand\paperIV{\textit{Paper IV}}

\title{ACES VIII: A Survey of Compact, High-Velocity Features Observed in \CS}

%%%%%%%%%%%%%%%%%%%%%%%%%%%%%%%
%%% Paper sprint team members
%%%%%%%%%%%%%%%%%%%%%%%%%%%%%%%

\author[0000-0002-5776-9473]{Dani R. Lipman}
\affiliation{Department of Physics, University of Connecticut, 196A Auditorium Road, Unit 3046, Storrs, CT 06269, USA}
\email{dani.lipman@uconn.edu}

\author[0009-0005-8744-9592]{Jakub Poznanski}
\affiliation{Department of Physics, University of Connecticut, 196A Auditorium Road, Unit 3046, Storrs, CT 06269, USA}
\email{jakub.poznanski@uconn.edu}

\author[0000-0002-1313-429X]{Savannah R. Gramze}
\affiliation{Department of Astronomy, University of Florida, Gainesville, FL 32611 USA}
\email{savannahgramze@ufl.edu}

\author[0009-0009-0414-8698]{Claire E. Cook}
\affiliation{University of Kansas, Department of Physics and Astronomy, 1251 Wescoe Hall Dr. Lawrence, KS 66045 USA}
\email{claire.e.cook67@gmail.com}

\author[0000-0002-6073-9320]{Cara Battersby}
\affiliation{Department of Physics, University of Connecticut, 196A Auditorium Road, Unit 3046, Storrs, CT 06269, USA}
\email{cara.battersby@uconn.edu}

\author[0009-0002-7459-4174]{Jennifer Wallace}
\affiliation{Department of Physics, University of Connecticut, 196A Auditorium Road, Unit 3046, Storrs, CT 06269, USA}
\email{jennifer.2.wallace@uconn.edu}

\author[0000-0002-8586-6721]{Pablo Garc\'ia}
\affiliation{Chinese Academy of Sciences South America Center for Astronomy, National Astronomical Observatories, CAS, Beijing 100101, China}
\affiliation{Instituto de Astronom\'ia, Universidad Cat\'olica del Norte, Av. Angamos 0610, Antofagasta, Chile}
\email{ }

\author[0009-0003-6268-660X]{Samantha Adams}
\affiliation{Department of Physics, University of Connecticut, 196A Auditorium Road, Unit 3046, Storrs, CT 06269, USA}
\email{zjd24001@uconn.edu}

\author[0000-0002-5582-8521]{Xinyu Mai}
\affiliation{University of Kansas, Department of Physics and Astronomy, 1251 Wescoe Hall Dr. Lawrence, KS 66045 USA}
\email{xinyu-mai@ku.edu}

\author[0009-0002-1465-1958] {Kai Smith}
\affiliation{University of Kansas, Department of Physics and Astronomy, 1251 Wescoe Hall Dr. Lawrence, KS 66045 USA}
\email{ksmith@scripps.edu}

\author[0000-0002-4013-6469]{Natalie Butterfield}
\affiliation{National Radio Astronomy Observatory, 520 Edgemont Road, Charlottesville, VA 22903, USA}
\email{nbutterf@nrao.edu}

\author[0009-0008-2494-4408]{Sophia Kempe}
\affiliation{Department of Physics, University of Connecticut, 196A Auditorium Road, Unit 3046, Storrs, CT 06269, USA}
\email{sophia.kempe@uconn.edu}

\author[0000-0001-6431-9633]{Adam Ginsburg}
\affiliation{Department of Astronomy, University of Florida, Gainesville, FL 32611 USA}
\email{adamginsburg@ufl.edu}

\author[0009-0004-0685-7678]{Rojita Buddhacharya}
\affiliation{Astrophysics Research Institute, Liverpool John Moores University, 146 Brownlow Hill, Liverpool L3 5RF, The UK}
\affiliation{Center for Astrophysics $\vert$ Harvard \& Smithsonian, 60 Garden Street, Cambridge, MA 02138, USA}
\email{r.buddhacharya@2024.ljmu.ac.uk}

\author[orcid=0000-0002-5566-0634,gname='Tomoharu',sname='Oka']{Tomoharu Oka}
\affiliation{Department of Physics, Institute of Science and Technology, Keio University, 3-14-1 Hiyoshi, Kohoku-ku, Yokohama, Kanagawa 223-8522, Japan}
\email{[tomo@phys.keio.ac.jp](mailto:tomo@phys.keio.ac.jp)}

%%%%%%%%%%%%%
%%%%%%%%%%%%%

\author[0009-0005-9578-2192]{Danya Alboslani}
\affiliation{University of Connecticut, Department of Physics, 196A Auditorium Road, Unit 3046, Storrs, CT 06269, USA}
\affiliation{University of Virginia, Department of Astronomy, 530 McCormick Rd, Charlottesville, VA 22904, USA}
\email{danya.alboslani@virginia.edu}

\author[0000-0003-3341-6144]{Jairo Armijos-Abenda\~no}
\affiliation{Observatorio Astron\'omico de Quito, Observatorio Astron\'omico Nacional, Escuela Polit\'ecnica Nacional, 170403, Quito, Ecuador}
\email{jairo.armijos@epn.edu.ec}

\author[0000-0001-8135-6612]{John Bally} 
\affiliation{Center for Astrophysics and Space Astronomy, Department of Astrophysical and Planetary Sciences,
University of Colorado, Boulder, CO 80389, USA} 
\email{}

\author[0000-0003-0410-4504]{Ashley T. Barnes}
\affiliation{European Southern Observatory (ESO), Karl-Schwarzschild-Stra{\ss}e 2, 85748 Garching, Germany}
\email{}

\author[0000-0002-0533-8575]{Nazar Budaiev}
\affiliation{Department of Astronomy, University of Florida, Gainesville, FL 32611 USA}
\email{}

\author[0000-0002-4407-885X]{Alyssa Bulatek}
\affiliation{Department of Physics and Astronomy, Haverford College, Haverford, PA 19041, USA}
\email{}

\author[0000-0001-8064-6394]{Laura Colzi}
\affiliation{Centro de Astrobiolog\'ia (CAB), CSIC-INTA, Ctra. de Ajalvir Km. 4, 28850, Torrej\'on de Ardoz, Madrid, Spain}
\email{lcolzi@cab.inta-csic.es}

\author[0000-0002-0706-2306]{Christoph Federrath}
\affiliation{Research School of Astronomy and Astrophysics, Australian National University, ACT 2611, Australia}
\email{christoph.federrath@anu.edu.au}

\author[0009-0004-0121-1560]{Zi-Xuan Feng}
\affiliation{Como Lake centre for AstroPhysics (CLAP), DiSAT, Universit{\`a} dell'Insubria, via Valleggio 11, 22100 Como, Italy}
\affiliation{Universit\"{a}t Heidelberg, Zentrum f\"{u}r Astronomie, Institut f\"{u}r Theoretische Astrophysik, Albert-Ueberle-Str.\ 2, 69120 Heidelberg, Germany}
\email{zixuan.feng@uninsubria.it}

\author[0000-0001-9656-7682]{Jonathan D. Henshaw}
\affiliation{Max-Planck-Institut f\"ur Astronomie, K\"onigstuhl~17, D-69117 Heidelberg, Germany}
\email{}

\author{Paul Ho}
\affiliation{Institute of Astronomy and Astrophysics, Academia Sinica, 11F of ASMAB, AS/NTU No. 1, Sec. 4, Roosevelt Road, Taipei 10617, Taiwan}
\affiliation{East Asian Observatory, 660 N. A'ohoku, Hilo, Hawaii, HI 96720, USA}
\email{}

\author[0000-0001-9155-3978]{Pei-Ying Hsieh}
\affiliation{National Astronomical Observatory of Japan, 2-21-1 Osawa, Mitaka, Tokyo 181-8588, Japan}
\email{}

\author[0000-0003-4140-5138]{Katharina Immer}
\affiliation{European Southern Observatory (ESO), Karl-Schwarzschild-Stra{\ss}e 2, 85748 Garching, Germany}
\email{}

\author[0000-0003-4493-8714]{Izaskun Jim\'enez-Serra}
\affiliation{Centro de Astrobiolog\'ia (CAB), CSIC-INTA, Ctra. de Ajalvir Km. 4, 28850, Torrej\'on de Ardoz, Madrid, Spain}
\email{}

\author[0000-0002-0560-3172]{Ralf S.\ Klessen}
\affiliation{Universit\"{a}t Heidelberg, Zentrum f\"{u}r Astronomie, Institut f\"{u}r Theoretische Astrophysik, Albert-Ueberle-Str.\ 2, 69120 Heidelberg, Germany}
\affiliation{Universit\"{a}t Heidelberg, Interdisziplin\"{a}res Zentrum f\"{u}r Wissenschaftliches Rechnen, Im Neuenheimer Feld 225, 69120 Heidelberg, Germany}
\email{}

\author[0000-0001-6353-0170]{Steven N. Longmore}
\affiliation{Astrophysics Research Institute, Liverpool John Moores University, IC2, Liverpool Science Park, 146 Brownlow Hill, Liverpool L3 5RF, UK}
\affiliation{Cosmic Origins Of Life (COOL) Research DAO, \href{https://coolresearch.io}{https://coolresearch.io}}
\email{}

\author[0000-0003-2619-9305]{Xing Lu}
\affiliation{Shanghai Astronomical Observatory, Chinese Academy of Sciences, 80 Nandan Road, Shanghai 200030, P.\ R.\ China}
\affiliation{State Key Laboratory of Radio Astronomy and Technology, A20 Datun Road, Chaoyang District, Beijing, 100101, P.\ R.\ China}
\email{xinglv.nju@gmail.com}

\author[0000-0001-8782-1992]{Elisabeth A.C. Mills}
\affiliation{University of Kansas, Department of Physics and Astronomy, 1251 Wescoe Hall Dr. Lawrence, KS 66045 USA}
\email{}

\author[0000-0002-6398-7530]{N. Bijas}
\affiliation{UK ALMA Regional Centre Node, Jodrell Bank Centre for Astrophysics, The University of Manchester, Manchester M13 9PL, UK}
\email{bijasthejas@gmail.com}

\author[0000-0002-6379-7593]{Francisco Nogueras-Lara}
\affiliation{Instituto de Astrofísica de Andalucía, CSIC, Glorieta de la Astronomía s/n, E-18008 Granada, Spain}
\email{fnogueras@iaa.es}

\author[0000-0002-5811-0136]{Dylan M. Par\'e}
\affiliation{Joint ALMA Observatory, Alonso de Cordova 3107, Vitacura, Casilla 19001, Santiago de Chile, Chile}
\affiliation{National Radio Astronomy Observatory, 520 Edgemont Road, Charlottesville, VA 22903, USA}
\email{dylan.pare@villanova.edu}

\author[0000-0002-6362-8159]{Maya A. Petkova}
\affiliation{Physics and Astronomy Department, Chalmers University of Technology, SE-412 96 Gothenburg, Sweden}
\email{}

\author[0000-0002-3972-1978]{Jaime E. Pineda}
\affiliation{Max-Planck-Institut f\"ur extraterrestrische Physik, Gie\ss enbachstra\ss e 1, 85748 Garching bei M\"unchen, Germany}
\email{jpineda@mpe.mpg.de}

\author[0000-0002-7269-342X]{Marc W. Pound}
\affiliation{University of Maryland, Department of Astronomy, College Park, MD 20742-2421, USA}
\email{}

\author[0000-0002-2887-5859]{V\'ictor M. Rivilla}
\affiliation{Centro de Astrobiolog\'ia (CAB), CSIC-INTA, Ctra. de Ajalvir Km. 4, 28850, Torrej\'on de Ardoz, Madrid, Spain}
\email{}

\author[0000-0002-3078-9482]{\'Alvaro S\'anchez-Monge}
\affiliation{Institut de Ci\`encies de l'Espai (ICE), CSIC, Campus UAB, Carrer de Can Magrans s/n, E-08193, Bellaterra, Barcelona, Spain}
\affiliation{Institut d'Estudis Espacials de Catalunya (IEEC), E-08860, Castelldefels, Barcelona, Spain}
\email{asanchez@ice.csic.es}

\author[0000-0002-3941-0360]{Miriam G. Santa-Maria}
\affiliation{Instituto de Física Fundamental (CSIC). Calle Serrano 121-123, 28006, Madrid, Spain}
\email{miriam.g.sm@csic.es}

\author[0000-0002-1730-8832]{Anika Schmiedeke}
\affiliation{Green Bank Observatory, 155 Observatory Rd, Green Bank, WV 24944, USA}
\email{ajschmiedeke@nrao.edu}

\author[0000-0002-4268-6499]{Yoshiaki Sofue}
\affiliation{Institute of Astronomy, The University of Tokyo, Mitaka, Tokyo 181-0015, Japan}
\email{sofue@ioa.s.u-tokyo.ac.jp}

\author[0000-0001-6113-6241]{Mattia C. Sormani}
\affiliation{Como Lake centre for AstroPhysics (CLAP), DiSAT, Universit{\`a} dell'Insubria, via Valleggio 11, 22100 Como, Italy}
\email{}

\author{Howard A. Smith}
\affiliation{Center for Astrophysics $\vert$ Harvard \& Smithsonian, 60 Garden Street, Cambridge, MA 02138, USA}
\email{hsmith@cfa.harvard.edu}

\author[0000-0002-9483-7164]{Robin G. Tress}
\affiliation{Institute of Physics, Laboratory for Galaxy Evolution and Spectral Modelling, EPFL, Observatoire de Sauverny, Chemin Pegasi 51, 1290 Versoix, Switzerland}
\email{}

\author[0000-0001-7330-8856]{Daniel Walker}
\affiliation{UK ALMA Regional Centre Node, Jodrell Bank Centre for Astrophysics, The University of Manchester, Manchester M13 9PL, UK}
\email{}

\author[0000-0003-2384-6589]{Qizhou Zhang}
\affiliation{Center for Astrophysics $\vert$ Harvard \& Smithsonian, 60 Garden Street, Cambridge, MA 02138, USA}
\email{}

\begin{abstract}

The extreme kinematics of the Milky Way's Central Molecular Zone (CMZ) are influenced by processes such as dynamical shearing, cloud collisions, and stellar feedback. These events are visible in molecular data as vertically spiked features in position-velocity (PV) diagrams referred to as high velocity dispersion compact clouds (HVCCs). Using ALMA CMZ Exploration Survey (ACES) CS (2--1) molecular data, we identify a total of 235 HVCC candidates, 163 of which are visually identified, and an additional 72 identified via automated dendrogram methods. For each HVCC we catalog and report the physical and kinematic properties, explore line ratios of the cold dense gas tracer \HNCO with C-shock tracers, classify the morphology of their PV diagrams, and view their position-position-velocity distribution. The sample includes structures which are compact ($d<5\,$pc) and have large velocity extents (\SI{20}{\kms}$<\Delta \mathrm{V} <$ \SI{140}{\kms}), with most structures showing thin, `spiked' PV morphologies. We highlight areas of high ratios between \HNCO and C-shock tracers along the edge of known orbital streams, implying a buildup of bar lane gas accreting onto the CMZ. We also find a collection of HVCCs overlapping with the \SI{50}{\kms} cloud and known circumnuclear disk features. This catalog will be used for future investigation of nuclear inflow and determining dominant mechanisms disrupting average CMZ gas flows.

\end{abstract}


\section{Introduction}\label{sec:Introduction}

The inner $\sim$\SI{300}{pc} of the Milky Way, the Central Molecular Zone (CMZ) is the central depot of the Galaxy \citep{Henshaw2023,Battersby2025a}, formed as gas in the Galactic disk is transported inwards via Galactic bar lanes \citep{Sormani_and_Barnes_2019,Hatchfield2021}, onto approximately elliptical x$_2$ orbits aligned perpendicular to the Galactic bar \citep{Binney1991,Sormani2018b,Tress2020}. The CMZ holds 3--10\% of the total molecular gas in the Galaxy \citep{RomanDuval2016}, accounting for up to 80\% of the Galaxy's dense molecular gas ($n > 10^{3-4}$~cm$^{-3}$) \citep{Longmore2013,Barnes2017}, and exhibits extreme properties, including elevated gas surface densities ($10^{3.2}~\Msun\mathrm{pc}^{-2}$), magnetic fields (10 -- 1000 $\mu$G), and temperatures ($T_{gas} \sim 50-100~\mathrm{K}$) \citep[][]{Ginsburg2016,Mills2018a,Pillai2015,Butterfield2024,Longmore2012} that are all 1--2 orders of magnitude higher compared to the solar neighborhood. Additionally, gas in the CMZ tends to be more turbulent than in the disk, leading to broader linewidths \citep[e.g. $\sigma \sim 12~\mathrm{\kms} $ on 10~pc scales in the CMZ vs $\sigma \sim 3~\mathrm{\kms}$ in the disk;][]{Shetty2012,Heyer2015,Kauffmann2017c,Federrath2016, Henshaw2016_gas_kinematics}. The elevated temperature, density, and velocity dispersion compared to local star-forming regions are comparable to those seen in starburst galaxies and galaxies at the peak of cosmic star formation \citep{Kruijssen2013,Henshaw2023}. The dynamic range in physical properties, as well as being our nearest galaxy center \citep[$d \approx 8.2~\rm{kpc}$;][]{GRAVITYCollaboration2021}, makes the CMZ an ideal laboratory for testing theories of star formation and the affects of gas dynamics in different environments \citep[e.g.][]{Hennebelle2011,Padoan2011,Federrath_Klessen_2012,Burkhart2018}.

On average, most of the dense gas in the CMZ follows elliptical x$_2$ orbits at a galactocentric radius of approximately 150~pc \citep{Kruijssen2015,Tress2020, Walker2025, Lipman2025}. Some of this gas may eventually form stars, or perhaps experience collisions and other disruptive events, which help funnel gas towards the circumnuclear disk \citep[CND;][]{Tress2020,Sormani2020,Lu2021,Feng2026}, a region of dense gas and dust orbiting the supermassive black hole, \sgra, with an inner radius of \SI{1.5}{pc} \citep{Mills2013} and outer extent between 3--5~pc \citep{Hsieh2017}. 
%The exact extents of both the CMZ orbital ring and the CND are not strictly defined, these areas are both morphologically and kinematically complex along the line of sight. 

The kinematic complexity of the CMZ and bar lane regions is highlighted in position-velocity (PV) space, where molecular gas tracers, such as CO, reveal bright vertically spiked structures, referred to as Extended Velocity Features \citep[EVFs;][]{Sormani2019b}. EVFs are massive gas features with extreme velocity dispersions ($\Delta V \geq$ \SI{100}{\kms}) located in the Galactic bar region, suspected to arise from high velocity cloud collisions ($\sim$\SI{200}{\kms}) due to material flowing toward or overshooting the CMZ \citep{Sormani2019b,Hatchfield2021}. 
% I think the magnetic loops thing would take too long to explain. I'm just going to leave it out - SG
%, or at the base of magnetic loops \citep{Fukui2006}. 
EVFs within our Galaxy include Bania's Clump 2 \citep[BC2;][]{stark1986, Sormani2019b} and G5 \citep{Liszt2006, Sormani2019b, Gramze2023}, both of which seem to be associated with the result of collisions between clouds of mass $\geq$ 10$^{4}$~\Msun \citep{Kumar1997, Riffert1997}. Extragalactic examples of EVFs have also been identified as giant molecular cloud complexes tens of parsecs in diameter \citep{Kolcu2025}.
%These EVFs are tens of parsecs in diameter, making up giant molecular cloud complexes.  Additionally, EVFs are limited to the centers of galaxies, with extragalactic examples recently identified \citep{Kolcu2025}. EVFs do not appear in the Galactic disk, as they are most likely associated with bar lane dynamics as gas accretes onto and overshoots the CMZ \citep{Sormani2019b,Hatchfield2021}. 

%\begin{figure*}[t!]
%
%\includegraphics[width=1\textwidth]{by_eye_identification.pdf}
%
%\caption{Showing the overlapping position-velocity cuts that were used to identify EVF 14}
%\label{fig: detection_example}
%\end{figure*}

While EVFs are associated with massive ($\sim 10^{4}$~\Msun) clouds in the bar regions, smaller vertically spiked velocity features ($\sim 10^{3}$~\Msun) are seen on sub-cloud scales within the CMZ, and are associated with High Velocity Dispersion Compact Clouds (HVCCs). HVCCs were first identified in \citet{Oka1998} as compact clouds ($d<$ \SI{10}{pc}) with large velocity widths ($\Delta V \geq$ \SI{30}{\kms}). HVCCs are often thought to be signatures of dynamical or stellar feedback in the CMZ, or have been associated with larger features such as BC2, which may consist of multiple smaller cloud collisions \citep{Sormani_and_Barnes_2019}. This connection suggests HVCCs are linked to, or form part of, the ultra-fast cloud–cloud collisions driven by the bar lanes and accretion flows onto x$_2$ orbits. 

A catalog of visually identified HVCCs in the CMZ was presented by \citet{Nagai2008} and expanded by \citet{Oka2012}, where they were interpreted as expanding feedback events, such as supernovae, disrupting average gas flows. More recently, \citet{Oka2022} identified HVCCs algorithmically in single-dish CO J = 3 -- 2 line data. 
%and introduce a different selective criteria using velocity dispersions (based on FWHM) $\sigma_{\rm{V}} \geq$ \SI{20}{\kms}. 
They visually characterized the morphology of the PV diagrams as simple, shell, wing, bridge, and complex shapes, arising from various physical processes, such as cloud-cloud collisions, stellar feedback, or shearing. Similar compact, broad velocity features have been identified in giant molecular clouds outside of the CMZ \citep[e.g. see][]{Makita2026}, and show evidence of strong dynamical shearing. Within the context of the CMZ, these turbulent events can disrupt gas flows, allow for transfer of gas between inner x$_2$ orbits \citep[e.g.][]{Lipman2026}, and instigate accretion flows toward the CND or \sgra.

%Among these, only a single clump was identified within the region surveyed with ACES, the well-studied HVCC CO 0.02-0.02 (or G0.02) \citep{Oka1999, Iwata2023, Pare2024, Sofue2025}, which is suggested to be dense gas following an eccentric orbit near Sgr A$^{\star}$ \citep{Sofue2025}. HVCCs were later explored further in \citet{Oka2012}, where they were given a more selective definition as High Velocity Dispersion Compact Clouds (HVCCs) in \citet{Oka2022} with $\Delta V \geq$ \SI{50}{\kms}. 

%Meanwhile, G0.02 may be its own class of object, as it is spatially more compact (d$<$\SI{10}{pc}) and in close proximity to Sgr A$^{\star}$, and it does not seem to be associated with a cloud-cloud collision \citep{Iwata2023}. 

%Of the four HVCCs identified in \citet{Oka1998}, the three outside of the ACES field of view are associated with cloud-cloud collisions \cite{Busch2022} or EVFs thought to contain cloud-cloud collisions \citep{Sormani2019}, with BC2 being associated with two. This hints toward the possibility of HVCCs being associated with, or components of, the ultra-fast cloud-cloud collisions associated with the bar lanes and accretion onto the CMZ. Meanwhile, G0.02 seems to be its own class of object, as it is spatially much smaller (d$<$\SI{10}{pc}) and far closer to Sgr A$^{\star}$, as well as seemingly not being associated with a cloud-cloud collision. 

In this study, we use data from the Atacama Large Millimeter/submillimeter Array (ALMA) CMZ Exploration Survey (ACES) \citep{Longmore2026_ACESI} to identify and categorize compact HVCCs seen in the \CS~molecular line data using a combination of visual and algorithmic methods to confirm features. The use of high resolution, interferometric data from ALMA provides enhanced spatial sensitivity and spectral information for CMZ-wide identification of HVCCs, and greatly improves upon previous single-dish observations. %We aim to present a catalog of HVCC candidate structures in ACES to provide a sample to study possible physical origins of these features.  

In addition to presenting a catalog of HVCC candidate structures in ACES, we aim to differentiate between properties used for identifying HVCCs, which vary considerably in the literature. Past studies use either observed velocity widths ($\Delta V$) or velocity dispersion from FWHM estimates ($\sigma_V$), with different thresholds for identification. Velocity widths of $\Delta V >$ \SI{30}{\kms} \citep{Oka2012} and $\Delta V >$ \SI{50}{\kms} \citep{Nagai2008} have both been used for identifying HVCCs, though reported dispersions are on average $25 < \sigma_{\rm{V}} <30 $~\SI{}{\kms} \citep{Oka2022}. Additionally, some HVCC analogs have smaller velocity extents of $\Delta V \sim 25$~\SI{}{\kms} \citep{Yokozuka2021}. Many of the features identified in single-dish data become resolved in interferometric data such as ACES, making these velocity width and dispersion definitions difficult to apply consistently. Thus we introduce the term `verticality' as the ratio of velocity extent to longitudinal extent ($\Delta V/\Delta\ell$), as a new metric for identification of HVCCs (See Section~\ref{subsec: identification_overview}).

%The ACES field covers the entire CMZ in high spatial resolution for multiple molecular gas tracers, such as \CS and \HNCO. 
%Viewing the data in PV space reveals a plethora of bright, vertically spiked features, which can be identified in post longitude-velocity and latitude-velocity space to locate on the full spatial mosaic. 
 %As the objects in our sample are restricted to the CMZ and are smaller than \SI{10}{pc} in size, we give them the classification of HVCCs, while keeping in mind that these objects may have a variety of origins.

\begin{table*}
\centering
\caption{ACES molecular line data used for this study. Columns include the species, velocity resolution, rest frequency, upper energy level (E$_\mathrm{u}$/k$_\mathrm{B}$), Einstein A coefficient (A$_{\rm ul}$), and key notes about each tracer as discussed in Sec.~\ref{sec: Data}.}
\label{tab:chem_table}
\begin{tabular}{lccccc}
\hline
Species & Resolution & Frequency & E$_\mathrm{u}$/k$_\mathrm{B}$ & A$_{\rm ul}$ & Tracer Notes \\
&  (\kms)  & (GHz)   & (K)    & 10$^{-5}\cdot$(s$^{-1}$) & \\

\hline

\CS      &    1.5     &  97.9810 & 7.1 &  1.7 & dense/shocked gas, resistant to photo-dissociation   \\
\HNCO & 0.2    &  87.9252 & 10.6 & 0.9  & dense gas, low-vel. shocks, sensitive to photo-dissociation \\
\SiO     &    1.7     &  86.8470 & 6.3 & 2.9 & high-velocity shocks, resistant to UV destruction\\
\HCtrN & 1.5 & 100.0764 & 28.8 & 7.8 & low-velocity shocks, susceptible to photo-dissociation \\

\hline
\end{tabular}
\end{table*}

In Section~\ref{sec: Data}, we describe the data used in this paper. Section~\ref{sec:Methods} covers the methods used for identification and analysis of the HVCCs. In Section~\ref{subsec: results_structure_properties} we present the physical, kinematic, and chemical properties of the structures, and in Section~\ref{Results: PPV} we present a view of the structures in PPV space. Sections~\ref{subsec: discuss_general_properties} and \ref{subsec:discuss_origins} discuss the general properties of the sources and the potential origins which may lead to these kinematic features. Sections ~\ref{subsec: discuss_50kmsC} and \ref{subsec: discuss_IMBH} present overlap of our sampled sources with structures associated with the Circumnuclear Disk (CND), or Intermediate Mass Black Hole (IMBH) candidates in the CMZ. We discuss our uncertainties in Section~\ref{sec: uncertainties}, including possible identification biases and foreground confusion. We end with a summary of our conclusions in Section~\ref{sec: Conclusions}. The analysis and figure-making code developed for this study are available publicly at \href{https://github.com/ACES-CMZ/ACES_EVF_HVCC}{https://github.com/ACES-CMZ/ACES\_EVF\_HVCC}.

\section{Data}\label{sec: Data}

\subsection{The ACES survey and \CS line data}

We utilize large survey data taken with the Atacama Large Millimeter/submillimeter Array (ALMA). The ALMA CMZ Exploration Survey (ACES) is a Band 3 ALMA large program completed in Cycle 8 \citep{Longmore2026_ACESI}. ACES observed a contiguous region in the Galaxy's Center covering a Galactic longitude of approximately $-0.6$\deg$<\ell<0.9$\deg~and a Galactic latitude of approximately $-0.3$\deg$<b<0.2$\deg~with a total area of 1286 square arcminutes. ACES observed both broad-band continuum at 3mm \citep{Ginsburg2026_ACESII} and spectral line windows with varying velocity resolution \citep{Walker2026_ACESIII,Lu2026_ACESIV,Hsieh2026_ACESV}, and synthesized images from the 12m, 7m, and total power array of ALMA. The resulting spatial resolution (full width at half maximum) is around 2\arcsec~(or 0.1~pc at the assumed CMZ distance used throughout this work of 8.2~kpc; \citealt{Reid2019,Gravity19,GRAVITYCollaboration2021}).

%The ACES data cover six spectral windows: two with narrow (59 MHz) width, two with medium (469 MHz) width, and two with broad width (1.875 GHz). The narrow-width spectral bands are described in \citet{Walker2026_ACESIII}, and highlight the HNCO (4-3) and \hcop (1-0) data, which were observed at high spectral resolution \SI{0.2}{\kms}. The medium-width spectral bands are described in \citep{Lu2026_ACESIV} and include \SiO, \SO, \HthCOp, \HthCN, \HNthC, and \HCFN, among other molecular line transitions, with a velocity resolution of \SI{1.7}{\kms}. The broad-width spectral bands are described in \citep{Hsieh2026_ACESV} and include the \CS, \SO, \chtcho, \HCtrN, and \ha lines at a velocity resolution of about \SI{1.5}{\kms}.

The ACES data cover six spectral windows: two with narrow (59 MHz) width, two with medium (469 MHz) width, and two with broad width (1.875 GHz) \citep[see data reduction papers for details][]{Walker2026_ACESIII,Lu2026_ACESIV,Hsieh2026_ACESV}. ACES data products are available at \url{https://almascience.org/alma-data/lp/aces} \citep[see][for details on the ACES survey, data reduction code, and processing pipelines]{Longmore2026_ACESI}.

In this study, we primarily use the ACES \CS data for structure identification. The top panel of Figure~\ref{fig: Identification} shows the area covered with ACES and the CS emission. \CS is an exemplary bright and abundant broad-width line in ACES, with a spectral resolution of \SI{1.5}{\kms}. CS has been used to trace dense, translucent gas \citep{Martin2008,Goldsmith2017,Humire2020,Rod-Baras2021}, and is not very sensitive to changes in cosmic ray ionization rates, temperatures, or densities \citep{Requena-Torres2006,Dutkowska2025}. Additionally, \CS may be used for identifying outflows \citep{Bouvier2024,Gramze2025} and velocity bridges associated with cloud-cloud collisions \citep{Busch2022, Barnes_Priestley2024}, which result in bright, spiked PV features. The association of \CS with high-velocity shocks from outflows and collisions, and its brightness throughout the CMZ, make it a particularly interesting line to search for HVCCs and explore their physical origins.

For the algorithmic identification of structures discussed in Section~\ref{subsec: dendro_identification}, we used a downsampled version of the ACES \CS cube \citep[see][]{Hsieh2026_ACESV} due to computational limitations and inefficiencies when using the full-resolution datacube. We downsampled the cube spatially by a factor of 9; the emission is then spatially smoothed to a resolution of 5\arcsec. The spectral resolution is maintained at \SI{1.5}{\kms}. We estimate the noise in the downsampled \CS ~data using the median absolute deviation, $\sigma_{\text{MAD}}\sim$ \SI{0.01}{Jy/beam}. Individual full-resolution subcubes were used after initial identification to obtain detailed physical and kinematic properties of structures. However, we note the spatial downsampling limits the minimum spatial size of detectable objects to $\sim$\SI{0.5}{pc} using the methods described (see Section~\ref{sec: uncertainties}).

\subsection{Additional ACES molecular lines}

Further exploration of the identified structures in \CS was done using three other ACES molecular tracers: \HNCO,  \SiO, and \HCtrN, which are observed with spectral resolutions of \SI{0.2}{\kms}, \SI{1.7}{\kms}, and \SI{1.5}{\kms}, respectively. We present brief descriptions of these species here, and direct the reader to \citet{Dutkowska2025} for an extensive analysis and modeling of these lines in the ACES survey.

\HNCO is a tracer of dense gas \citep{Martin2008,Kelly2017}, often associated with extinction magnitudes A$_v\sim10$~mag in simplified models \citep[e.g.][]{Tideswell2010}. It is formed and enriched by sublimation off icy dust grains due to soft shocks in the gas phase, and thus traces weaker shocked regions \citep[see][for a detailed description of \HNCO formation and its behavior in the CMZ]{Tideswell2010}. While HNCO is abundant in shocked and post-shock gas, it shows decreasing trends with cosmic ray ionization rates in protostellar environments \citep{Dutkowska2025}, and is quite sensitive to UV radiation compared to \SiO \citep{Martin2008}. 

\HCtrN, forms efficiently within the gas phase at warmer temperatures in turbulent high-pressure clumps \citep{Mills2018a}, and is enhanced in hot, dense cores \citep{Caselli1998,Chapman2009,Tanaka2018,Taniguchi2019,Dutkowska2025}. It is associated with low-velocity shocks in the CMZ \citep{Tanaka2018}, and is fairly susceptible to photo-dissociation \citep{Martin2012}.

\SiO serves as a tracer for mid- to high-velocity shocks ($>$ \SI{20}{\kms}) within molecular clouds \citep{Martin-Pintado1992}, and is widespread throughout the CMZ \citep{Martin-Pintado1997,Takekawa2024}. SiO may be released directly from grains in high-velocity shocks \cite[e.g.][]{Martin-Pintado1992} or formed in the gas phase following the liberation of atomic Si \citep{Schilke1997,Gusdorf2008,Miettinen2014}. Additionally, SiO abundance is observed to be enhanced by several orders of magnitude in shock-affected regions \citep{Martin-Pintado1992} due to sputtering of dust grains \citep[e.g. see][]{Jimenez2008}. \SiO~also possesses a lower photo-dissociation rate \citep[e.g. see][]{Dutkowska2025} compared to the bulk dense gas tracer HNCO. %SiO is also very difficult to eject from the cores of dust grains, requiring high velocity shocks. 
Conversely, HNCO results from the release of ice mantles, which are more easily sputtered in low-velocity shocks \citep[e.g.][]{Tideswell2010} and destroyed by UV radiation \citep{Rico-Villas2020}. Because SiO is resistant to UV destruction, as well as changes in temperature or density, it is useful for tracing stronger, faster shocks compared to HNCO, which can only be enhanced by slow shocks without being destroyed \citep{Kelly2017, Huang2023}. 
\\

We present a summary of the lines used in this study in Table~\ref{tab:chem_table}, including the species name with transitions, spectral resolution, rest frequency, upper-level energy (E$_\mathrm{u}$/k$_\mathrm{B}$), and Einstein A coefficient (A$_{\rm ul}$).

%The initial ACES paper series (ACES I-V) describes an overview of the ACES survey \citep{Longmore2026_ACESI}, the continuum data \citep{Ginsburg2026_ACESII}), narrow spectral windows \citep{Walker2026_ACESIII}, intermediate spectral windows \citep{Lu2026_ACESIV}, and broad spectral windows \citep{Hsieh2017}. 

%\section{Defining HVCCs and Size Scales}
%Previous definitions of EVF vs HVCC. Are HVCCs just smaller versions? EVFs are usually large-scale GMCs (~1e4Msun) falling into galactic centers. HVCC seems to be observed within or between cloud collisions. 
%HVCCs as described in Oka+2011 (investigate first instance) (Oka 1998) %https://ui.adsabs.harvard.edu/abs/1998ApJS..118..455O/abstract)
%Compact d$<$ 10pc
%Broad deltaV $>$ 30km/s (FWHM? Sigma? extents?) 

\section{Methods}\label{sec:Methods}

\subsection{Methods of identification}

The identification of thin, high-velocity structures in the CMZ has largely relied on visual inspection of prominent PV features \citep[e.g.][]{Sormani_and_Barnes_2019,Oka1998,Nagai2008,Oka2012}. Automated approaches for identifying HVCCs in previous CO (J = 3 -- 2) data have been introduced by applying velocity and spatial masks to highlight extended features, and then using a modified clumping algorithm to extract separate structures \citep{Oka2022}.

%In addition to the variety of identification methods, the definitions used to categorize HVCCs vary considerably in the literature. Past studies use either observed velocity widths ($\Delta V$) or velocity dispersion from FWHM estimates ($\sigma_V$), with different thresholds for identification. Velocity widths of $\Delta V >$ \SI{30}{\kms} \citep{Oka2012} and $\Delta V >$ \SI{50}{\kms} \citep{Nagai2008} have both been used for identifying HVCCs, though reported dispersions may be on average $25 < \sigma_{\rm{V}} <30 $~\SI{}{\kms} \citep{Oka2022}. Additionally, some HVCC analogs have smaller velocity extents of $\Delta V \sim 25$~\SI{}{\kms} \citep{Yokozuka2021}. Many of the features identified in single-dish data may become resolved in interferometric data such as ACES, making these velocity width and dispersion definitions difficult to apply consistently. Thus, we introduce the term `verticality' as the ratio of velocity extent to longitudinal extent ($\Delta V/\Delta\ell$), as a new metric for identification of HVCCs (See Section~\ref{subsec: identification_overview}).

We began identification of structures in the ACES \CS data via visual inspection of thin longitude and latitude slices of the full-resolution cube (Section~\ref{subsec: identification_overview}). We then present 3D dendrograms (Section~\ref{subsec: dendro_identification}) as a potential automated identification procedure by confirming the visual detections and appending additional structures from the dendrogram survey that were missed from the visual detections. The use of both methods verifies which visually identified structures are significant detections above the noise level, as well as confirms the utility of dendrograms as an automated detection method, which utilizes the full spectral cube to identify structures in a single step without excess masking. %The following sections present detailed explanations of the visual (Section~\ref{subsec: identification_overview}) and dendrogram (Section~\ref{subsec: dendro_identification}) identification of sources.

\subsubsection{Visual identification}\label{subsec: identification_overview}

\begin{figure*}[t!]
\begin{minipage}{\textwidth}
\includegraphics[width=1\textwidth]{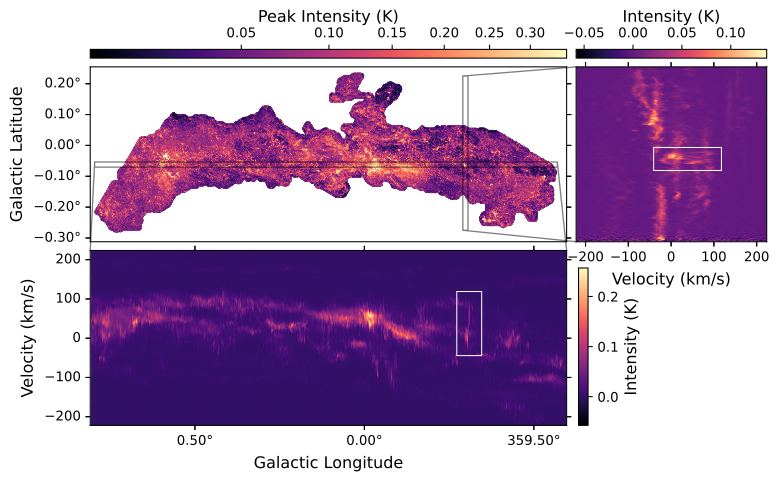}
\end{minipage}
\caption{We identify extended structures via both visual and algorithmic methods. Example schematic of visual identification for HVCC structures in the ACES \CS peak intensity map (top left) using LV (bottom) and BV (top right) slices. The LV diagrams are taken at a specific latitude slice across all longitudes, and the BV diagrams are taken at a specific longitude slice across all latitudes. The white boxes in the LV and BV diagrams showcase an example spiked feature that was identified as a prominent HVCC candidate. By locating a spike on the LV diagram, we open the corresponding BV diagram at the longitude slice where the feature was found. If the V$_{LSR}$ and velocity extents match, these plots can be used to pinpoint an HVCC on the ACES \CS peak intensity map in LB space.}
\label{fig: Identification}
\end{figure*}

We identify spiked, vertical features in the ACES \CS data by visually inspecting thin PV slices of the ACES \CS peak intensity map. An example of the identification strategy is shown in Figure~\ref{fig: Identification}. Longitude-velocity (LV) diagrams were made over thin slices at different latitudes of 1\arcmin width (corresponding to a spatial size of 2--3~pc), ensuring the visual inspection recovers compact structures in latitude ranges enclosing the typical size-scales for HVCCs. Features that are compact in longitude and extended in velocity were then searched for in the corresponding latitude-velocity (BV) diagram. If a given feature matched an evident source in position space, and showed corresponding central velocity and intensity, it was identified on the plane-of-sky in longitude-latitude (LB) space and added to the catalog. 
%The visual identification team included 12~people (hereafter, ``inspectors") who were divided into 5~groups. 
%Each group was responsible for searching through a given 
To break up the visual identification, we split the ACES field of view into multiple tiles of longitude range, where: tile~1 includes longitudes from $-$0.590\deg to $-$0.307\deg, tile~2 is from $-$0.307\deg to 0.007\deg, tile~3 is from 0.007\deg to 0.293\deg, tile~4 is from 0.250\deg to 0.660\deg, and tile~5 is from 0.602\deg to 0.827\deg. 
Two researchers were assigned to each tile. For each structure identified in their respective tiles, they reported the central ($\ell$,$b$) coordinate, spatial extents ($\Delta \ell$ and $\Delta b$), central radial velocity (\Vlsr), and velocity width ($\Delta V$). 
%Inspectors reported for each structure identified in their respective tiles.

The initial catalog was filtered to remove any duplicates within  0.01\deg and \SI{15}{\kms} of the central position and velocity, respectively; keeping the feature with the larger velocity extent. The catalog was also filtered by the ratio of velocity extent to longitudinal extent, effectively quantifying the vertical nature of each source, which we term `verticality'. Note, the verticality term is distinct from a `velocity gradient', which implies directly measuring the $dv/d\ell$ slope of each source, rather than the more generalized procedure used here. 

We set a minimum verticality ratio of \SI{1000}{\kms deg^{-1}}, meaning a feature with an angular size $<$0.02\deg in Galactic longitude must exhibit a velocity width of at least \SI{20}{\kms}. %comparable to the average velocity width of \SI{30}{\kms} identified in lower-resolution CO (J = 1 -- 0) data from \citet{Oka1998}.This verticality threshold was applied to the by-eye detections to filter out any candidates that are not visually extended in velocity. 
Overall, the visual inspection yielded a total of 163~by-eye detections. % An initial visual inspection returned 125 candidates. %However, it was apparent that the by-eye inspection missed several candidates in the denser regions in tile 3. We therefore assigned another inspector to resample tile~3, and added 38~additional candidates to the catalog. Overall, the visual inspection yielded a total of 163~by-eye detections. 

While the visual identification recovered a number of HVCCs with velocity and spatial extents comparable to previous studies \citep[e.g.][]{Oka2022}, we note that the visual identification is not fully complete, as it misses less distinct features. Additionally, some visual identifications may not be significant detections above local noise levels, or can consist of separate smaller structures reported as one. We thus explore 3D dendrograms as an automated detection method to both confirm and supplement the visual identifications. %The next section presents the use of a 3D dendrogram algorithm to search for structures which were not detected by the visual inspection. %which confirms the noise-significance of visually identified structures and computes physical properties more accurately than the by-eye method allows. 

\begin{figure*}[t!]
\includegraphics[width=1\textwidth]{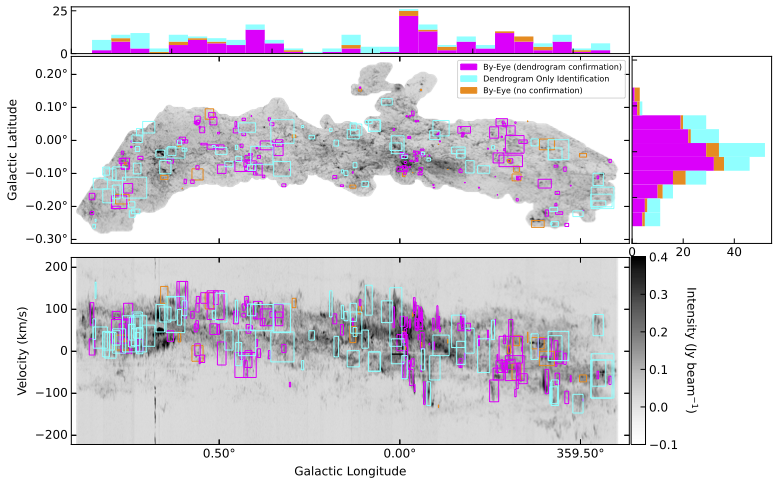}
\caption{We utilize a dendrogram method to both confirm visual detections and add complementary sources to our catalog. (Top) An overview figure showing the spatial distribution of candidates identified in our catalog overlaid on a peak intensity map of the \CS ~cube. The by-eye sources that are confirmed by the low-resolution dendrograms are indicated with the magenta rectangles. The cyan rectangles represent the sources that were detected by the low-resolution dendrogram, but not from visual inspection. The extents of the regions are taken from the dendrogram contour corresponding to the peak intensity velocity slice of the structures from the downsampled \CS ~cube. The orange rectangles indicate the HVCCs that were identified by eye, but had no matching dendrogram contour within the visually reported region. The marginal histograms show the distribution of sources in Galactic longitude and latitude. (Bottom) The \CS ~peak intensity longitude-velocity map. The magenta and cyan rectangular regions represent the $\ell, v$ extents at the peak intensity latitude slice of the extracted dendrogram structures. The orange regions correspond to the by-eye longitude and velocity extents for candidates without a dendrogram match. } 
\label{fig: dendro}
\end{figure*}

\subsubsection{Complementary dendrogram identification}
\label{subsec: dendro_identification}

In addition to the visual inspection, we aim to confirm that dendrograms are a suitable automatic detection method for HVCCs. The goals of the complementary dendrogram analysis are to confirm the structures that were identified visually, as well as filter noisy detections and assess if dendrograms identify features that were missed by the visual inspection. 

The automated identification of structures was performed using the \verb|astrodendro| Python package\footnote{Detailed information on the \texttt{astrodendro} Python package can be found at \url{http://www.dendrograms.org/}.}, which generates dendrograms from astronomical data. Dendrograms are a tree-like representation of the hierarchical structure present in a given data set, including ``branch" and ``leaf" structures, which represent isocontours containing substructure, and those at the top of the structure hierarchy, with no additional substructure, respectively \citep{Rosolowsky2008b}. 

%To identify structures that the visual detection may have missed, we used \verb|astrodendro| to 
We generate 3-dimensional dendrograms from the downsampled \CS ~data cube (see Section~\ref{sec: Data}), providing an automated prescription for identifying coherent structures in position-position-velocity (PPV) space. The noise of the cube is estimated using the mean absolute deviation $\sigma_{\text{MAD}} \sim$\SI{0.01}{Jy/bm}. %We used the downsampled \CS ~cube as this is computationally inexpensive and the identification of structures that correspond to by-eye detections does not require the high-resolution data product. 

The \verb|astrodendro| algorithm requires three basic input parameters to generate a dendrogram:

\begin{enumerate}
    \item \verb|min_val|: the minimum allowed pixel value for a given structure
    \item \verb|min_sig|: the difference a peak leaf pixel value must be in comparison to an adjacent structure to be considered an independent structure.
    \item \verb|min_pix|: the minimum number of pixels that a structure can contain.
\end{enumerate}

In our implementation, we use \verb|min_val| = 15$\sigma_{\text{MAD}}$, \verb|min_sig| = 10$\sigma_{\text{MAD}}$, and a \verb|min_pix| = 2, corresponding to the number of pixels within one synthesized beam in the downsampled \CS ~cube. The final parameters were chosen based on a comparison to the visual identification, ensuring sensitivity to at least 80\% of detections in the visually identified sample. The high significance values ensure we only kept robust candidates in the catalog. The dendrogram output was then filtered using the same verticality threshold as the visual catalog. 

We then determined the overlap between the dendrogram detections and the visually identified candidates. The dendrogram returned 3D contours including the volume pixels (voxels) associated with each coherent structure in ($\ell$,$b$,v) space. If a region reported by the visual inspection contained the central voxel identified by the low-resolution dendrogram, it was considered a match between the two methods. 

After filtering for verticality and velocity widths, we find that 140~regions from the visual identification contained a coherent structure identified by the dendrogram of the downsampled cube. 23 visually identified regions did not contain a dendrogram identification, having failed to meet thresholds for the filtering or input parameters (e.g. verticality, velocity dispersion, min\_val, min\_sig, or min\_pix). These non-identified regions appear either noisy or spatially smoothed in the low-resolution cube. We keep these structures in the catalog to investigate their morphology and local noise further in the high-resolution data, and better understand the limitations of both the visual and dendrogram detection methods. 

The dendrogram identified a total of 20192~unique structures that were not found by visual inspection. Since the dendrogram aimed to identify coherent features in PPV space and did not initially filter based on verticality or spatial size, most of these structures are spatially extended or have too narrow velocity ranges to be considered HVCCs. We filter the output to ensure only the most confident detections, keeping any structures with a velocity width (i.e. velocity extents of the enclosed structure) of at least \SI{50}{\kms}, a size of $d<10$~pc, and verticality of $>1000~\mathrm{km\,s^{-1}}\mathrm{deg}^{-1}$, selecting the largest structures within a given nested sub-structure (see Section~\ref{sec: uncertainties} for discussion of confusion effects on identification and filtering). The filtering resulted in 72~additional candidates with no corresponding visually identified structure, which were added to the catalog. The visual inspection and dendrogram identification resulted in a combined total of 235 sources.\\

\begin{figure*}[t!]
\begin{center}
\includegraphics[width=\textwidth]{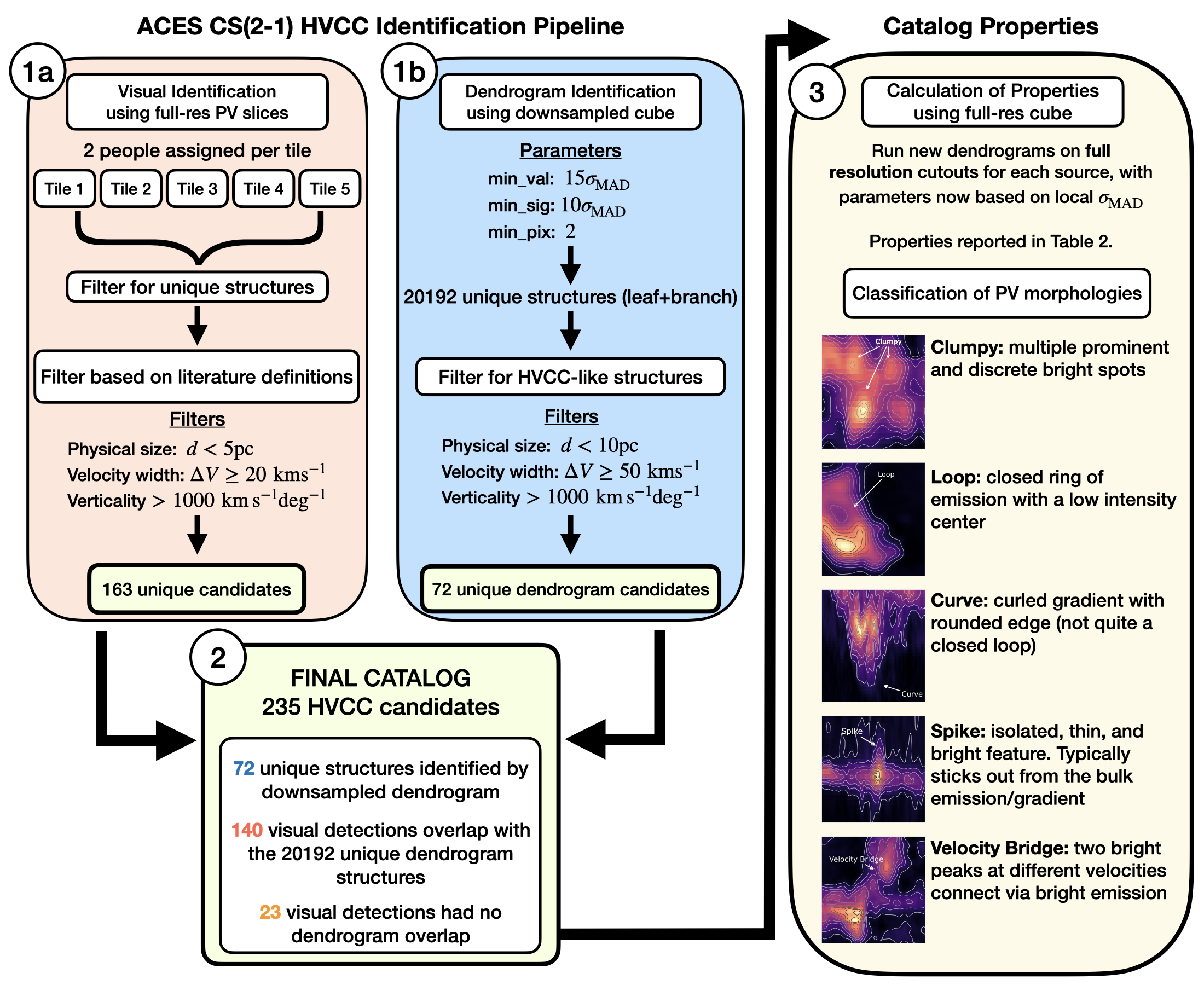}\label{fig:flowchart}
\caption{The identification of HVCC candidates and catalog classifications are summarized in the flowchart above. We use both visual identification of the full-resolution \CS cube (1a, Section~\ref{subsec: identification_overview}) and algorithmic dendrogram identification of the downsampled data using dendrograms (1b, Section~\ref{subsec: dendro_identification}) to create the final catalog of HVCCs (2). We then use full-resolution subcubes to calculate physical properties and classify the PV morphologies of each source (3, see Sections~\ref{final_cat_ppv} -~\ref{sec:catalog_morph_classification}).}
\end{center}
\end{figure*}

The left side of Figure~\ref{fig:flowchart} provides a schematic summary of the pipeline for HVCC identification, including the visual (block 1a) and dendrogram (block 1b) methods, leading to a total of 235 candidates (block 2). In the following subsections, we describe the right side of the schematic (block 3) for creating the final catalog, including calculation of physical and kinematic properties for the sources and classifying the morphologies of emission in their PV diagrams.

%The next section details the use of dendrograms on high-resolution subcube cutouts of all identified candidates to confirm their significance above local noise levels, and the creation of the finalized catalog.

%The dendrogram identified many more structures in the 0.25 \deg $< \ell <$ 0.50 \deg region than the by eye identification. We resampled tiles 3 and 4 and returned an additional 38 s. 

\begin{figure*}[t!]
\begin{center}
\includegraphics[width=.9\textwidth]{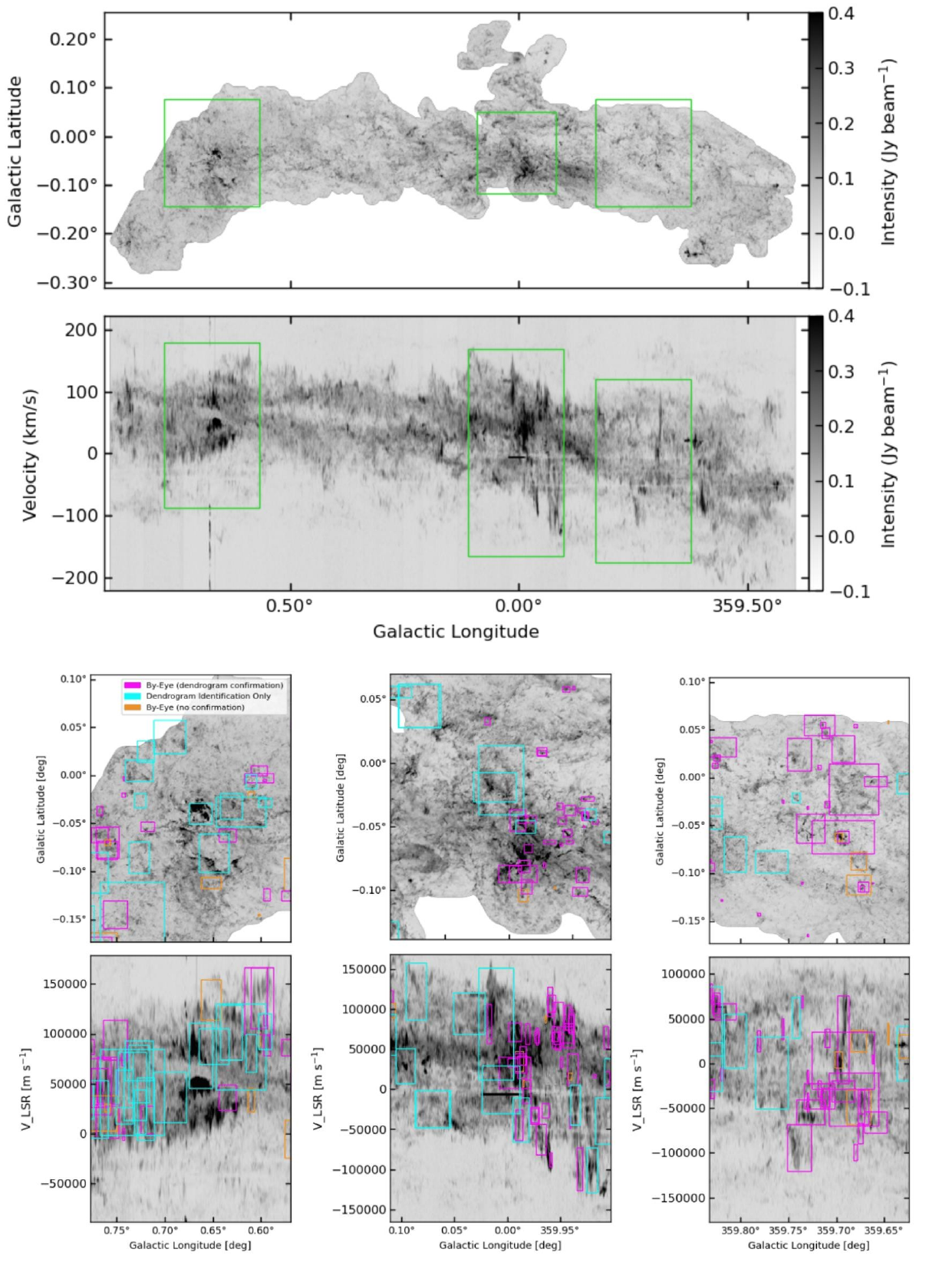}\label{fig:zooms}
\caption{The above figure highlights three zoomed-in regions of our final catalog shown in Figure~\ref{fig: dendro}. The top two panels showcase the \CS~ peak intensity LB and LV maps with green regions corresponding to the zoomed-in LB and LV cutouts in the bottom two rows. The bottom two panels highlight samples of HVCCs near the complex Sgr B2 region (left), Sgr A region where many visibly extended candidates were found (middle), and a more diffuse region at negative longitudes (right).}
\end{center}
\end{figure*}

\subsection{Creation of the final catalog and PPV masks from high-resolution subcubes}
\label{final_cat_ppv}

The 235 identified sources are included in the final catalog, reported in Table~\ref{tab:physical_properties}. Each structure is assigned a name using the identified molecule and central ($\ell$,$b$) position, as well as an ID number from the identification procedures (e.g. ID 1, CS $-$0.261+0.024). To create the final catalog, we investigated each structure in a high-resolution cutout of the \CS data to extract central ($\ell$,$b$) coordinates, \Vlsr, and velocity widths. The coordinates from the subcube analysis are the final values used in the catalog. Additionally, the purpose of the subcube dendrograms is to confirm that detections are above the local noise in the full resolution version of the data.

%Of the 163 by-eye sources, 23 are not confirmed by the confirmation dendrogram. The by-eye detections without a corresponding dendrogram detection are marked as not meeting one of the requirements (verticality, velocity dispersion, min\_val, min\_sig, or min\_pix). 

We used the \texttt{spectral\_cube} Python package \footnote{\url{https://spectral-cube.readthedocs.io/en/latest/}} to create subcubes from the high-resolution \CS~data cube by specifying a spatial region either from the dendrogram detections or visual inspection (for the 23 nondendrogram matches). The subcubes are spatial cutouts based on the $\Delta \ell$ and $\Delta \rm{b}$ extents of each object, with the full velocity range of the \CS mosaic.  

Using the high-resolution subcubes, we ran the same dendrogram procedure from Section \ref{subsec: dendro_identification}. However, the min\_sig parameter was now determined locally for each region, i.e. we use the $\sigma_{\text{MAD}}$ value of the individual subcube as a measure of the local noise, rather than the noise of the full \CS ~mosaic. The local noise of each cube is vastly different than the generalized noise in the whole CMZ, and a dim source may be identified in the high-resolution subcube if its emission is significant compared to the local noise. 

All ``leaf" structures from the dendrogram represent the most compact and densest material in the tree, and are considered a part of the identified structure (See Appendix~\ref{appendix_dendro_example} for an example subcube with final dendrogram contours). We then utilize the 3D ($\ell$,$b$,v) pixel mask outputs from \verb|astrodendro| to create PPV masks of the structures. Each PPV mask saves the leaves with intensities above a given source's median to be used for visualizing the brightest parts of each HVCC in PPV space.

Of the original sample of 235 HVCCs, there are only 11~visually detected structures for which the high-resolution dendrogram returned no contours above the local noise levels, 2 of which were also not confirmed by the low-resolution dendrogram. All other features identified in the high-resolution data fulfill the conditions for verticality, minimum pixels, and minimum intensity to not be excluded from the original sample.

The overall LB and LV distribution of HVCCs identified from the different methods is summarized in Figure~\ref{fig: dendro}. In total, we report 235~coherent structures in ($\ell$,$b$,v) space. Of these, 140 were identified both via visual inspection and dendrogram analysis of the downsampled cube (magenta regions), 23 were identified only by visual inspection (orange regions), and 72 were identified only by the dendrogram analysis of the downsampled cube (cyan regions). The detection methods for each structure are summarized in Table~\ref{tab:physical_properties}.

The top panel of Figure \ref{fig: dendro}, shows the ($\ell, b$) positions and extents of structures overlaid on the full-resolution \CS peak intensity map. Marginal histograms show the latitude and longitude distributions of identified structures. The ($\ell$,$b$) extents of the magenta and cyan regions are taken from the dendrogram contour corresponding to the peak intensity velocity slice of the structures from the downsampled ~\CS~cube. The bottom panel shows where the same cataloged HVCCs are located in LV space, as overlaid on the \CS LV peak intensity map, with extents corresponding to those of the peak intensity latitude slice for the extracted dendrogram structures. The ($\ell$,$b$,v) extents for the non-dendrogram detections are the quantities reported by the visual inspection. 

We also highlight three zoomed-in regions in both LB and LV space in the bottom panels of Figure~\ref{fig:zooms}. The top two panels showcase the \CS peak intensity LB and LV maps with green regions corresponding to the zoomed-in cutouts in the bottom two rows. We highlight samples of HVCCs near the complex Sgr B2 region (left), the Sgr A region where many visibly extended candidates were found (middle), and a more diffuse region at negative longitudes (right). 

While the visual inspection was effective in picking out the brightest sources, the smallest of the narrow structures were often missed by eye, but found by the dendrogram. These structures are identifiable by eye when re-examined, but were missed in the visual inspection, which motivates the use of the dendrograms for automated detection. Conversely, some visual identifications reported very faint or noisy peaks (i.e. the 10 with no high-resolution dendrogram contours discussed above), which returned zero dendrogram contours. In these cases, the dendrograms provide a useful filtering system for spurious by-eye detections.

%Fig~\ref{fig: dendro} presents an overview of the spatial and velocity extents of sources in our final catalog. The extents of the by-eye sources are represented by red rectangles, and the 2-dimensional contours of the dendrogram structures that overlap with them in $\ell, \mathrm{b}$, $v$  space are shown in cyan. In the top panel of Figure \ref{fig: dendro}, we show the $\ell, \mathrm{b}$ positions and extents of structures overlaid on the $\ell, \mathrm{b}$ peak intensity map for the \CS cube. Marginal histograms show the Galactic latitude and longitude distributions of cataloged candidates. In the \textcolor{red}{middle} panel of \ref{fig: dendro}, we also show where the same cataloged candidates are located in $\ell, v$ space, as overlaid on the $\ell, v$  peak intensity map for the \CS cube. We also provide two $\ell, \mathrm{b}$ and $\ell, v$  zoom-in panels at the bottom of \ref{fig: dendro} to showcase the general morphology of these structures across all three dimensions. \textcolor{red}{Will add more details about these zoom-ins and key takeaways once this figure has been officially completed.}

\begin{figure*}[t!]
\includegraphics[width=1\textwidth]{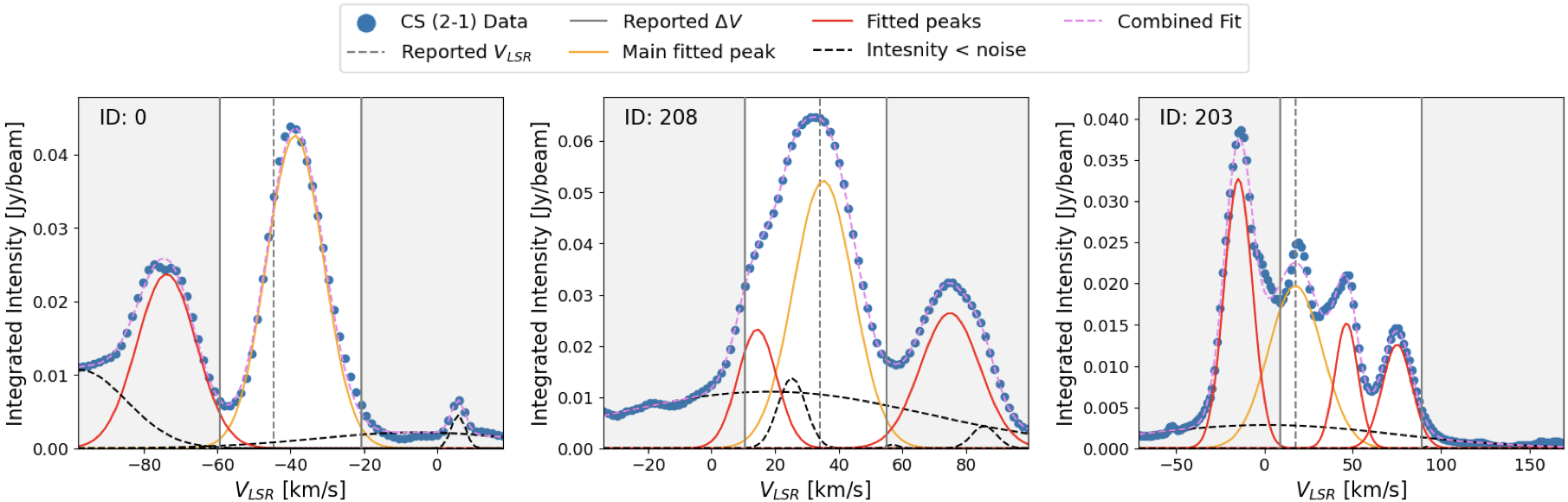}

\caption{Examples of fitted spectra for determining FWHM. Based on the averaged \CS spectra (blue points) for a given HVCC we fit a Gaussian mixture based on the number of identified peaks. The unshaded region indicates the reported $\Delta V$ range for the HVCC. We report the main peak of the fit closest to the reported \Vlsr (orange line), with an offset closest to the reported \Vlsr (grey dashed line) that is above the noise level of the subcube. Other fitted peaks above the noise level are shown in red. Fitted peaks with intensities below the noise of the subcube are plotted as black dashed lines. The combined fitted Gaussian mixture is shown as a dashed violet line.}
\label{fig: Fitted_FWHM_examples}
\end{figure*}

\subsection{Calculation of physical properties}\label{subsec: calculate_physical_properties}

We calculate various fundamental physical properties of the HVCCs and include them in our final catalog, reported in Table~\ref{tab:physical_properties}. %An excerpt of the full catalog is shown in Table~\ref{tab:excerpt_physical_properties}.

For each structure with a high-resolution dendrogram contour, we report the central ($\ell$,$b$) coordinates of the leaf contour, the velocity of the peak intensity value within the leaf mask (\Vlsr), and the velocity range of the leaf contour ($\Delta V$) calculated by taking the difference between the maximum and minimum velocities found by the dendrogram. For the 11~structures with no high-resolution dendrogram contour, we report the by-eye velocity widths and \Vlsr. We also report the verticality for each structure, calculated by dividing the velocity width by the longitude extent of the structure.

%The position of each source in Galactic coordinates was determined during the by-eye detection using PV diagrams, as described in Section \ref{subsec: identification_overview}. The radial velocity ($v_{\text{lsr}}$) we report for each source is the velocity containing the peak intensity value within the dendrogram mask. 
We calculate the physical size of each structure in parsecs, assuming a common distance to the CMZ of 8.2~kpc\footnote{Since distances to the individual objects are not known, assuming a common distance to all structures may impact the calculated effective radii of structures with errors of 5-10\%, assuming the structures could lie between 8-8.2~kpc of the center.}, using the PPVStatistic class in astrodendro to estimate its effective radius ($R_{\text{eff}} = \sqrt{A_{\text{leaf}}/\pi}$), where $A_{\text{leaf}}$ is the area of the identified dendrogram leaf, at the velocity slice where the dendrogram mask has the largest surface area in LB space. Similarly, we obtain the median intensity for each structure using the pixels contained in its dendrogram leaf and report this in our catalog in units of Jy beam$^{-1}$.

%Characterizing the structures in our catalog as HVCCs is dependent on the properties of their averaged spectra. %While we use velocity widths to identify and filter our catalog, similar to the HVCCs reported by \citet{Nagai2008}, \citet{Oka2022} used velocity dispersions to categorize structures as HVCCs. Additionally, 
Many of the structures in our catalog show complex spectra within their reported velocity ranges. To investigate the various velocity peaks, we obtain a fitted FWHM for their average \CS~ spectra. Examples of the fitting, for both single and multi-peaked spectra, are shown in Figure~\ref{fig: Fitted_FWHM_examples}. The fitting is done as follows: using \texttt{spectral-cube}, we create a spectral slab of the individual subcube, using a broad velocity range on either side of the reported minimum and maximum velocity extents, and average along the spatial axes to produce the spectrum (blue points in Figure~\ref{fig: Fitted_FWHM_examples}). The number of peaks in the averaged spectra was identified using \texttt{scipy.signal.find\_peaks\_cwt} to convolve the spectra with a wavelet, using widths $\ge$ \SI{1}{\kms}. We then use the \texttt{LMFIT} library's \texttt{GaussianModel} model \citep{LMFIT_newville_2025_16175987} to create and fit a Gaussian mixture based on the number of identified peaks.

We calculate the FWHM of the resulting Gaussians and report the fitted FWHM of the main peak (orange line) corresponding to the peak closest to the \Vlsr (vertical grey dashed line). We require the main peak to be above the $\sigma_{\mathrm{MAD}}$ noise threshold of the cube in the reported velocity range. Peaks with fitted intensities below the subcube noise are noted by black dashed lines. We also report the number of fitted peaks within the reported velocity range (the unshaded area) above the local noise threshold, which include other peaks above the noise level (red lines). The combined fitting from the Gaussian mixture is shown as a violet dashed line.

\subsection{Morphological classifications}\label{sec:catalog_morph_classification}

\begin{figure*}[t!]
\begin{minipage}{\textwidth}
\includegraphics[width=1\textwidth]{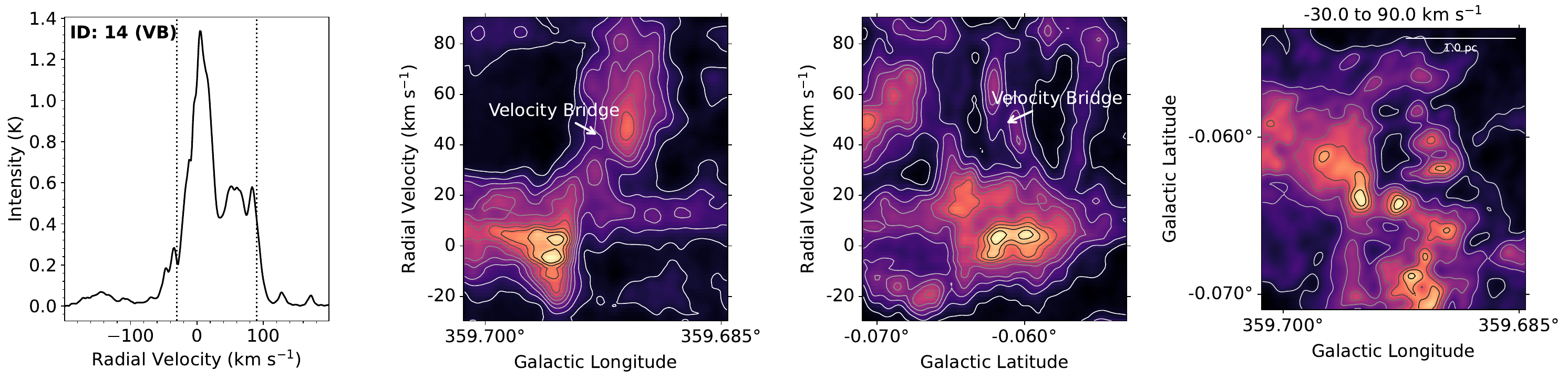}
\end{minipage}
\begin{minipage}{\textwidth}
\includegraphics[width=1\textwidth]{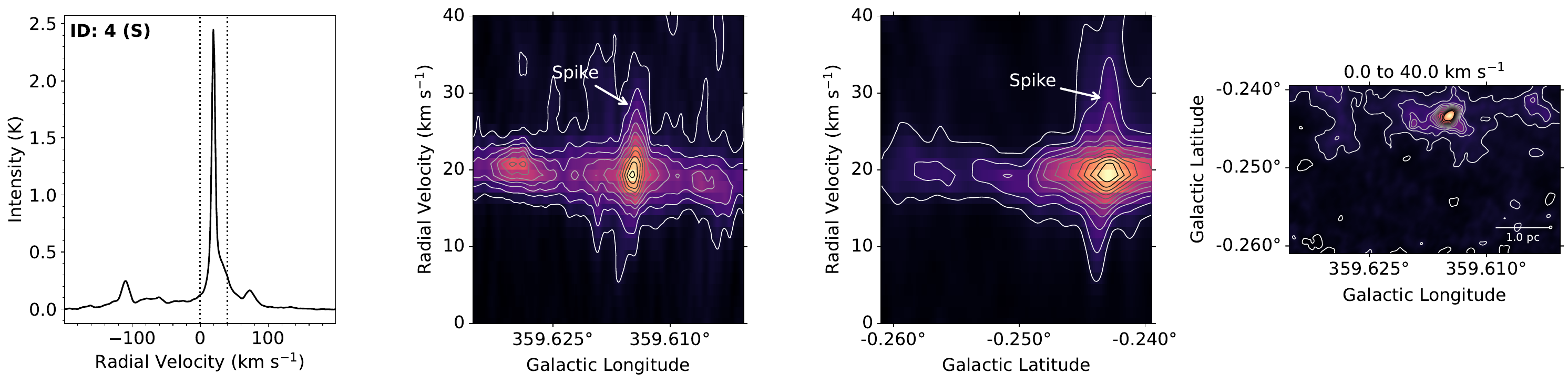}
\end{minipage}
\begin{minipage}{\textwidth}
\includegraphics[width=1\textwidth]{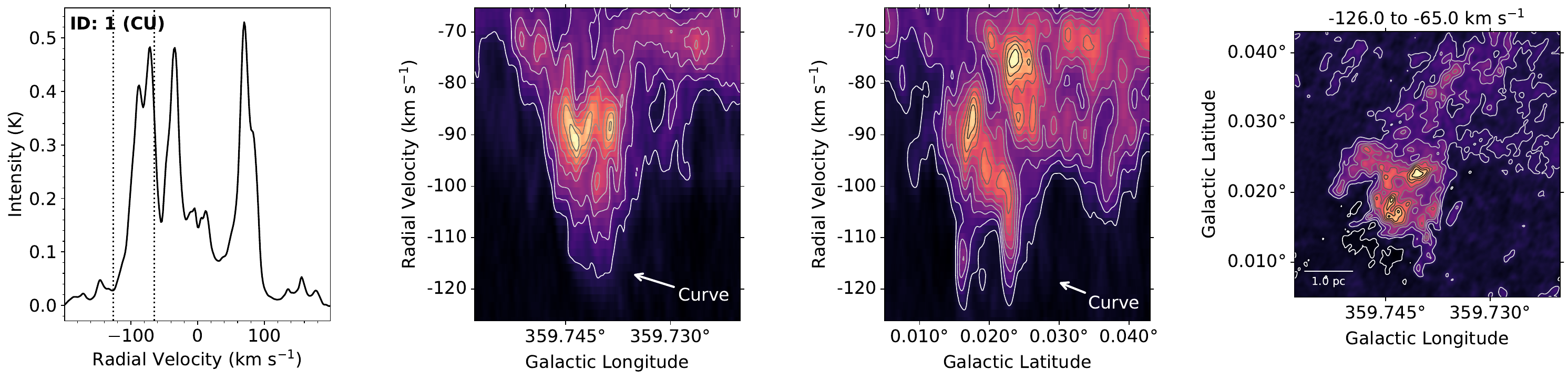}
\end{minipage}
\begin{minipage}{\textwidth}
\includegraphics[width=1\textwidth]{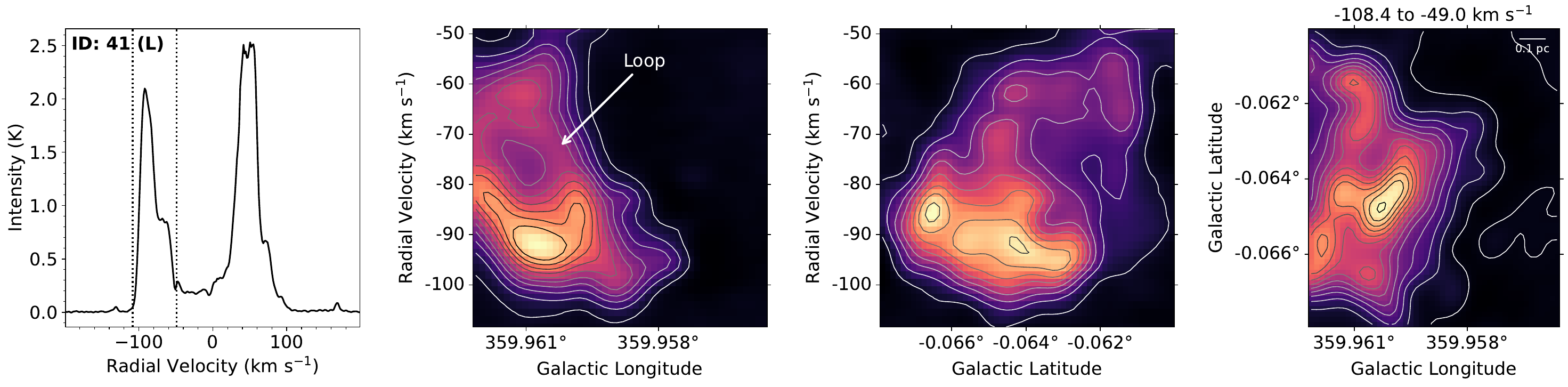}
\end{minipage}
\begin{minipage}{\textwidth}
\includegraphics[width=1\textwidth]{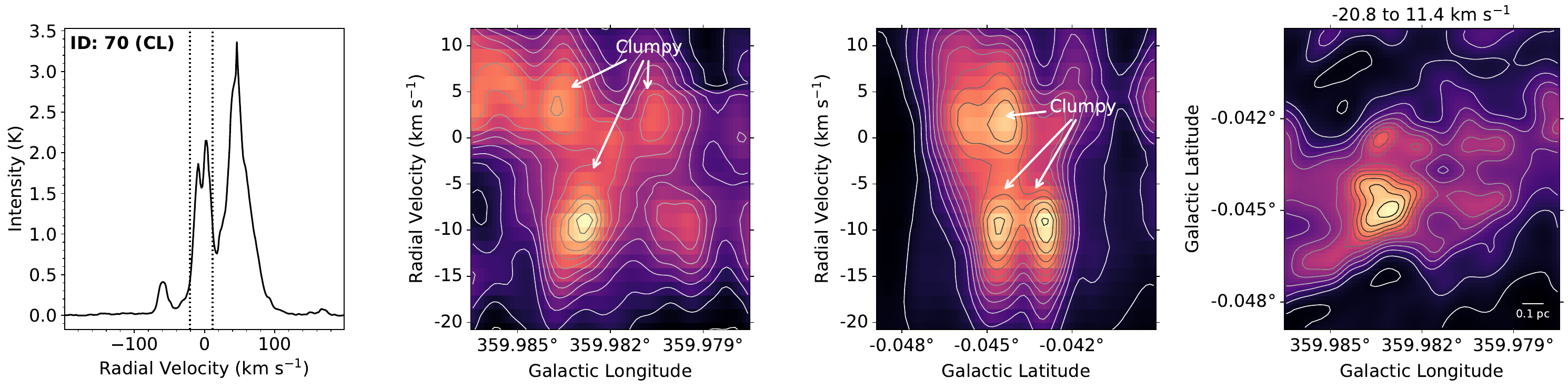}
\end{minipage}
\caption{The HVCCs in the catalog show various morphologies in PV space. Here we show examples of the PV Features categorized in the catalog. From top to bottom: velocity bridge (VB), spike (S), curve (CU), loop (L), and clumpy (CL) morphologies. For each row, we show from left to right: average \CS~spectrum, LV diagram, BV diagram, and integrated intensity map. The characteristic radial velocity range of the HVCCs are shown by the vertical dashed lines in the left column of panels.}
\label{fig:classification}
\end{figure*}

We identify different morphologies seen in the PV diagrams of each structure, created by taking the associated spectral slab and averaging along the latitude or longitude axes to create LV and BV plots, respectively. Averaging slices along different angles could lead to different morphologies in PV space; however, we only explore parallel longitude and latitude cuts for efficiency and consistency. We assign five PV classifications: ``spiked" (S) for vertical features which are sharply extended in velocity, ``clumpy" (CL) for structures which show several clumped areas of peaked emission, ``loop" (L) for well defined circular structures with dim centers, ``curve" (CU) for curled features that are not quite closed loops, and ``velocity bridge" (VB) features where two emission peaks exist in the same spatial region at different velocities with a dimmer source connecting them. Morphological classifications are reported in Table~\ref{tab:physical_properties} using notations S, CL, L, CU, and VB. Descriptions of each classification are summarized on the right side of the schematic in Figure~\ref{fig:flowchart}. Detailed examples of each PV classification are shown in Figure ~\ref{fig:classification}, with columns from left to right showing the average \CS~spectrum, LV diagram, BV diagram, and integrated intensity map for each classification type. %Three inspectors assigned a PV Feature to each 's shape in the PV diagrams, which were then synthesized to report the most agreed-upon description.

\begin{figure*}[t!]
\includegraphics[width=1\textwidth]{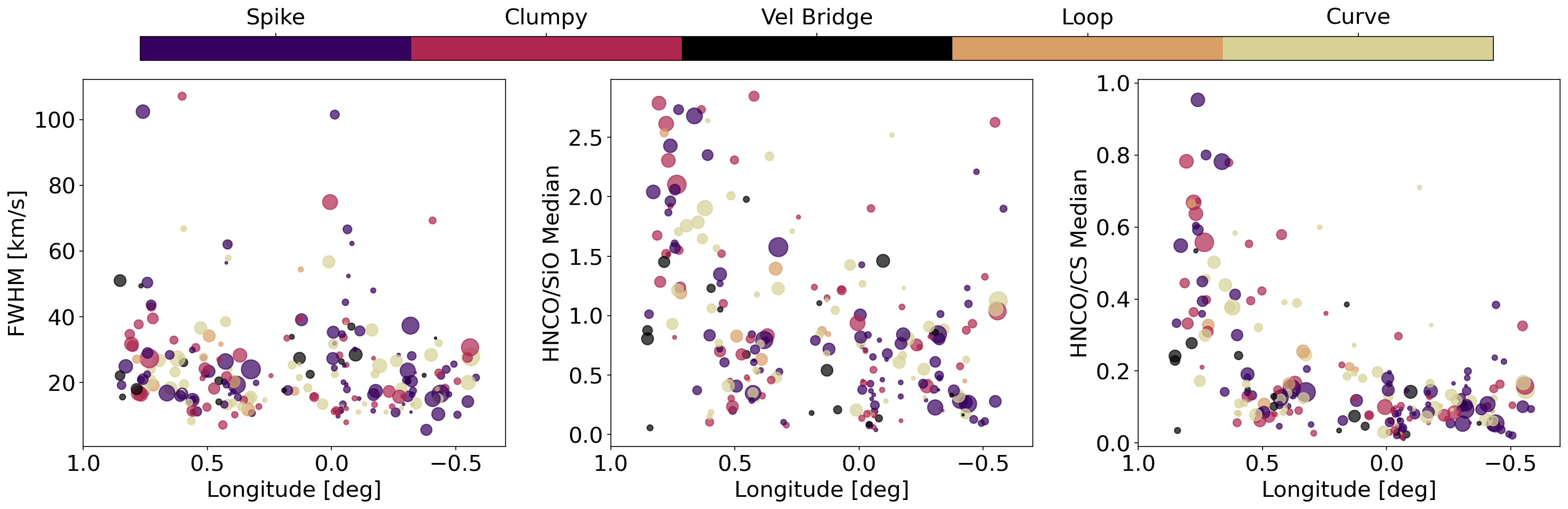}

\caption{We find weak trends in structure properties across galactic longitude, including the main fitted FWHM peak (left), HNCO/SiO line ratio (middle), and HNCO/CS line ratio (right). Each plot presents each HVCC as a point colored based on the five PV morphologies, including: spike (purple), clumpy (pink), velocity bridges (black), loop (orange), and curve (yellow) features. The size of each point is scaled by the reported dendrogram contour size from Table~\ref{tab:physical_properties} in parsecs. Separate plots for each PV class can be found in Appendix~\ref{appendix_PV_distributions}.}
\label{fig: scatter_stats}
\end{figure*}

\subsection{Creation of line ratio maps and statistics}\label{subsec: method_lineratios}

%\CS, \HthCN, \HthCOp, \SiO, \SO, SO(2-1), \HNthC, \HCtrN, and \HNCO.

We explore the chemistry surrounding our identified HVCCs by creating line ratio maps from integrated intensity maps for all possible ratios of \CS, \HNCO, \HthCN, and \SiO from ACES, as introduced in Section~\ref{sec: Data}. The mean and median line ratio values were extracted from all HVCCs to provide a comprehensive summary of which molecules were dominant across the source. %After initial exploration, we chose to further analyze ratios between \CS, \HNCO, \SiO, and \HCtrN. 

Integrated intensity maps of each line were made from slabs of the high-resolution subcubes and calculated in units of Jy beam$^{-1}$ $\mathrm{\kms}$. For a given structure, the slab for each molecular line was integrated between that object's minimum and maximum velocity based on the $V_{\mathrm{LSR}}$ and $\Delta V$ reported in Table~\ref{tab:physical_properties}. We mask noise below 3 times the local $\sigma_{\mathrm{MAD}}$ and calculate line ratio maps, further removing any contributions from absorption by excluding negative values. 

We focus on four line ratios used as diagnostics for shock tracers and dense gas: \HNCO/\CS, \HNCO/ \SiO, \HNCO/ \HCtrN, and \HCtrN/ \SiO~(hereafter referred to as HNCO/CS, HNCO/SiO, HNCO/HC$_3$N, and HC$_3$N/SiO). The ratios calculated here are based on the intensity of line emission rather than abundance, and do not consider opacity or extinction effects. Although the local densities, temperature, and UV field can affect the abundances, intensity ratios are useful probes of the relative strength of shocks in different areas. We therefore focus mainly on the velocity range observed toward the sources, within the bounds of the low-intensity line wings of the CS emission.

%First, we create moment 0 maps of the molecular lines noted above, i.e. an integrated intensity map representative of the total emission from a specific spectral line across its entire spatial region. The maps were created from slabs of the high-resolution subcubes taken between the minimum and maximum velocity ranges for each, and were calculated in units of Jy beam$^{-1}$ $\mathrm{\kms}$. Each moment 0 map was then spatially masked using the local median absolute deviation of the slab to define a noise threshold. All noise below 3 times the noise level was masked for each moment 0 map. Line ratio maps were then calculated from the masked moment 0 maps for each line, further removing any contributions from absorption by excluding negative moment 0 map values. 

\section{Results}\label{sec: Results}

\subsection{Structure statistics and properties}\label{subsec: results_structure_properties}

Through the visual identification procedure detailed in Section \ref{subsec: identification_overview}, and subsequent high-confidence dendrogram detections, we present a total of 235 HVCC candidate structures in the ACES \CS ~ data. We choose to include 10 sources not identified in the high-resolution dendrograms, and 2 which were not confirmed detections from the dendrogram analysis of the downsampled cube, in order to further compare various methodological limitations. We report on the physical properties of our candidates in Table \ref{tab:physical_properties}. %An excerpt of the Table is shown in Table~\ref{tab:excerpt_physical_properties}. 

Collectively, our cataloged structures cover  3.5\% of all ACES voxels, and 21\% of ACES voxels above a flux threshold of $3\times \sigma_{\rm{MAD}}$ in the downsampled ACES \CS ~data cube. The identified structures span a wide range of physical scales with effective radii (R$_{\text{eff}}$) from 0.14 to 9.8 pc, with a median R$_{\text{eff}}$ of $\sim$1.6 pc. A sample of spatially compact features exists near \sgra, seen in Figure \ref{fig: dendro} and the bottom middle panels of Figure ~\ref{fig:zooms}. These structures have a smaller median size ($\rm{R}_{\text{eff}} \sim 0.9$ pc) than those found elsewhere.

The CMZ has a highly asymmetric gas distribution \citep[e.g.][]{Bally2010,Longmore2013a,Battersby2025a}, with 60.8\% of ACES voxels above $3\sigma_{\rm{MAD}}$ existing at positive longitudes. 123 HVCCs (71\% of the sample) are found at positive longitudes, consistent with the location of the bulk of CMZ gas. Of the positive longitude voxels in the ACES field with intensities $> 3\sigma_{\rm{MAD}}$, 25\% are associated with HVCCs. 

The majority of our cataloged HVCCs have large linewidths for their main CS spectra peaks, with a mean FWHM for the sample of \SI{25.8}{\kms}, a significant deviation from the bulk CMZ gas dispersion of $\sim$\SI{10}{\kms} \citep[e.g.][]{Henshaw2016_gas_kinematics}. The mean linewidth is also comparable to the velocity widths of the HVCC sample from previous CO studies \citep[e.g.][with \SI{0.43}{\kms} spectral resolution]{Oka2012}, and slightly smaller than the average dispersions reported in more recently identified samples \citep[e.g.][with \SI{0.897}{\kms} spectral resolution]{Oka2022}. The fitted FWHM of the main peaks span a wide range of values, with a minimum of \SI{5.6}{\kms} and a maximum of \SI{107.1}{\kms}. While the velocity resolution of ACES is coarser than \citet{Oka2012,Oka2022}, we expect the enhanced spatial resolution of ACES to resolve larger HVCC candidates, splitting seemingly contiguous structures into finer spatial features, which would be missed by the dendrogram and likely disregarded by visual inspection. 

Many HVCCs in the catalog (58\%) have spectra with multiple peaks above local noise levels. When cataloging each structure's linewidth, we report the linewidth for the main peak in the spectra, closest to the \Vlsr. There are 108 features with FWHM $<$ \SI{20}{\kms}, 68 of which are confirmed visual detections. Despite the filters used for initial identification, both visually and algorithmically identified HVCCs are often composed of several smaller linewidth features. Algorithmic detection provides better initial filtering for complex spectra, though it is possible that broader spectral features could be split apart by significant optical depth absorption and erroneously filtered out (see Section~\ref{sec: uncertainties}). 
%Of the visually detected candidates, 5~structures (3\% of visual identifications) have FWHM $<$~\SI{10}{\kms}, and 56~structures (34\%) have FWHM $<$~\SI{20}{\kms}. 

Most of the structures have spiked PV features (37\%), with clumpy (27\%) and curved (22\%) making up the majority of the observed shapes. There are only a handful of structures showing velocity bridge (8.5\%) or loop (4.7\%) features. We assess trends in the kinematic and physical properties of each PV class in Figure~\ref{fig: scatter_stats}, which presents the distribution of HVCCs across Galactic longitude compared to their main fitted FWHM (left), HNCO/SiO median (middle), and HNCO/CS median (right). Each panel shows individual HVCCs colored by their PV class, with marker sizes scaled by the dendrogram contour size from Table~\ref{tab:physical_properties} (See Appendix~\ref{appendix_PV_distributions} Figure~\ref{fig:app_pvmorph_scatters} for individual plots of each PV class).

\subsection{Longitudinal distribution of HVCCs in the CMZ}\label{subsec:longitudinal_dist_of_HVCCs}

While the sizes of HVCCs do not necessarily correlate with line ratios or FWHM, there exist weak trends in line ratios across Galactic longitude. Line ratios tend to increase toward more positive longitudes, especially near the Sgr B region. Both HNCO/SiO and HNCO/CS are higher in this area, decreasing towards the Sgr A region where there exists a collection of HVCCs with smaller radii and very low line ratios. The line ratios then slightly increase around $\ell = -0.5$\deg towards the edge of the ACES footprint. Due to the low variation in CS abundance across the CMZ \citep{Requena-Torres2006,Dutkowska2025}, the changes in HNCO/CS directly correspond to differences in HNCO abundance, implying a large buildup of cold, dense gas on the edges of the observed CMZ. However, we note that the Sgr B region contains residual artifacts from continuum subtraction, especially in the broad spectral windows \citep[see][]{Walker2026_ACESIII}. The HNCO/CS and HNCO/SiO line emission ratios also show an overall positive correlation within the sample, which suggests similar physical processes affecting their intensities (see Appendix~\ref{appendix_PV_distributions} Figure~\ref{fig:appendix_comp_HNCO_CS_HC3N}).

Out of the five morphological classifications, the spiked, curved, and clumpy features show particularly low HNCO/CS and HNCO/SiO ratios. While CS abundance across the CMZ is rather constant \citep{Requena-Torres2006}, enrichment of SiO compared to HNCO could arise from lower velocity C-shocks ($v_s<$\SI{30}{\kms}) \citep{Jimenez2008,Martin2008}. Low \HNCO\ to \SiO\ ratios have been used to indicate higher velocity shocks \citep{Kelly2017,He2021,Huang2023}, as the molecules are chemically distinct and may be enriched by different mechanisms. For example, HNCO is known to only be enhanced by slow shocks \citep{Tideswell2010,Huang2023}, and should be comparatively less abundant in areas where SiO would be enriched. Similarly, both HNCO and HC$_3$N are easily photo-dissociated compared to CS, making them helpful probes for regions of increased UV radiation \citep[e.g][]{Martin2006,Rico-Villas2020}. In particular, these molecules are often seen with low abundance around the inner CND, where the molecular gas is destroyed by strong radiation from young stars in the nuclear cluster \citep{Martin2012,Mills2026}.

\begin{figure*}[t!]%
\begin{center}
    
\includegraphics[width=\textwidth]{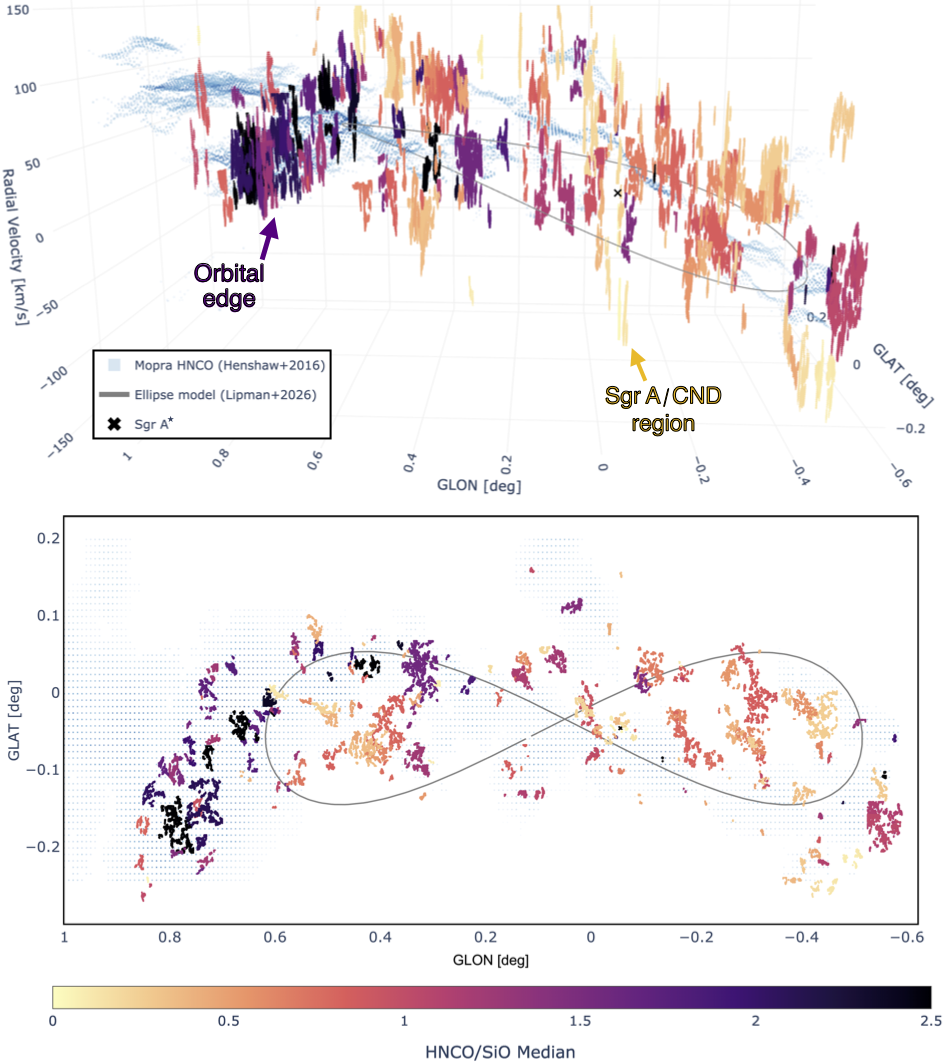}

\caption{We find our sample spans the entire CMZ, with a collection of sources near \sgra in projection. This figure shows a position-position-velocity (PPV) (top) and LB (bottom) view of the identified structures on top of the kinematic decomposition of Mopra HNCO data created by \citet{Henshaw2016_gas_kinematics} (gray background points). Only voxels with intensities above the median for a given HVCC are plotted. The closed ellipse orbital model from \citet{Lipman2026} is shown as a gray line. The black star indicates the location of the supermassive\ black hole, \sgra. The HVCCs are colored by their median HNCO/SiO~line ratios. Annotations in the top panel denote the locations of the orbital edge (dark purple text) and the Sgr A region (yellow text), including the CND, as discussed in Section~\ref{subsec: discuss_general_properties}. The top panel PPV figure is available as an interactive webapp \href{https://aces-hvcc-ppv-interactive-app.streamlit.app/\#aces-high-velocity-clouds-ppv}{here}.}
\end{center}

\label{fig: PPV}
\end{figure*}

 \subsection{Viewing the HVCCs in PPV space}\label{Results: PPV}

We use the 3D pixel masks generated from \verb|astrodendro| to visualize the PPV extent of each candidate. 3D masks are shown in the top panel of Figure \ref{fig: PPV}. For visualization purposes, each structure's mask is filtered to only plot voxels with intensities greater than the median of that HVCC. The bottom panel displays a 2D LB projection. Each colored point corresponds to the ($\ell$,$b$,v) pixels within a given dendrogram mask. We also compare the distribution of our sample to molecular gas streams obtained from spectral decomposition of MOPRA HNCO data from \citet{Henshaw2016_gas_kinematics} (gray background points), as well as the proposed median orbital model from \citet{Lipman2026} (gray line), which describes the motion of CMZ gas on an x2 ellipse. The structures are colored according to their median HNCO/SiO line ratios. The strong chemical distinction between HNCO and SiO makes their ratio useful for identifying areas of higher-velocity shocks where SiO becomes enriched compared to HNCO \citep[e.g.][]{Schilke1997,Martin2008,He2021}. Though we note that differing excitation mechanisms may also affect emission ratios.  

While the overall sample follows the rotational velocity gradient of the CMZ, a few clusters of points lie off of the bulk gas streams in velocity space. The top-panel PPV plot of Figure ~\ref{fig: PPV} highlights where the HVCCs have significant spatial and velocity deviations. In particular, the Sgr A area contains HVCCs which appear at more extreme velocities than the main orbital streams, and are consistent with the velocity and spatial ranges of CND gas. %There are also many structures with quite low HNCO/SiO line ratio values at higher velocities overlapping with a far-sided orbital stream (i.e. behind \sgra). 
While many of the HVCCs seem to occur close to the edges of the x$_2$ orbital ellipse, there are groupings of structures scattered inside of the orbit in LB space near regions of more diffuse material. In general, the sample shows a large range in HNCO/SiO values, with the highest line ratios clustered at the outer edges of the elliptical orbit within the largest concentration of dense gas. The lowest line ratios are clustered near \sgra and positive longitudes.

\section{Discussion}\label{sec: Discussion}

\subsection{General properties and distribution of structures throughout the CMZ}\label{subsec: discuss_general_properties}

The HVCCs in our \CS sample are distributed throughout the CMZ, with particularly large concentrations near the Sgr~A region and the longitudinal edges of the map, as seen in the marginal histograms of Figure~\ref{fig: dendro} and in Figure~\ref{fig: PPV}. The HVCCs also overlap with samples from previous studies. The automated procedure for single-dish CO J = 3 -- 2 data from \citet{Oka2022} algorithmically identified 184 HVCCs in the CMZ. Of the 154 CO-identified HVCCs from \citet{Oka2022} within the ACES field, 37~overlap with our cataloged \CS regions. There are also 9 matches from the \citet{Oka2012} catalog, which was created via visual identification of single-dish CO data. The low recovery of sources from previous catalogs is largely due to the difference in resolution between surveys, as many single-dish sources become resolved in the ACES data, leading to larger line widths being broken into multiple spatial or spectral components which fail to pass the velocity thresholds to be placed in the catalog. Additionally, differences in identification parameter thresholds (e.g. the use of velocity extents, FWHM dispersions, or verticality) impact the selection, as cataloging studies often use different definitions for filtering candidates. The use of the downsampled cube for automated detection also limits the dendrogram's ability to detect particularly compact sources $<$ \SI{0.5}{pc}, as these structures can be smoothed out completely by the spatial downsampling (e.g., see Section~\ref{subsec: discuss_IMBH}).

The automated detection method from \citet{Oka2022} recovered 43.3\% of HVCCs from \citet{Nagai2008} in single-dish CO data, noting some HVCCs being split into multiple components by the automated method. However, they fail to recover a few prominent HVCCs, including HCN–0.009–0.044 \citep{Takekawa2017,Takekawa2019}, which we successfully find with the dendrogram. Faintness of structures in different molecular tracers or confusion with nearby sources impact both automated methods.

While our analysis differs in resolution and molecular line tracers from previous studies, the dendrogram analysis recovers both visibly distinct features and contiguous features which may not be recognized as a single object on visual inspection. We also recover thin, faint structures above local noise thresholds. All of the cataloged sources pass size and velocity width thresholds to be considered HVCC candidates. Thus, we confirm dendrogram analysis as an automated approach for identifying HVCC-like structures in interferometric data.

The fitted FWHM values are broad compared to linewidths of foreground structures at these spatial resolutions. The ACES \CS data have a FWHM spatial resolution of $\sim2$\arcsec, and an assumed characteristic CMZ linewidth of $\sigma = 0.8~\mathrm{\kms}$ or FWHM $= 1.8~\mathrm{\kms}$ \citep[e.g. see][]{Shetty2012}. The broad linewidths, central velocities, and locations of the HVCCs in our sample are consistent with those of known CMZ structures, rather than foreground features. 

%The average linewidths of the HVCCs in this survey are lower compared to the velocity dispersions previously cataloged with CO in \citet{Oka2022}. However, we see the same variety in morphologies in position-velocity space, with many features appearing as spikes, arcs, and bubbles as previously observed. The lower velocity ranges could be due to a significant difference in angular resolution (\SI{16}{"} in \citet{Oka2022}  vs \SI{2}{"} in this work). The previous CO-focused surveys utilized single-dish observations, with an order of magnitude coarser angular resolution compared to ACES. Using data with a higher angular resolution can have significant impacts on both the size and linewidths for these structures. 

In addition to being confidently placed within the CMZ, the HVCCs align well with the dense areas traced by single-dish HNCO \citep{Henshaw2016_gas_kinematics} in PPV space (see Figure ~\ref{fig: PPV}). 
 %The velocity extents of the HVCCs are quite broad compared to the gas streams, making it clear that the sample includes areas of high velocity widths which can be separated from the average orbital streams. 
We also note interesting trends in the SiO and CS for the cataloged structures, which seem to indicate higher velocity C-shocks are less prevalent in the densest regions on the edges of the gas streams. The spatial distribution of HNCO/SiO median line ratios show two interesting clustered populations of HVCCs: 1) a cluster of high-velocity, but low ratio structures near the Sgr A region, which shows as a vertical strip of golden features in the PPV plot in the top panel of Figure ~\ref{fig: PPV}; and 2) an area of high HNCO/SiO ratios on the left-hand edge of the elliptical orbit near Sgr B2, a high-density and prolifically star-forming cloud \citep{Schmiedeke2016}, which appear as dark purple and black clusters in Figure~\ref{fig: PPV}. Another cluster of higher emission ratios also appears on the negative longitude edge of the CMZ. 

Similar trends in both emission and abundance ratios of HNCO/SiO are noted in \citet{He2021}, with higher ratios on the positive and negative longitude edges of the CMZ, and lower HNCO/SiO ratios near the Sgr A region (see their Figure 10). They suggest the high ratio region near Sgr B2 marks a likely point where material from the bar lanes accretes onto the CMZ, indicating an interaction between x$_1$ and x$_2$ orbits. The Sgr B2 region is a known location of cloud-cloud collisions and SiO enrichment consistent with C-type shocks \citep{Tsuboi2015,Armijos-Abendano2020,Zeng2020,Colzi2022,Colzi2024}, though many dynamical interpretations are possible. Alternate dynamical explanations include buildup of gas on the slower-orbiting edges of the x$_2$ orbit \citep{Kruijssen2015,Tress2020}, making it more likely for collisions to occur due to the increased density of material; or disruptions between nested x$_2$ orbits \citep[e.g.][]{Lipman2026} to produce similar shocks. However, shocks are not the only possible explanation for the variable HNCO emission. The lower line ratios in the central region can also result from photodissociation, which would erase the HNCO signature in the strong UV radiation field near Sgr A compared to the edges of the orbit \citep{Martin2008}. This trend is also seen in the center of the starburst galaxy NGC 253 \citep{Meier_Turner_2012,Meier2015}. Likewise, the higher column densities on the edges of the CMZ may help shield HNCO from destruction and maintain a higher HNCO/SiO ratio in these regions.

Lastly, the area around the dust ridge ($0.2^{\circ}< \ell < 0.6^{\circ}$) contains a mix of median HNCO/SiO ratios between approximately 0.1 -- 0.2. However, there also appears to be a notable cluster of higher ratio points near ($\ell$,$b$) = (0.45$^{\circ}$, 0.05$^{\circ}$) between velocities of $0 < \Vlsr < 50~\mathrm{\kms}$, overlapping with a prominent molecular cloud complex (commonly referred to as clouds e\&f). The dust ridge region is noted to have a lack of star formation, despite its quantity of dense gas \citep{Walker2018}. However, the collection of HVCCs in this region could point toward the beginnings of pre-stellar environments, as the increased abundance of cold, dense gas (e.g. higher HNCO/CS ratios) due to cloud collisions would result in large velocity dispersions from turbulent, shocked material. %The higher HNCO/SiO ratios in this region also indicate disruption in the 3D dynamics of the dust ridge, where some clouds have turbulent properties dominated by either higher- or lower-velocity collisions compared to the surrounding gas. %  The lack of star formation present in the dust ridge in addition to its distance from \sgra could explain the lack of so a lack of star formation and distance from  HVCCs identified in this part of the CMZ. 

%DEATH TO THE STACKED HISTOGRAM--I AM SO SORRY EVERYONE!
\begin{comment}
\begin{figure*}[t!]
\includegraphics[width=1\textwidth]{Histogram of Line Ratio Map Medians HVDCC Lighter Purples.png}

\caption{The structures show evidence for both high and low velocity shocks. This figure shows stacked histograms of the median values of the structure's line ratio maps, colored by PV classification: Spiked, Loopy, Clumpy, Velocity Bridge, and Curved. If an object is categorized as having more than one of these classifications (e.g. exhibiting both a curve and a spike), it is counted here for both classifications. Line ratios were created from moment 0 maps with units of Jy beam$^{-1}$ km$^{-s}$. The number of bins for each histogram is 20.}
\label{fig: ratio_histogram}
\end{figure*}
\end{comment}

\subsection{Potential origins based on PV descriptions}\label{subsec:discuss_origins}

We classify the HVCCs in PV space and discuss the physical processes from which different morphologies originate. PV classes of HVCCs are summarized in Figure~\ref{fig:classification}, including velocity bridge (VB), spiked (S), curved (CU), loop (L), and clumpy (CL) emission features. Common physical interpretations for these features are cloud-cloud collisions, star formation and feedback processes, or shearing motions from dynamical interactions. 

Velocity features associated with cloud-cloud collisions (CCCs) have been previously observed in the bar regions of the Milky Way \citep{Zeng2020,Gramze2023,Henshaw2023} and nearby barred galaxies \citep{Kolcu2025}, and studied via hydrodynamical simulations  \citep{Haworth2015a,Haworth2015b,Barnes_Priestley2024}. While CCCs in the Galactic disk are associated with clouds with small velocity differences ($\sim$ few \SI{}{\kms}) \citep{Jimenez-Serra2010,Cosentino2018,Cosentino2019}, the highly dynamic environment of the CMZ allows for collisions between clouds with extreme velocity differences ($\sim$ few tens \SI{}{\kms}) \citep{Higuchi2014, Busch2022, Gramze2023} and significant shear \citep[e.g.][]{Federrath2016}. In this context, velocity bridges (VB) in PV space are physically interpreted as two clouds at different velocities colliding, with the material between them forming a stretched band of emission connecting the bulk cloud emission \citep[e.g.][Figures~5 and~7, respectively]{Haworth2015b, Haworth2015a}. We note these as key signatures of early-stage CCCs in our HVCC sample (e.g., ID~14 in Figure~\ref{fig:classification}). Other PV trends resemble various stages of collisions, as seen in hydrodynamical simulation studies of the evolutionary sequences of CCCs \citep{Haworth2015b, Haworth2015a}. For example, vertical spiked (S) emission (e.g., ID~4 in Figure~\ref{fig:classification}) or curved (CU) emission (e.g., ID~1 in Figure~\ref{fig:classification}) arise after the clouds have been interacting for some time as they merge. Also, while they may not have well defined velocity bridges, HVCCs with a clumpy (CL) morphology in PV space (e.g., ID~70 in Figure~\ref{fig:classification}) may be related to the early stages of multiple cloud collisions within a small area or even cloud fragmentation. %We also note that IDs~14, 4, and~1 share similar linear scales ($\sim$ few pc), and their PV features resemble an evolutionary sequence traced by different time snapshots in the CCCs HD simulations in \citet{Haworth2015b, Haworth2015a}.

Massive star formation and stellar feedback are also known to produce vertically extended PV morphologies, as these processes affect the distribution of ionized \citep{Luisi2021,Bonne2022,Keilmann2025} and molecular \citep{Feddersen2018} emission in PV space. The GC harbors some of the most massive OB associations in the Milky Way \citep[e.g. Arches, Quintuplet, and Nuclear stellar cluster; Sgr B1 and Sgr C, see][]{Figer1999_nov,Figer1999_mar,Figer2009,Clarkson2012,Nogueras-Lara2022,Nogueras-Lara2024} with several runaway young stars scattered across its volume \citep{Dong2015,Clark2021}, Supernova Remnants (SNRs) and expanding \HII~regions \citep[and references therein]{Garcia2016}. In particular, spiked (S) and looped (L) PV morphologies are noted trends associated with stellar feedback processes in the form of expanding bubbles or CO shells \citep{Luisi2021,Bonne2022,Keilmann2025,Feddersen2018} and molecular outflows \citep{Ohashi2014,Lopez2025}, respectively. Curved features also arise from interactions between expanding material from supernovae and nearby clouds, as seen in sources outside of the CMZ \citep[e.g.][]{Cosentino2019,Makita2026}. Thus, the observed curved and looped features in the HVCC sample could be physically interpreted as stellar feedback processes causing rapid expansion of the gas from supernovae or stellar outflows such as winds from massive stars plowing through the ISM. Likewise, the spiked morphologies resemble features often interpreted as a molecular outflow towards massive protostars \citep[e.g.][their Figure~3]{Lopez2025}. Though it is beyond the scope of this initial cataloging study, HVCCs showing overlap with X-ray or IR sources would imply the physical origins of these features are likely related to wind feedback from massive stars or evolved stars, respectively.
%loop in ID~41 (projected diameter $\sim$ \SI{0.6}{pc}) resembles known [CII] 157~$\mu$m expanding shells in massive star-forming regions \citep{Luisi2021,Bonne2022,Keilmann2025} as well as expanding CO shells in the Orion~A molecular cloud \citep{Feddersen2018}, with linear scales from \SI{0.5}{pc} to \SI{4.5}{pc}. Therefore, massive star formation feedback could impact part of our interpretation of PV space information in terms of the CCCs scenario.

\begin{figure*}[t!]
\begin{centering}
\includegraphics[width=\textwidth]{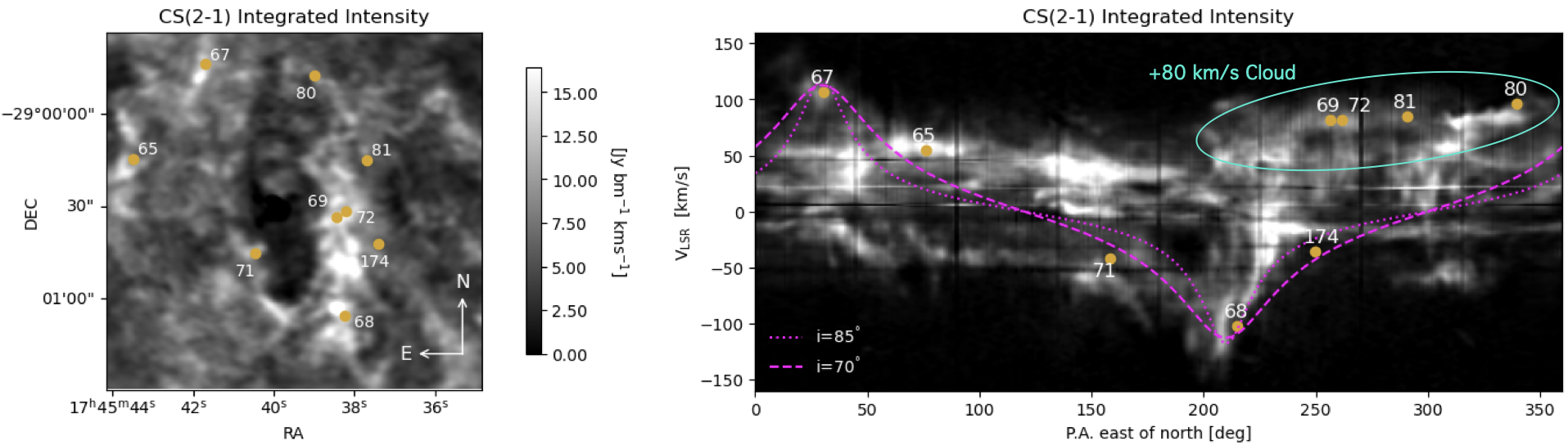}
\caption{A sample of HVCCs within an arcminute of \sgra overlap with well-studied CND structures and theoretical models for the rotational velocity of CND gas. Left: integrated intensity map of the CND region from the ACES \CS data with overlapping HVCCs (golden circles). Right: a plot of position angle vs radial velocity, including two theoretical models for rotating rings with inclinations of $i=85^\circ$ (magneta dotted line) and $i=70^\circ$ (magenta dashed line), both with a peak rotational velocity of \SI{120}{\kms}. The grayscale is the integrated emission from the region in the left panel, with the same HVCCs noted by gold circles. The position angle is oriented east of Galactic north. The \SI{+80}{\kms} Cloud feature is noted by the cyan ellipse in the top right.
}
\end{centering}
\label{fig:CND_PA_vel}
\end{figure*}

\begin{figure*}[t!]
\begin{centering}
\includegraphics[width=\textwidth]{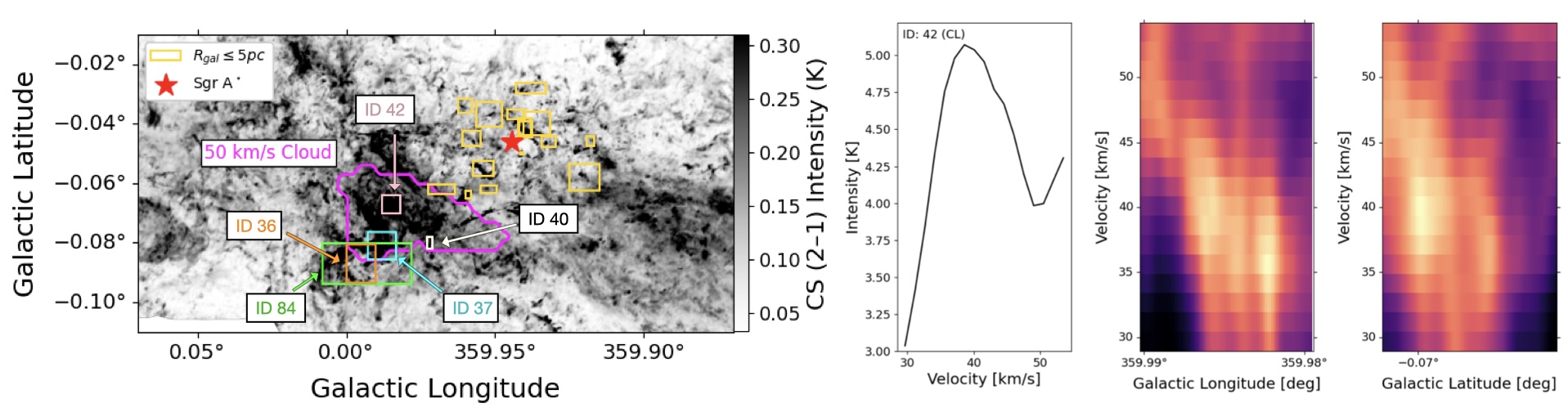}
\caption{
(Left panel) A $0.15^\circ \times 0.15^\circ$ cutout of the \CS integrated intensity map around the \SI{50}{\kms} molecular cloud. HVCC regions: 37 (cyan), 42 (blue), 40 (light blue), 84 (light green), and 36 (orange) are overlaid as colored boxes. HVCC regions within the central \SI{5}{pc} are shown as yellow boxes. The magenta contour shows the shape of the \SI{50}{\kms} Cloud based on Herschel column density maps from \citep{Battersby2025b}. \sgra is shown as a red star. (Right three panels) We highlight ID 42 as a candidate with a particularly strong velocity gradient in its PV structure, potentially indicative of acceleration along the line of sight. We show the spectrum (left), LV diagram (middle), and BV diagram (right).
}
\end{centering}
\label{fig:50kmsC}
\end{figure*}

Lastly, interactions with a compact object such as \sgra or large-scale shear motions \citep[e.g.][]{Federrath2016} can produce extended PV features. This has been suggested for CO~0.02-0.02 by \citet{Sofue2025}, which proposed this structure lies on an eccentric orbit around the GC to explain its large velocity dispersion and proximity to \sgra. This source overlaps with ID 206 (CS 0.011-0.019) from our catalog. ID 206 is a much larger area of size $\sim$\SI{4.01}{pc} containing the central coordinates of CO 0.02-0.02, with similar \Vlsr and velocity width, though a significantly smaller fitted FWHM for the main peak than seen from the CO emission in \citet{Oka1999}. The PV morphology is curved, supporting the physical interpretation of accelerated shearing motions near the GC, as is seen by molecular gas streamers flowing toward the CND \citep[e.g.][]{coil1999, Mills2026}. 
%HVCCs near \sgra could be on similar orbits, while those farther away (R$_{gal} >$ \SI{3}{pc}) are likely not influenced by the supermassive black hole's gravitational potential \citep{Scheodel2018}. 
Violent shearing motions are also experienced by molecular clouds near the apocenter of the x2~orbits around the GC \citep[e.g][]{Tress2020,Sormani2020,Tress2024,Pare2025}, leading to extended PV features over large velocity ranges.  %{\color{red} Here some text about shear usin this references \citep[ D. M. Par\'e et al. 2025, in prep.]{Sormani2020,Tress2020,Tress2024}: Molecular clouds can be sheared apart by the strong Galactic shear as they approaches apocenter of its X$_2$ orbit around the GC.}

\subsection{Overlap with potential nuclear inflow features}\label{subsec: discuss_50kmsC}

A cluster of HVCCs appears in projection near the Sgr A region, as shown in the bottom middle panels of Figure~\ref{fig:zooms} and the right two panels of Figure~\ref{fig: scatter_stats}. Many of these structures lie within the CND, which is often identified as either a single or small collection of HVCCs in previous studies \citep{Oka2012,Oka2022}. 

Massive spiked and curved velocity features have previously been used to estimate the inflow of bar lane material onto CMZs \citep[e.g.][]{Sormani_and_Barnes_2019,Kolcu2025}. Similarly, the HVCCs in the CND region are indicative of disruptions from typical gas flows, signaling potential nuclear inflow candidates from material leaving inner x$_2$ orbits towards the CND and \sgra \citep[e.g.][]{Lipman2026}, or gas following the nuclear orbits of the CND itself \citep[e.g.][]{Sofue2025,Sofue2025_GCarms}. 

We identify 9 HVCC candidates within an arcminute of \sgra, shown as gold points overlaying the grayscale \CS integrated intensity map in the left panel of Figure~\ref{fig:CND_PA_vel}. The right panel shows a map of position-angle vs velocity for the region, with angles measured east of north. Several previous studies have characterized the rotational motion of CND gas \citep[e.g.][]{Gusten1981,Guesten1987,Martin2012,Goicoechea2018,Tsuboi2018,Hsieh2021,Tanaka2026}, often noting its clumpy and irregular structure leading to a series of related orbital rings. We plot two examples of theoretical ring model orbits similar to those in \citet{Guesten1987,Martin2012}, with inclinations of $i=85^\circ$ (dotted line) and $i=70^\circ$ (dashed line), and peak rotational velocity of $v_{\rm{rot}}~\rm{sin}\theta = ~$\SI{120}{\kms}. All of the HVCCs in this region follow the CND rotational motion and are associated with well-known features such as the northeast lobe (67), southern extension (71), southwest lobe (e.g. 68, 174), and \SI{+80}{\kms} cloud (69, 72,81, 80). ID 65 overlaps with the eastern arm region, which has many proposed interpretations, including overlap with part of the \SI{50}{\kms} Cloud \citep{Tanaka2026,Mills2026}, inflowing material toward \sgra due to tidal shearing \citep{Tsuboi2018}, or the existence of an outflow source \citep{Yusef-Zadeh1987}.

16 HVCCs lie within \SI{5}{pc} of \sgra, including the 9 within the CND region, shown as yellow boxes in the left panel of Figure~\ref{fig:50kmsC}. This subset has small spatial sizes (d~$<1.5$~pc) and low line ratios of HNCO/CS $\lesssim 0.10 $ and HC$_3$N/CS $ \lesssim 0.15 $. Gas in the CND also shows low HNCO/CS ratios and HC$_3$N abundance compared to the surrounding gas \citep{Martin2008,Mills2026}, likely due to CS molecules resisting destruction from the strong UV radiation pervaded by the central cluster \citep{Goicoechea2006}, while HNCO and HC$_3$N are destroyed \citep{Woodall2007,Martin2012}. SiO is also enriched in the CND area due to C-shocks from cloud collisions \citep{Martin2008}. However, UV fields in this region are quite extended and potentially dominate the molecular abundances. The proximity to \sgra and chemical signatures support further investigation of these features for potential association with the CND or infalling material.

%Molecular clouds in and around the CND have been extensively studied using ALMA \citep[e.g.][]{Tsuboi2018,Goicoechea2018,Moser2017,Hsieh2021}. The overall morphology of the CND is often noted as a clumpy distributions of gas with rotational velocities of $V_{\rm{rot}}\sim$\SI{115}{\kms}\citep{Guesten1987,Christpher2005,Montero2009,Martin2012}. However, many anomalous cloudlets have been reported in the region with deviating rotational velocities from the CND, possibly due to tidal shearing and infalling motions from outside the region. \citep{Tsuboi2018,Goicoechea2018}. 

%Several HVCCs overlap with the \SI{50}{\kms} Cloud The cloud also lies in the path of the Sgr~A East supernova remnant, where the SNR expanding shock compresses and heats the molecular gas into dense, high velocity clumps \citep{Lee2003, McGary2001}. 
The left panel of Figure~\ref{fig:50kmsC} shows five HVCC candidates that lie in LB projection against the \SI{50}{\kms} Cloud, which is often regarded as a potential origin of multiple dense gas streamers from a Galactocentric radius of $\sim50$~pc towards the CND \citep{Hsieh2017,Nogueras-Lara2026,Lipman2026}. The cloud spans velocities between 20 -- \SI{50}{\kms} \citep[e.g.][]{Uehara2019}. Of the five spatially overlapping candidates, ID~37 and~42 have peak velocities within this range. In particular, ID~42 (right three panels of Figure~\ref{fig:50kmsC}) shows a particularly steep velocity gradient in LV space, implying acceleration along the line of sight. IDs 36 and 84 exhibit peak \Vlsr of 21 and \SI{40}{\kms}, respectively, lying just inside the \SI{50}{\kms} Cloud envelope, whereas ID 40 shows a minimum \Vlsr of 55, exceeding the cloud's maximum. The spatial alignment and consistent \Vlsr values of HVCCs 36, 37, 42, and 84 with the \SI{50}{\kms} Cloud suggest that although not fully encased by the cloud, there is evidence for a dynamical connection to infalling structures. Further exploration of HVCCs within the central \SI{10}{pc} in projection is planned to explore inflowing candidates towards the CND and \sgra.

\subsection{Possibility of IMBH candidate searches in ACES}\label{subsec: discuss_IMBH}

HVCCs have previously been suggested as candidates for intermediate black holes \citep[IMBHs; e.g.][]{Oka2017,Takekawa2017,Takekawa2019,Takekawa2020}. As a first-order approximation for the expected properties of an IMBH candidate in the ACES data, we consider a Keplerian disk of molecular material orbiting a 10$^4$ \Msun~ IMBH. Using simple Keplerian rotation ($v = \sqrt{GM/R}$), the approximate rotation speed expected at the ACES resolution of 0.1~pc is about \SI{21}{\kms}. 

Previously, \citet{Ginsburg2024_MUBLO} reported an IMBH candidate in the ACES data, G0.02467–0.0727 (also referred to as the millimeter ultra-broad-line object, MUBLO), which appears as a compact ($<1$\arcsec) 3mm continuum source at \Vlsr$\approx 40~-~$\SI{50}{\kms} with FWHM $\approx$ \SI{160}{\kms}. Material orbiting the central object at $70~-~$ \SI{80}{\kms} would occur at $10^3 - 10^4$~AU for a $10^4$~\Msun~IMBH. The MUBLO also appears in multiple ACES line transitions, including CS, making it a good benchmark for the limitations on the current identification method.

The HVCCs we identify have average radii of $r \approx 1-$\SI{2}{pc}, implying a rotational velocity of $V_{rot} \approx 4-$\SI{7}{\kms}. This orbital velocity is much lower than the typical $\Delta V$ reported in Table~\ref{tab:physical_properties}. We do not identify the MUBLO in our sample either from the visual inspection of the full-resolution PV diagrams or dendrogram analysis. The MUBLO is very compact and is smoothed out entirely in the downsampled data, despite maintaining the original spectral resolution. Thus, our current methodology using the 9$\times$ downsampled data is not sensitive to objects $< $\SI{0.5}{pc} in size. A version of the methodology using the full-resolution data would be more complete to objects of these spatial scales.

While we do not recover the MUBLO in our catalog, a couple HVCCs overlap with IMBH candidates from single-dish observations: ID 66 matches candidate HCN–0.009–0.044 from \citet{Takekawa2017}. ID 66 (CS $-$0.011$-$0.044) has a spatial size of \SI{0.95}{pc} and a \Vlsr of \SI{-43}{\kms}, compared to reported values in \citet{Takekawa2017} of \SI{0.33}{pc} and \SI{-40}{\kms}. We also examine other IMBH candidates which do not have a matching detection in our sample \citep[e.g][]{Oka2017,Takekawa2019,Kaneko2023}. In many cases, the reported IMBH candidates either become spatially smoothed out in the downsampled cube (as with the MUBLO), become spatially resolved in the ACES data, or are spectrally resolved in ACES, such that the velocity structures seen in single-dish data are broken into much smaller ($\Delta V<$\SI{30}{
\kms}), finer features that are missed by visual inspection or filtered from dendrogram identification. %In particular, \citet{Kaneko2023} report a cloud at ($\ell$,$b$) $= (-0.090\degree, -0.014\degree)$ found using the CO J = 3-2 emission line with JCMT, with corresponding \CS emission from the Nobeyama Radio Observatory (NRO) 45 m Telescope. While the object itself seems to appear spatially in the ACES \CS data, the velocity gradient appears to be broken up into 

\begin{comment}
    
We note one source of interest in our catalog, ID~104, which coincides with a \SI{0.1}{pc} continuum peak centered on strong CS and SiO emission, embedded in a \SI{\sim3.62}{pc} radius cloud with a velocity of \SI{20}{\kms} less than the surrounding gas. The FWHM of the source's \CS spectrum is \SI{127}{\kms}, and a strong gradient is seen in the LV diagram, which is also indicative of gas rotating around a compact object. However, more information is needed to confidently determine this structure as an IMBH candidate, such as follow-up observations to search for masers and X-ray emission in this region to rule out other origins for the shocked emission, and detailed modeling of the Keplerian orbits.

A formal prediction of the linewidth in \CS~due to the presence of an IMBH would require more detailed modeling, consideration of excitation, size scales, and observational effects. The simple Keplerian model used here assumes a constant radius and a perfectly circular orbit. In reality, the measured linewidth arises from many factors, including the temperature, density of the surrounding environment, and blending of gas along the line of sight. Therefore, a more detailed model that accounts for these irregularities is necessary. We expect that our catalog is not a complete search for IMBH candidates in ACES. HVCCs have also been found in CO, and we believe there are many more within the ACES data that are less obvious or hidden within the noise.
\end{comment}

\section{Potential sources of uncertainty from detection and classification}\label{sec: uncertainties}

The sample of HVCCs presented here includes those detectable in the ACES \CS data, meaning that HVCCs that are not bright in this molecule are not identified. Our initial identification of HVCCs was done visually and confirmed with dendrogram analysis. %Because of this, the visual portion of the catalog is incomplete where HVCCs that are dim or have large size scales are unnoticed or filtered. 
Although the sample is incomplete for dim sources or objects within noisy fields, the inclusion of dendrogram leaves with sizes below \SI{10}{pc} and velocity widths over \SI{50}{\kms} ensures that overall, our sample is complete to HVCCs of these scales in the downsampled \CS map. However, there are various sources of confusion which could obfuscate the properties of the identified HVCCs. 

First, the ACES \CS data contains horizontal absorption features at $-$52, $-$28, and \SI{-3}{\kms} \citep{Hsieh2026_ACESV} due to gas in the spiral arms along the line of sight \citep{Jones2012,Reid2016,Eden2020}. It is possible for foreground material to cut across emission associated with HVCCs and absorb the background light, splitting the continuous spectrum and potentially obfuscating the true linewidths of the structures. While this confusion could impact the dendrogram detection, it should not affect the visual identification, which is less sensitive to small deviations in the spectra. 

Second, the many different velocity components in the GC often overlap, making it difficult or impossible to disentangle them into independent structures. CMZ gas has line-of-sight velocities ranging from \SI{-200}{\kms} to \SI{200}{\kms}, and velocity dispersions on average much wider than disk gas \citep[e.g.][]{Bally1987,Henshaw2016_gas_kinematics}. Bright bulk gas with large linewidths and large spatial extents can hide evidence of HVCCs.

Third, large optical depth effects from high-density compact objects can artificially create ``clumpy" morphologies. A combination of line saturation broadening the velocity range of the main peak, and self-absorption causing dips in intensity could create discontinuous or clumpy features. These effects could also result in a single source being identified as multiple HVCCs. 

Lastly, another possible origin of HVCCs not related to physical processes affecting their gas directly is coincidental alignment along the line-of-sight. The CMZ is full of molecular gas at a variety of velocities, some of which appear close in projection and position-velocity space. Similar alignment effects are also seen within the solar neighborhood \citep[see][]{Beaumont2013}. We cannot discard that the ``clumpy" morphology could be related to such coincidental alignment.  

We present the cataloged HVCC candidates as an initial survey of high-velocity features in the ACES \CS molecular line. The catalog may contain structures which would not necessarily be classified as HVCCs based on previous definitions 
in the literature from studies of lower-resolution CO data. However, the candidates presented here provide interesting features tracing both dense and shocked gas in the CMZ, which can be used for comparisons with previously cataloged HVCCs, potential in-falling features, and IMBH candidates.

\section{Summary and Conclusions}\label{sec: Conclusions}

The ALMA large program ACES (ALMA CMZ Exploration Survey) provides high spatial (full resolution: 2\arcsec; 0.1~pc, downsampled: 5\arcsec; 0.5~pc) and spectral (\SI{1.5}{\kms}) resolution \CS molecular line data, which glows brightly in the CMZ, and exemplifies a population of extended vertical features in PV space with properties comparable to HVCCs previously identified with CO molecular tracers. In this paper, we use both visual identification and a dendrogram algorithm to detect HVCC candidates in the ACES \CS data, with size scales (d $<$ \SI{10}{pc}) and velocity extents ($\Delta \mathrm{V} \gtrapprox$ \SI{20}{\kms}). We further filter our detections by verticality (ratio of velocity extent/longitudinal extent of a given structure), and use dendrograms to confirm the visually identified sources. We present a catalog of 235~HVCCs in the ACES \CS data. We report the physical and kinematic properties of each structure, summarize the chemistry of the population via line ratios of \HNCO, \CS, \SiO, and \HCtrN as useful tracers of C-shocks, and classify the morphology of their PV features into five classes: spiked, loop, curved, clumpy, and velocity-bridges. The code developed for the identification and analysis of HVCCs, and the creation of figures in this study are available publicly at \href{https://github.com/ACES-CMZ/ACES_EVF_HVCC}{https://github.com/ACES-CMZ/ACES\_EVF\_HVCC}.

We highlight the following main conclusions from our sample:

\begin{itemize}
    \item We find HVCC candidates throughout the CMZ, with a high concentration near the Sgr~A region. The structures in our sample overlap with previous surveys and studies of HVCCs previously identified and cataloged with CO molecular tracers \citep{Oka2012,Oka2022}. The identified structures show lower velocity widths than previous studies due to differences in resolution between the previous single-dish data used for the CO analysis versus the interferometric data used in this paper.

    \item When viewed in PPV space, we find two distinct areas of high \HNCO/\SiO line ratios: one at the positive velocity edge of modeled x2~orbits where dense gas is more concentrated, and another overlapping with the dust ridge clouds e\&f. The high \HNCO/\SiO line ratios may indicate lower velocity shocks in these regions compared to the surrounding gas. These areas result from slow-moving gas collecting as it moves towards the apogee of the orbit, allowing for a large number of cloud collisions at lower velocities than seen in highly shocked regions towards the center near the \sgra region.

    \item We find a handful of HVCCs near \sgra in projection with spatial and kinematic overlap, as well as similar chemical signatures, with CND gas. Additionally, we find some HVCCs associated with the \SI{50}{\kms} Cloud. These structures have morphologies and line ratios similar to proposed inflowing material towards the GC arising from shearing and collisions, and provide a sample of candidates for material falling off of inner x$_2$ orbits towards the nuclear region.

    \item Two HVCCs in our sample overlap with two previous IMBH candidates in the CMZ, while other candidates become spectrally resolved in the ACES survey and are not undetected. Our algorithmic detection method is complete to sources $>0.5$~pc in size, with improvements possible with the use of the full-resolution data cube for future IMBH surveys.
    
\end{itemize}

We present a catalog of \CS  candidates in the CMZ. Future work is planned to further investigate the physical origins of the various PV structures, as well as determine nuclear inflow candidates from the average orbital streams towards the central few parsecs of the GC. The catalog created of high-velocity features in shocked molecular gas from this analysis is a statistical sample which promises to provide key insights into the disruption of gas orbits in the CMZ, and the mechanisms which may quench or drive star formation in dense regions.

\section*{Acknowledgments}

This paper makes use of the following ALMA data: ADS/JAO.ALMA\#2021.1.00172.L. ALMA is a partnership of ESO (representing its member states), NSF (USA) and NINS (Japan), together with NRC (Canada), MOST and ASIAA (Taiwan), and KASI (Republic of Korea), in cooperation with the Republic of Chile. The Joint ALMA Observatory is operated by ESO, AUI/NRAO and NAOJ.
\\

This work reports original results made with National Radio Astronomy Observatory instrument(s). The National Radio Astronomy Observatory is a facility of the National Science Foundation operated under cooperative agreement by Associated Universities, Inc.
\\
This work was written as part of a ``paper sprint'', in which a dedicated team of researchers engaged in an intense, two-week collaborative research and draft writing process. Please reach out to Dani Lipman for more details on the organization and outcome of this paper sprint.
\\

D.\ Lipman gratefully acknowledges funding from the National Science Foundation under Award Nos. 1816715, 2108938, and CAREER 2145689; and NASA FINESST Award No: 80NSSC24K1474.

C.\ Battersby  gratefully  acknowledges  funding  from  National  Science  Foundation  under  Award  Nos. 2108938, 2206510, 2414862, and CAREER 2145689, as well as from the National Aeronautics and Space Administration through the Astrophysics Data Analysis Program under Award ``3-D MC: Mapping Circumnuclear Molecular Clouds from X-ray to Radio,” Grant No. 80NSSC22K1125 as well as participation in the PRIMA project under Grant No. 80NSSC25K7944.

J. Wallace gratefully acknowledges funding from National Science Foundation under Award Nos. 2108938 and 2206510.

P. Garc\'ia is supported by Chinese Academy of Sciences South America Center for Astronomy (CASSACA) Key Research Project E52H540201. This work is supported by the China-Chile Joint Research Fund (CCJRF No. 2312). CCJRF is provided by Chinese Academy of Sciences South America Center for Astronomy (CASSACA) and established by National Astronomical Observatories, Chinese Academy of Sciences (NAOC) and Chilean Astronomy Society (SOCHIAS) to support China-Chile collaborations in astronomy.

S. Gramze and A. Ginsburg acknowledge support from the NSF under grants AAG 2206511 and CAREER 2142300.

C.F.~acknowledges funding by the Australian Research Council (Discovery Projects grant~DP230102280 and DP250101526), and the Australia-Germany Joint Research Cooperation Scheme (UA-DAAD).

A.S-M.\ acknowledges support from the PID2023-146675NB grant funded by MCIN/AEI/10.13039/501100011033, and by the programme Unidad de Excelencia Mar\'{\i}a de Maeztu CEX2020-001058-M, the MaX-CSIC Excellence Award MaX4-SOMMA-ICE, as well as support from the RyC2021-032892-I grant funded by MCIN/AEI/10.13039/501100011033 and by the European Union `Next GenerationEU'/PRTR.

L.C. acknowledges support from grant no. PID2022-136814NB-I00 by MICIU/AEI/10.13039/501100011033 and by ERDF, UE. The project that gave rise to these results received the support of a fellowship from the ”la Caixa” Foundation (ID 100010434). The fellowship code is LCF/BQ/PR25/12110012.

F.N.-L. gratefully acknowledges financial support from grant PID2024-162148NA-I00, funded by MCIN/AEI/10.13039/501100011033 and the European Regional Development Fund (ERDF) “A way of making Europe”, from the Ramón y Cajal programme (RYC2023-044924-I) funded by MCIN/AEI/10.13039/501100011033 and FSE+, and from the Severo Ochoa grant CEX2021-001131-S, funded by MCIN/AEI/10.13039/501100011033.

M.G.S.-M. thanks the Spanish MCINN for funding support under grant PID2023-146667NB-I00 funded by MCIN/AEI/10.13039/501100011033. 

I.J.-S and R.L acknowledge funding from the ERC grant OPENS, GA No. 101125858, funded by the European Union, and from grant PID2022-136814NB-I00 funded by the Spanish Ministry of Science, Innovation and Universities/State Agency of Research MICIU/AEI/ 10.13039/501100011033 and by "ERDF/EU".

V.M.R. acknowledges support from the grant PID2022-136814NB-I00 by the Spanish Ministry of Science, Innovation and Universities/State Agency of Research MICIU/AEI/10.13039/501100011033 and by ERDF, UE, the grant RYC2020-029387-I funded by MICIU/AEI/10.13039/501100011033 and by "ESF, Investing in your future", and from the Consejo Superior de Investigaciones Cient{\'i}ficas (CSIC) and the Centro de Astrobiolog{\'i}a (CAB) through the project 20225AT015 (Proyectos intramurales especiales del CSIC), and from the grant CNS2023-144464 funded by MICIU/AEI/10.13039/501100011033 and by "European Union NextGenerationEU/PRTR".

R.S.K. acknowledges financial support from the ERC via Synergy Grant ``ECOGAL'' (project ID 855130) and from the German Excellence Strategy via the Heidelberg Cluster ``STRUCTURES'' (EXC 2181 - 390900948). In addition RSK is grateful for funding from the German Ministry for Economy and Energy (BMWE) in project ``MAINN'' (funding ID 50OO2206), and from DFG and ANR for project ``STARCLUSTERS'' (funding ID KL 1358/22-1).

M.C.S. and Z.F. acknowledge financial support from the European Research Council under the ERC Starting Grant “GalFlow” (grant 101116226). M.C.S. further acknowledges financial support from the Fondazione Cariplo under the grant ERC attrattivit\`{a} n. 2023-3014.

This research made use of astrodendro, a Python package to compute dendrograms of Astronomical data (http://www.dendrograms.org/)

\software{Astropy \citep{astropy:2013, astropy:2018, astropy:2022}, astrodendro \citep{astrodendro:2019}, Numpy, Scipy, LMFIT \citep{LMFIT_newville_2025_16175987},}
\facility{ALMA}

\newpage
\appendix 
%\counterwithin{figure}{section}
%\counterwithin{table}{section}

\section{List of properties for cataloged structures} \label{properties_table}
We present a catalog of all 235 identified HVCC structures in Table~\ref{tab:physical_properties}, and report properties as described in Section~\ref{final_cat_ppv}.  %We include the HVCC name using the identified molecule (CS) and the central position of the source; ID Number; central Galactic longitude ($\ell$); central Galactic latitude ($b$); central velocity determined by the peak channel from the dendrogram detection ($V_{\mathrm{LSR}}$); velocity width determined by the dendrogram ($\Delta V$); verticality; calculated effective size of the dendrogram structure; median integrated intensity of the dendrogram structure, fitted FWHM of the brightest spectral component within the identified velocity range; the number of detected peaks from the spectral fitting; PV feature classification including curve (CU), spike (S), clumps (CL), loop (L), and velocity-bridge (VB); and detection method including by-eye (e) or via dendrogram (d). We also note by-eye detections which had no dendrogram match from the downsampled cube (n). We show cross-matched ids from previous studies (Previous ID) and provide references. The dendrogram detection is complete to structures with $\mathrm{size} < 10~\mathrm{pc}$ and $\Delta V > 50~\mathrm{km/s}$. 11 by-eye detections (denoted by $\dagger$) did not produce any matching contours from the dendrogram analysis due to the noise threshold. For these rows, we report the by-eye velocity widths and $V_{\mathrm{LSR}}$, and do not report any physical properties.

\section{Example of dendrogram contours on full resolution subcube}\label{appendix_dendro_example}

The low resolution dendrogram was used to confirm the spatial and velocity extents for visually identified HVCC candidates, as well as identify extents for candidates which were initially missed by-eye. We then make a cutout of the full resolution \CS data cube, and rerun the 3D dendrogram using the local noise levels to identify the true structure of the HVCC (See Sec.~\ref{final_cat_ppv}). 

\begin{figure*}[h!]
\begin{minipage}{\textwidth}
\centering
\includegraphics[width=1\textwidth]{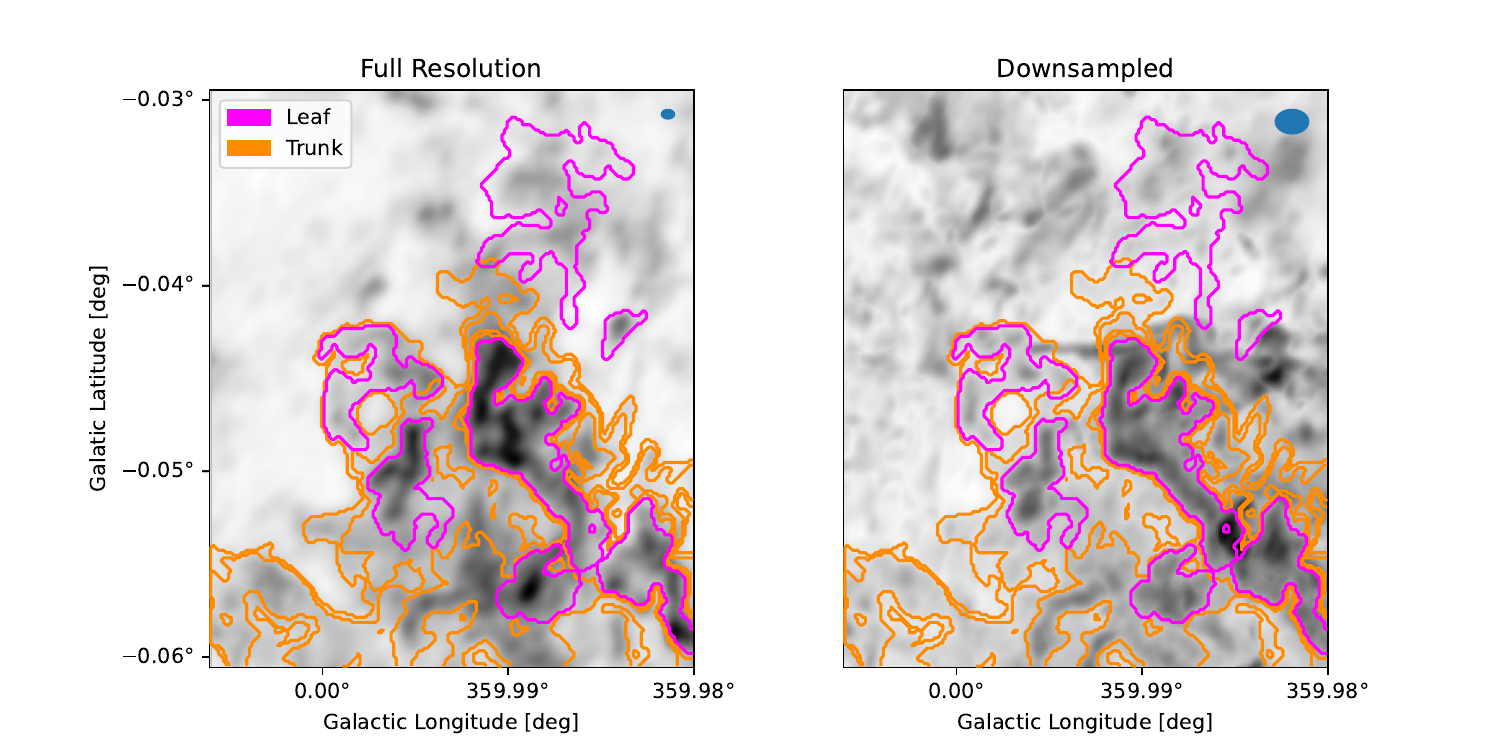}
\end{minipage}
\begin{minipage}{\textwidth}
\centering
\includegraphics[width=1\textwidth]{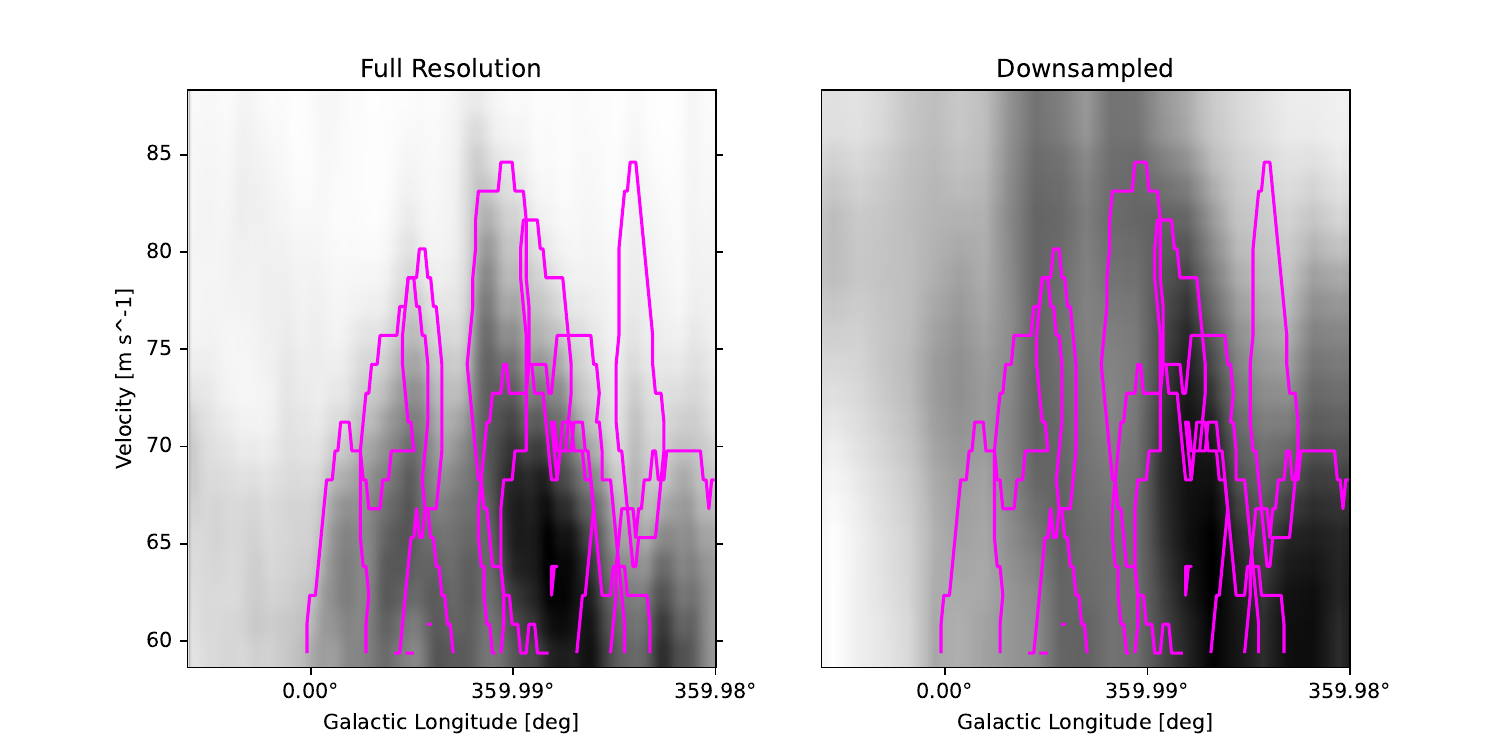}
\end{minipage}
\caption{Example of dendrogram contours from the full resolution subcube as described in Sec.~\ref{final_cat_ppv} for HVCC ID 89 (CS -0.007-0.045). The grey background image shows the integrated intensity map of the cutout for the full resolution cube (top left) and the downsampled cube (top right). We show the dendrogram trunk (orange) and composite of the associated leaf contours from the 3D dendrogram (pink). The corresponding beam size is shown a a blue ellipse in the top right corner of each panel of the top row. The bottom row shows the leaf contours for the source in LV space on the full resolution (bottom left) and downsampled (bottom right) data. The pink leaf contours from the dendrogram are used as the final mask containing the HVCC.}

\label{fig:appendix_dendro_ex}
\end{figure*}

An example output of the full resolution dendrogram is shown in the top row panels of Figure ~\ref{fig:appendix_dendro_ex} for ID 89 (CS -0.007-0.045). The grayscale backgrounds are integrated intensity maps of the full resolution (left) and downsampled (right) data, with the beam sizes shown as a blue ellipse in the top right corner of each panel. We plot the dendrogram branch structures (orange contours) and the leaves associated with the 3D dendrogram (pink contours). The bottom row shows the corresponding LV diagrams of the full resolution and downsampled cubes (bottom left and right, respectively), with the leaf contours plotted in pink. Note the spatial extents of the grayscale background directly correspond to the box used for displaying the spatial extents in Figure~\ref{fig: dendro}.

The final contours for the HVCC are the collection of pink leaf contours, which are a composite of the largest leaf structures across the subcube's velocity channels. We count the entire leaf structure as a single HVCC.

\section{Distributions of PV Morphological Classes}\label{appendix_PV_distributions}

\begin{figure*}[h!]
\begin{centering}
    
\includegraphics[width=\textwidth]{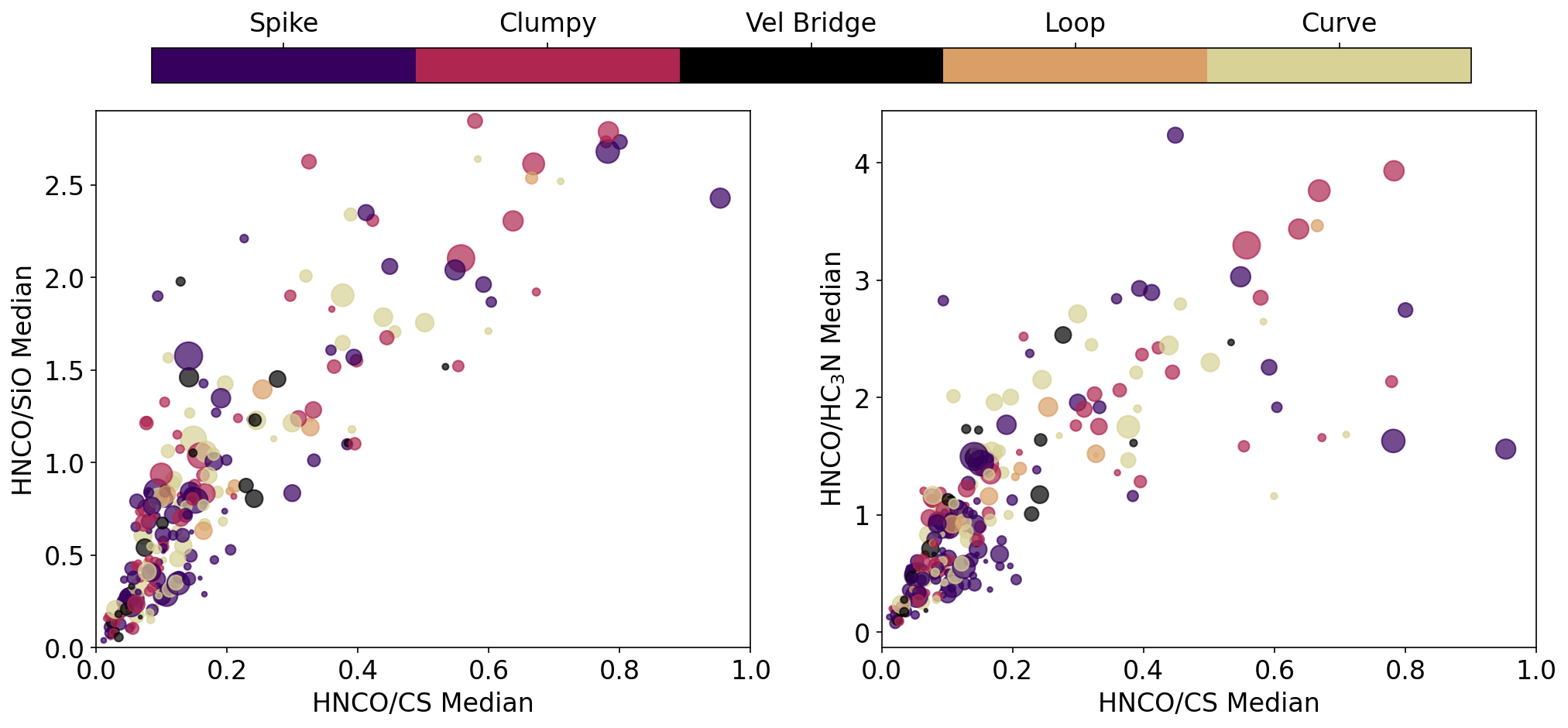}
\caption{Comparison of three line ratios discussed in the text, including HNCO/CS vs HNCO/SiO (left) and HNCO/CS vs HNCO/HC$_3$N (right). Colors correspond to the same PV morphological classes as Figure~\ref{fig:app_pvmorph_scatters}.}

\end{centering}

\label{fig:appendix_comp_HNCO_CS_HC3N}
\end{figure*}

While the trends in structure properties across longitude appear weak, we see a correlation between line emission ratios. Figure~\ref{fig:appendix_comp_HNCO_CS_HC3N} shows comparisons between the emission ratios of HNCO/CS vs HNCO/SiO (left) and HNCO/CS vs HNCO/HC$_3$N (right) for the cataloged structures. Both distributions show overall positive correlation between line ratios, particularly for the lower values, which are typically seen in the highly irradiated areas such as the Sgr A region.

To assess trends between different PV types, we separate the longitudinal trends in Figure~\ref{fig: scatter_stats} into individual morphological classes, shown in Figure~\ref{fig: scatter_stats}. The panels of Figure~\ref{fig: scatter_stats}  are the same for each PV morphological class. Each row shows Galactic longitude compared to the main fitted FWHM peak (left), HNCO/SiO line ratio (middle), and HNCO/CS line ratio (right). The morphological classes are separated by row, from top to bottom: (1) spike, (2) clumpy, (3) velocity bridges, (4) loop, and (5) curve features. Colors for PV class are the same as in Figure~\ref{fig: scatter_stats}. Likewise, marker sizes are scaled by the reported dendrogram contour size from Table~\ref{tab:physical_properties} in parsecs. We note the sample of spiked and clumpy morphologies with small sizes collected near \sgra, which also show comparatively low ratios of HNCO/HC$_3$N (see Section~\ref{subsec: results_structure_properties}).

\begin{longrotatetable}

\setlength\LTcapwidth{\textwidth}
\setlength{\tabcolsep}{3pt}
\footnotesize
%\startlongtable

\begin{longtable}{lcccccccccccccc}
\caption{List of cataloged structures. Columns include the HVCC name, ID Number; central Galactic longitude ($\ell$); central Galactic latitude ($b$); central velocity determined by the peak channel from the dendrogram detection ($V_{\mathrm{LSR}}$); velocity width determined by the dendrogram ($\Delta V$); verticality (i.e. the ratio between  velocity width and longitudinal extent); size of the dendrogram structure; median integrated intensity of the dendrogram structure, fitted FWHM of the brightest spectral component within the identified velocity range; the number of detected peaks from the spectral fitting; PV feature classification including curve (CU), spike (S), clumps (CL), loop (L), and velocity-bridge (VB); and detection method including by-eye (e) or via dendrogram (d). We also note by-eye detections which had no dendrogram match from the downsampled cube (n). We show cross-matched ids from previous studies (Previous ID) and provide references. The dendrogram detection is complete to structures with $\mathrm{size} < 10~\mathrm{pc}$ and $\Delta V > 50~\mathrm{km/s}$. 10 by-eye detections (denoted by $\dagger$) did not produce any matching contours from the dendrogram analysis due to the noise threshold. For these rows, we report the by-eye velocity widths and $V_{\mathrm{LSR}}$, and do not report any physical properties.}\label{tab:physical_properties}\\

\hline
HVCC Name & ID  & $l$ & $b$ & $V_{\mathrm{LSR}}$ & $\Delta V$ & Verticality & Size & Median &  Fitted & Peaks & PV & Detection &Other & Ref. \\

&&&&&&&&Intensity&FWHM&&Feature&Method&IDs&\\
\hline
 &  & $\mathrm{deg}$ & $\mathrm{deg}$ & $\mathrm{km\,s^{-1}}$  & $\mathrm{km\,s^{-1}}$ & $\mathrm{km\,deg^{-1}\,s^{-1}}$ & $\mathrm{pc}$ & $\mathrm{Jy\,beam^{-1}}$ & $\mathrm{km\,s^{-1}}$ &&&&&\\ \hline
 
\\
CS-0.582-0.062 & 18 & -0.582 & -0.062 & -71.3 & 20.8 & 1995.5 & 1.31 & 0.165 & 21.5 & 2 & S & e & & \\
CS -0.561-0.174 & 170 & -0.561 & -0.174 & -59.4 & 109.9 & 1252.5 & 8.61 & 0.081 & 27.8 & 2 & CU & d & 30 & b\footnote{b \citep{Oka2022}} \\
CS -0.558-0.182 & 169 & -0.558 & -0.182 & -59.4 & 90.6 & 1548.1 & 8.07 & 0.087 & 30.7 & 2 & CL & d & 36 & b \\
CS -0.551-0.192 & 171 & -0.551 & -0.192 & -23.8 & 63.8 & 3294.9 & 5.43 & 0.089 & 20.0 & 3 & CU & d & & \\
CS -0.549-0.118 & 216 & -0.549 & -0.118 & 66.8 & 54.9 & 1549.9 & 3.71 & 0.072 & 14.2 & 2 & S & d & 37 & b \\
CS -0.548-0.112 & 12 & -0.548 & -0.112 & -50.5 & 35.6 & 1433.3 & 2.59 & 0.149 & 27.6 & 3 & CL & e & & \\
CS -0.507-0.045 & 19 & -0.507 & -0.045 & -62.3 & 29.7 & 1027.9 & 1.4 & 0.084 & 10.2 & 1 & S & n & 38 & a\footnote{a \citep{Oka2012}}  \\
CS -0.507-0.212 & 6 & -0.507 & -0.212 & -117.3 & 23.8 & 1401.9 & 1.17 & 0.163 & 20.4 & 1 & CL & e & & \\
CS -0.493-0.222 & 164 & -0.493 & -0.222 & -127.7 & 63.8 & 3387.5 & 1.07 & 0.09 & 25.6 & 1 & S & d & & \\
CS -0.473-0.137 & 8 & -0.473 & -0.137 & -71.3 & 22.3 & 1471.0 & 0.82 & 0.129 & 16.4 & 1 & S & e & & \\
CS -0.462-0.256 & 3 & -0.462 & -0.256 & -96.5 & 31.2 & 1700.6 & 1.66 & 0.088 & 24.0 & 2 & S & e & & \\
CS -0.457-0.087 & 15 & -0.457 & -0.087 & -26.7 & 35.6 & 2565.5 & 1.9 & 0.121 & 18.2 & 1 & CL & e & & \\
CS -0.441-0.22 & 22 & -0.441 & -0.22 & -10.4 & 29.7 & 1228.7 & 1.48 & 0.105 & 22.1 & 1 & S & e & & \\
CS -0.439-0.028 & 185 & -0.439 & -0.028 & -22.3 & 69.8 & 3160.0 & 7.99 & 0.072 & 16.6 & 4 & S & d & & \\
CS -0.436-0.159 & 9 & -0.436 & -0.159 & -65.3 & 29.7 & 1811.8 & 0.78 & 0.117 & 16.5 & 2 & S & e & & \\
CS -0.432-0.195 & 7 & -0.432 & -0.195 & -26.7 & 23.8 & 2552.7 & 1.27 & 0.136 & 32.0 & 2 & CU, S & e & & \\
CS -0.432-0.078 & 13 & -0.432 & -0.078 & -13.4 & 44.5 & 3959.0 & 1.71 & 0.16 & 22.9 & 1 & CL & e & & \\
CS -0.43-0.041 & 193 & -0.43 & -0.041 & -22.3 & 59.4 & 4373.0 & 4.53 & 0.085 & 10.4 & 4 & S & d & & \\
CS -0.429-0.253 & 165 & -0.429 & -0.253 & -118.8 & 66.8 & 4615.5 & 2.78 & 0.093 & 15.1 & 4 & CU & d & & \\
CS -0.426-0.209 & 168 & -0.426 & -0.209 & -111.3 & 50.5 & 1372.8 & 0.39 & 0.071 & 18.1 & 2 & L & d & & \\
CS -0.419-0.252 & 167 & -0.419 & -0.252 & -120.2 & 50.5 & 2189.4 & 0.16 & 0.083 & 33.5 & 4 & VB & d & & \\
CS -0.407-0.22 & 20 & -0.407 & -0.22 & -77.2 & 66.8 & 3758.0 & 1.29 & 0.188 & 69.3 & 2 & CL & e & & \\
\hline
HVCC Name & ID  & $l$ & $b$ & $V_{\mathrm{LSR}}$ & $\Delta V$ & Verticality & Size & Median &  Fitted & Peaks & PV & Detection &Other & Ref. \\

&&&&&&&&Intensity&FWHM&&Feature&Method&IDs&\\
\hline
 &  & $\mathrm{deg}$ & $\mathrm{deg}$ & $\mathrm{km\,s^{-1}}$  & $\mathrm{km\,s^{-1}}$ & $\mathrm{km\,deg^{-1}\,s^{-1}}$ & $\mathrm{pc}$ & $\mathrm{Jy\,beam^{-1}}$ & $\mathrm{km\,s^{-1}}$ &&&&&\\ \hline
 
\\

CS -0.407-0.012 & 21 & -0.407 & -0.012 & -22.3 & 40.1 & 893.5 & 6.23 & 0.056 & 15.0 & 1 & S & n & & \\
CS -0.401-0.131 & 5 & -0.401 & -0.131 & 10.4 & 43.0 & 863.5 & 4.38 & 0.077 & 28.4 & 2 & CU & e & 52 & b \\
CS -0.382-0.253 & 4 & -0.382 & -0.253 & 20.8 & 34.1 & 979.5 & 3.3 & 0.136 & 5.6 & 2 & S & n & & \\
CS -0.373-0.006 & 183 & -0.373 & -0.006 & -4.5 & 60.9 & 1104.2 & 0.43 & 0.07 & 22.2 & 1 & VB & d & & \\
CS -0.354+0.058$^{\dagger}$ & 60 & -0.354 & 0.058 & 33.5 & 25.0 & nan & nan & nan & 6.1 & 3 & S & n & & \\
CS -0.341-0.004 & 23 & -0.341 & -0.004 & -68.3 & 29.7 & 1194.3 & 3.24 & 0.073 & 13.6 & 2 & CU & e & & \\
CS -0.338-0.11 & 75 & -0.338 & -0.11 & 4.5 & 22.3 & 13361.4 & 0.14 & 0.192 & 23.0 & 2 & CL & e & & \\
CS -0.328-0.114 & 30 & -0.328 & -0.114 & -37.1 & 47.5 & 2164.9 & 1.32 & 0.074 & 17.5 & 2 & S & e & 56 & b \\
CS -0.327-0.116 & 73 & -0.327 & -0.116 & -37.1 & 38.6 & 6316.3 & 0.38 & 0.191 & 13.2 & 2 & CU & e & 56 & b \\
CS -0.325-0.18 & 24 & -0.325 & -0.18 & -77.2 & 22.3 & 3143.8 & 0.82 & 0.094 & 16.3 & 1 & S & e & 57 & b \\
CS -0.323-0.087 & 11 & -0.323 & -0.087 & 20.8 & 29.7 & 1068.9 & 1.99 & 0.102 & 28.1 & 1 & S & n & & \\
CS -0.323-0.112 & 10 & -0.323 & -0.112 & -37.1 & 40.1 & 1612.3 & 3.52 & 0.067 & 20.4 & 1 & S & n & 56 & b \\
CS -0.32+0.054 & 59 & -0.32 & 0.054 & -100.9 & 17.8 & 4137.7 & 0.16 & 0.067 & 11.0 & 1 & S & e & & \\
CS -0.318-0.012 & 16 & -0.318 & -0.012 & -68.3 & 59.4 & 1082.4 & 7.85 & 0.105 & 37.3 & 1 & S & e & & \\
CS -0.307-0.062 & 14 & -0.307 & -0.062 & 1.5 & 118.8 & 7009.2 & 0.98 & 0.148 & 20.4 & 7 & VB & e & & \\
CS -0.307-0.062 & 77 & -0.307 & -0.062 & 1.5 & 63.8 & 984.2 & 6.28 & 0.068 & 23.6 & 2 & S & e & 58 & b \\
CS -0.307+0.03 & 2 & -0.307 & 0.03 & -32.7 & 35.6 & 1215.8 & 3.04 & 0.071 & 15.8 & 1 & S & e & 60 & b \\
CS -0.302-0.062 & 88 & -0.302 & -0.062 & 1.5 & 20.8 & 1720.1 & 1.69 & 0.261 & 20.7 & 1 & S & n & & \\
CS -0.29-0.028 & 52 & -0.29 & -0.028 & 28.2 & 20.8 & 4044.5 & 0.2 & 0.131 & 13.4 & 1 & S & e & & \\
CS -0.289+0.047$^{\dagger}$ & 55 & -0.289 & 0.047 & -38.0 & 20.0 & nan & nan & nan & 17.9 & 1 & S & e & & \\
CS -0.286+0.055 & 56 & -0.286 & 0.055 & -44.5 & 19.3 & 3563.0 & 0.45 & 0.138 & 15.8 & 1 & CL & e & & \\
CS -0.282+0.011 & 44 & -0.282 & 0.011 & -34.1 & 17.8 & 4750.6 & 0.46 & 0.124 & 12.2 & 1 & CU & e & & \\
CS -0.282+0.055 & 0 & -0.282 & 0.055 & -44.5 & 38.6 & 1208.3 & 3.21 & 0.116 & 18.2 & 1 & CU & e & & \\
CS -0.273-0.053 & 17 & -0.273 & -0.053 & -59.4 & 40.1 & 1149.8 & 4.71 & 0.053 & 15.7 & 1 & CL & e & 62 & b \\
CS -0.27-0.031 & 51 & -0.27 & -0.031 & -13.4 & 19.3 & 3656.7 & 0.54 & 0.124 & 9.0 & 1 & S & e & & \\
CS -0.27-0.165 & 25 & -0.27 & -0.165 & -37.1 & 20.8 & 11511.1 & 0.32 & 0.108 & 28.4 & 0 & S & e & & \\
CS -0.264-0.11 & 74 & -0.264 & -0.11 & 69.8 & 19.3 & 7719.8 & 0.28 & 0.146 & 10.8 & 1 & CU & e & & \\
CS -0.261+0.024 & 1 & -0.261 & 0.024 & -93.5 & 59.4 & 1566.1 & 3.67 & 0.059 & 26.6 & 2 & CU & e & 55, 63 & a, b \\
CS -0.258-0.021 & 207 & -0.258 & -0.021 & 43.0 & 53.4 & 937.6 & 2.17 & 0.086 & 10.9 & 4 & S & d & & \\
CS -0.244-0.025 & 50 & -0.244 & -0.025 & 46.0 & 28.2 & 4144.7 & 0.77 & 0.059 & 12.7 & 2 & CL & e & & \\
CS -0.232-0.088 & 177 & -0.232 & -0.088 & -49.0 & 80.2 & 4781.7 & 3.78 & 0.089 & 17.3 & 5 & CL & d & & \\
CS -0.219-0.143 & 27 & -0.219 & -0.143 & 17.8 & 23.8 & 3800.5 & 0.44 & 0.116 & 11.0 & 2 & CU & e & & \\
CS -0.194-0.08 & 192 & -0.194 & -0.08 & 43.0 & 60.9 & 1495.0 & 5.32 & 0.124 & 25.1 & 2 & CU & d & & \\
CS -0.18-0.128 & 29 & -0.18 & -0.128 & -8.9 & 14.8 & 2812.8 & 0.28 & 0.085 & 17.9 & 1 & CU & e & & \\
\hline
HVCC Name & ID  & $l$ & $b$ & $V_{\mathrm{LSR}}$ & $\Delta V$ & Verticality & Size & Median &  Fitted & Peaks & PV & Detection &Other & Ref. \\

&&&&&&&&Intensity&FWHM&&Feature&Method&IDs&\\
\hline
 &  & $\mathrm{deg}$ & $\mathrm{deg}$ & $\mathrm{km\,s^{-1}}$  & $\mathrm{km\,s^{-1}}$ & $\mathrm{km\,deg^{-1}\,s^{-1}}$ & $\mathrm{pc}$ & $\mathrm{Jy\,beam^{-1}}$ & $\mathrm{km\,s^{-1}}$ &&&&&\\ \hline
 
\\

CS -0.178+0.032 & 86 & -0.178 & 0.032 & 62.3 & 17.8 & 430.4 & 4.84 & 0.119 & 17.4 & 1 & S & e & & \\
CS -0.175+0.021 & 46 & -0.175 & 0.021 & 62.3 & 28.2 & 2256.5 & 1.72 & 0.232 & 22.6 & 1 & CU & e & 67 & b \\
CS -0.173+0.013 & 87 & -0.173 & 0.013 & 65.3 & 29.7 & 5089.8 & 0.88 & 0.126 & 13.3 & 2 & CL & e & & \\
CS -0.169+0.038 & 49 & -0.169 & 0.038 & 66.8 & 22.3 & 3910.5 & 0.37 & 0.116 & 27.4 & 1 & CL & e & & \\
CS -0.169-0.041 & 180 & -0.169 & -0.041 & -16.3 & 54.9 & 3312.0 & 3.74 & 0.103 & 16.3 & 2 & S & d & & \\
CS -0.168-0.087 & 38 & -0.168 & -0.087 & -34.1 & 13.4 & 3435.6 & 0.58 & 0.207 & 11.4 & 1 & S & e & & \\
CS -0.168-0.093 & 34 & -0.168 & -0.093 & -34.1 & 20.8 & 2413.6 & 0.69 & 0.152 & 48.0 & 1 & S & e & & \\
CS -0.162-0.03 & 203 & -0.162 & -0.03 & 17.8 & 80.2 & 1582.1 & 4.02 & 0.066 & 36.0 & 3 & CU & d & & \\
CS -0.133-0.088 & 85 & -0.133 & -0.088 & 16.3 & 25.2 & 1996.8 & 0.5 & 0.216 & 13.5 & 3 & CU & n & & \\
CS -0.114+0.037 & 188 & -0.114 & 0.037 & 19.3 & 52.0 & 1138.3 & 2.7 & 0.086 & 35.7 & 1 & S & d & 71 & b \\
CS -0.111-0.003 & 54 & -0.111 & -0.003 & 53.4 & 11.9 & 3717.8 & 0.36 & 0.165 & 15.1 & 1 & S & e & & \\
CS -0.108+0.021 & 47 & -0.108 & 0.021 & 72.7 & 26.7 & 6413.1 & 0.55 & 0.269 & 11.0 & 4 & CU & e & & \\
CS -0.107-0.012 & 83 & -0.107 & -0.012 & -133.6 & 26.7 & 6206.3 & 0.54 & 0.103 & 8.6 & 2 & CU & n & & \\
CS -0.104+0.017$^{\dagger}$ & 48 & -0.104 & 0.017 & 80.9 & 54.9 & nan & nan & nan & 20.5 & 2 & CL & e & & \\
CS -0.097+0.016 & 173 & -0.097 & 0.016 & -59.4 & 57.9 & 3534.1 & 4.53 & 0.071 & 28.5 & 3 & VB & d & & \\
CS -0.094-0.143 & 26 & -0.094 & -0.143 & 14.8 & 17.8 & 2672.1 & 0.53 & 0.09 & 30.1 & 0 & S & e & & \\
CS -0.093-0.135 & 28 & -0.093 & -0.135 & 69.8 & 28.2 & 4835.3 & 0.71 & 0.136 & 15.2 & 2 & CL & e & & \\
CS -0.082-0.046 & 82 & -0.082 & -0.046 & 8.9 & 20.8 & 2536.3 & 0.49 & 0.202 & 62.3 & 1 & S & e & & \\
CS -0.08-0.058 & 166 & -0.08 & -0.058 & -108.4 & 56.4 & 6730.3 & 1.37 & 0.13 & 37.0 & 1 & VB & d & & \\
CS -0.068-0.046 & 68 & -0.068 & -0.046 & -102.4 & 53.4 & 10688.5 & 0.35 & 0.154 & 52.4 & 2 & S & e & 76 & b \\
CS -0.064-0.04 & 174 & -0.064 & -0.04 & -35.6 & 60.9 & 1191.3 & 2.06 & 0.129 & 66.6 & 3 & S & d & & \\
CS -0.062-0.028 & 79 & -0.062 & -0.028 & 17.8 & 38.6 & 2259.4 & 1.23 & 0.148 & 17.9 & 4 & S & e & & \\
CS -0.06-0.041 & 72 & -0.06 & -0.041 & 81.6 & 37.1 & 5453.3 & 0.69 & 0.311 & 25.4 & 1 & CL & e & & \\
CS -0.06-0.042 & 69 & -0.06 & -0.042 & 81.6 & 34.1 & 3902.1 & 0.43 & 0.318 & 12.1 & 2 & CL & e & & \\
CS -0.059-0.05 & 71 & -0.059 & -0.05 & -13.4 & 41.6 & 9069.0 & 0.38 & 0.2 & 19.4 & 5 & S & e & & \\
CS -0.059+0.153 & 64 & -0.059 & 0.153 & 17.8 & 19.3 & 2573.2 & 0.6 & 0.251 & 7.9 & 1 & CL & n & & \\
CS -0.058-0.088 & 39 & -0.058 & -0.088 & 78.7 & 29.7 & 2969.0 & 1.26 & 0.227 & 38.6 & 1 & CL & e & & \\
CS -0.057-0.037 & 81 & -0.057 & -0.037 & 84.6 & 35.6 & 4500.4 & 0.21 & 0.258 & 38.0 & 1 & L & e & & \\
CS -0.056-0.101 & 31 & -0.056 & -0.101 & 66.8 & 20.8 & 1802.9 & 1.11 & 0.138 & 44.4 & 2 & S & e & & \\
CS -0.051+0.059 & 58 & -0.051 & 0.059 & 31.2 & 16.3 & 3094.0 & 0.38 & 0.127 & 25.2 & 1 & S & e & & \\
CS -0.048-0.062 & 78 & -0.048 & -0.062 & 54.9 & 8.9 & 1644.4 & 0.59 & 0.215 & 20.1 & 2 & S & e & & \\
CS -0.048-0.037 & 80 & -0.048 & -0.037 & 96.5 & 80.2 & 6273.7 & 1.52 & 0.084 & 33.7 & 1 & CL & e & & \\
CS -0.046-0.055 & 65 & -0.046 & -0.055 & 54.9 & 20.8 & 1847.4 & 1.19 & 0.337 & 13.4 & 2 & S & e & & \\
CS -0.045+0.058$^{\dagger}$ & 57 & -0.045 & 0.058 & 69.0 & 19.3 & nan & nan & nan & 12.5 & 1 & CL & e & & \\
\hline
HVCC Name & ID  & $l$ & $b$ & $V_{\mathrm{LSR}}$ & $\Delta V$ & Verticality & Size & Median &  Fitted & Peaks & PV & Detection &Other & Ref. \\

&&&&&&&&Intensity&FWHM&&Feature&Method&IDs&\\
\hline
 &  & $\mathrm{deg}$ & $\mathrm{deg}$ & $\mathrm{km\,s^{-1}}$  & $\mathrm{km\,s^{-1}}$ & $\mathrm{km\,deg^{-1}\,s^{-1}}$ & $\mathrm{pc}$ & $\mathrm{Jy\,beam^{-1}}$ & $\mathrm{km\,s^{-1}}$ &&&&&\\ \hline
 
\\

CS -0.042-0.045 & 67 & -0.042 & -0.045 & 106.9 & 46.0 & 4141.8 & 1.31 & 0.086 & 34.7 & 1 & S & e & 81 & b \\
CS -0.041-0.064 & 41 & -0.041 & -0.064 & -83.1 & 44.5 & 15269.2 & 0.45 & 0.382 & 12.7 & 3 & L & e & & \\
CS -0.04-0.034 & 53 & -0.04 & -0.034 & 108.4 & 37.1 & 5453.3 & 0.8 & 0.094 & 26.4 & 1 & VB & e & & \\
CS -0.036-0.098 & 35 & -0.036 & -0.098 & 83.1 & 16.3 & 3562.8 & 0.33 & 0.225 & 11.2 & 1 & CL & n & & \\
CS -0.036+0.082 & 63 & -0.036 & 0.082 & -26.7 & 17.8 & 2035.9 & 1.16 & 0.094 & 11.5 & 1 & S & e & & \\
CS -0.032-0.062 & 76 & -0.032 & -0.062 & -78.7 & 56.4 & 5207.2 & 1.18 & 0.114 & 28.3 & 3 & CL & e & & \\
CS -0.028-0.08$^{\dagger}$ & 40 & -0.028 & -0.08 & 66.0 & 22.0 & nan & nan & nan & 30.0 & 0 & CL & e & & \\
CS -0.028+0.073$^{\dagger}$ & 62 & -0.028 & 0.073 & 65.3 & 14.8 & nan & nan & nan & 21.6 & 0 & CU & e & & \\
CS -0.026+0.008 & 45 & -0.026 & 0.008 & -35.6 & 17.8 & 2729.0 & 0.45 & 0.16 & 9.8 & 1 & S & e & & \\
CS -0.026+0.009 & 43 & -0.026 & 0.009 & -35.6 & 20.8 & 2580.0 & 0.62 & 0.131 & 10.7 & 2 & CL & e & & \\
CS -0.019+0.073$^{\dagger}$ & 61 & -0.019 & 0.073 & 69.0 & 25.2 & nan & nan & nan & 15.2 & 1 & VB & e & & \\
CS -0.018-0.044 & 70 & -0.018 & -0.044 & -10.4 & 29.7 & 3750.3 & 1.08 & 0.299 & 11.7 & 2 & CL & e & & \\
CS -0.015-0.067$^{\dagger}$ & 42 & -0.015 & -0.067 & 42.0 & 24.0 & nan & nan & nan & 25.4 & 1 & CL & e & & \\
CS -0.015-0.045 & 90 & -0.015 & -0.045 & -10.4 & 25.2 & 3708.2 & 0.27 & 0.345 & 11.1 & 2 & CL & n & & \\
CS -0.014-0.097 & 33 & -0.014 & -0.097 & 34.1 & 28.2 & 3442.0 & 1.24 & 0.187 & 22.5 & 1 & CU & e & & \\
CS -0.013-0.05 & 172 & -0.013 & -0.05 & -10.4 & 60.9 & 1869.3 & 2.17 & 0.09 & 101.5 & 4 & S & d & & \\
CS -0.012-0.081 & 37 & -0.012 & -0.081 & 23.8 & 23.8 & 2514.9 & 0.36 & 0.432 & 31.9 & 3 & S & e & & \\
CS -0.011-0.044 & 66 & -0.011 & -0.044 & -43.0 & 46.0 & 5813.0 & 0.95 & 0.137 & 24.4 & 3 & S & e & HCN–0.009–0.044 & d\footnote{d \citep{Takekawa2019}} \\
CS -0.011-0.104 & 32 & -0.011 & -0.104 & 37.1 & 25.2 & 3364.9 & 1.05 & 0.127 & 11.2 & 3 & CL & n & & \\
CS -0.007-0.045 & 89 & -0.007 & -0.045 & 59.4 & 23.8 & 1194.5 & 2.34 & 0.278 & 31.7 & 1 & CU & e & & \\
CS -0.007-0.087 & 84 & -0.007 & -0.087 & 23.8 & 38.6 & 2746.7 & 3.89 & 0.336 & 35.3 & 0 & S & e & 84 & b \\
CS -0.005-0.087 & 36 & -0.005 & -0.087 & 28.2 & 10.4 & 644.0 & 3.88 & 0.347 & 27.3 & 1 & S & e & 84 & b \\
CS 0.006-0.013 & 182 & 0.006 & -0.013 & 26.7 & 62.3 & 3086.3 & 6.04 & 0.136 & 74.9 & 5 & CL & d & 71, 85 & a, b \\
CS 0.011-0.019 & 206 & 0.011 & -0.019 & 99.5 & 139.5 & 3346.1 & 4.01 & 0.131 & 56.7 & 4 & CU & d & 85, CO 0.02-0.0 & b, c\footnote{c \citep{Oka1999}} \\
CS 0.017+0.033$^{\dagger}$ & 93 & 0.017 & 0.033 & 81.5 & 49.0 & nan & nan & nan & 16.4 & 2 & CL & e & & \\
CS 0.036+0.117 & 227 & 0.036 & 0.117 & 75.7 & 57.9 & 4251.2 & 3.04 & 0.13 & 13.4 & 3 & CU & d & & \\
CS 0.07+0.045 & 175 & 0.07 & 0.045 & -43.0 & 53.4 & 1225.4 & 2.16 & 0.09 & 15.6 & 3 & CL & d & & \\
CS 0.071+0.045 & 176 & 0.071 & 0.045 & -43.0 & 53.4 & 2228.6 & 1.26 & 0.088 & 15.8 & 3 & CL & d & & \\
CS 0.086+0.056 & 231 & 0.086 & 0.056 & 141.0 & 83.1 & 4129.8 & 1.77 & 0.104 & 22.5 & 4 & VB & d & 94 & b \\
CS 0.096-0.132 & 196 & 0.096 & -0.132 & 40.1 & 50.5 & 2018.0 & 1.81 & 0.167 & 18.4 & 2 & CU & d & 95 & b \\
CS 0.11+0.16 & 92 & 0.11 & 0.16 & 118.8 & 22.3 & 1842.8 & 0.0 & 0.099 & 35.9 & 1 & S & e & & \\
CS 0.11+0.153$^{\dagger}$ & 91 & 0.11 & 0.153 & 98.5 & 19.0 & nan & nan & nan & 11.7 & 2 & S & n & & \\
\hline
HVCC Name & ID  & $l$ & $b$ & $V_{\mathrm{LSR}}$ & $\Delta V$ & Verticality & Size & Median &  Fitted & Peaks & PV & Detection &Other & Ref. \\

&&&&&&&&Intensity&FWHM&&Feature&Method&IDs&\\
\hline
 &  & $\mathrm{deg}$ & $\mathrm{deg}$ & $\mathrm{km\,s^{-1}}$  & $\mathrm{km\,s^{-1}}$ & $\mathrm{km\,deg^{-1}\,s^{-1}}$ & $\mathrm{pc}$ & $\mathrm{Jy\,beam^{-1}}$ & $\mathrm{km\,s^{-1}}$ &&&&&\\ \hline
 
\\

CS 0.121+0.033 & 215 & 0.121 & 0.033 & 59.4 & 56.4 & 1607.8 & 3.89 & 0.089 & 39.1 & 1 & S & d & & \\
CS 0.124-0.075 & 199 & 0.124 & -0.075 & 43.0 & 31.2 & 1229.5 & 0.68 & 0.167 & 54.4 & 2 & L & d & & \\
CS 0.129+0.042 & 229 & 0.129 & 0.042 & 77.2 & 62.3 & 2180.5 & 0.9 & 0.069 & 39.4 & 4 & CL & d & & \\
CS 0.129+0.019 & 179 & 0.129 & 0.019 & -8.9 & 54.9 & 2340.1 & 3.71 & 0.124 & 27.4 & 1 & VB & d & & \\
CS 0.13-0.097 & 96 & 0.13 & -0.097 & 74.2 & 65.3 & 9597.7 & 0.96 & 0.148 & 21.5 & 4 & CU & e & & \\
CS 0.13-0.14 & 97 & 0.13 & -0.14 & 44.5 & 43.0 & 1657.6 & 0.4 & 0.143 & 25.5 & 2 & CU & n & & \\
CS 0.149-0.099 & 214 & 0.149 & -0.099 & 62.3 & 47.5 & 1253.6 & 1.93 & 0.163 & 17.4 & 2 & L & d & & \\
CS 0.159-0.094 & 218 & 0.159 & -0.094 & 99.5 & 41.6 & 945.1 & 1.63 & 0.16 & 25.3 & 2 & CU & d & 103 & b \\
CS 0.16-0.137 & 94 & 0.16 & -0.137 & 46.0 & 26.7 & 3053.8 & 0.65 & 0.073 & 33.9 & 2 & VB & e & & \\
CS 0.176-0.008 & 186 & 0.176 & -0.008 & -10.4 & 50.5 & 2061.3 & 2.51 & 0.061 & 17.6 & 4 & S & d & & \\
CS 0.18-0.111 & 95 & 0.18 & -0.111 & 81.6 & 23.8 & 2164.7 & 0.93 & 0.106 & 33.5 & 2 & CL & e & 86 & a \\
CS 0.191+0.032 & 233 & 0.191 & 0.032 & 130.6 & 62.3 & 8746.3 & 0.62 & 0.064 & 17.6 & 2 & VB & d & & \\
CS 0.244+0.01 & 208 & 0.244 & 0.01 & 34.1 & 44.5 & 2402.1 & 0.43 & 0.146 & 22.1 & 2 & CL & d & & \\
CS 0.269+0.042 & 178 & 0.269 & 0.042 & -34.1 & 50.5 & 1296.6 & 0.55 & 0.12 & 14.7 & 3 & CU & d & & \\
CS 0.293+0.013 & 106 & 0.293 & 0.013 & 118.8 & 35.6 & 2788.3 & 0.92 & 0.054 & 12.7 & 2 & CL & n & & \\
CS 0.304-0.089 & 126 & 0.304 & -0.089 & -77.2 & 23.8 & 2279.2 & 0.82 & 0.083 & 19.9 & 2 & S & e & & \\
CS 0.321-0.136 & 136 & 0.321 & -0.136 & 10.4 & 20.8 & 1817.1 & 1.28 & 0.064 & 10.7 & 1 & S & e & & \\
CS 0.325+0.027 & 184 & 0.325 & 0.027 & 7.4 & 74.2 & 1219.4 & 9.8 & 0.087 & 24.0 & 3 & S & d & 112 & b \\
CS 0.326-0.086 & 224 & 0.326 & -0.086 & 93.5 & 68.3 & 2147.7 & 4.14 & 0.109 & 15.5 & 3 & CU & d & 113 & b \\
CS 0.334-0.008 & 145 & 0.334 & -0.008 & 72.7 & 38.6 & 1312.6 & 3.12 & 0.072 & 15.0 & 2 & CU & e & & \\
CS 0.336-0.026 & 154 & 0.336 & -0.026 & 96.5 & 31.2 & 1760.5 & 4.47 & 0.083 & 11.8 & 3 & L & e & & \\
CS 0.354+0.049 & 149 & 0.354 & 0.049 & 95.0 & 28.2 & 1403.6 & 1.53 & 0.11 & 14.7 & 2 & CL & e & & \\
CS 0.361+0.064 & 130 & 0.361 & 0.064 & -11.9 & 31.2 & 3079.8 & 2.01 & 0.092 & 12.7 & 2 & CU & e & & \\
CS 0.369-0.015 & 143 & 0.369 & -0.015 & 65.3 & 54.9 & 1985.1 & 5.05 & 0.082 & 28.3 & 1 & CL & e & 101 & a \\
CS 0.377-0.109 & 146 & 0.377 & -0.109 & 66.8 & 20.8 & 1969.0 & 1.34 & 0.089 & 15.5 & 1 & S & e & & \\
CS 0.382-0.045 & 219 & 0.382 & -0.045 & 81.6 & 62.3 & 2308.9 & 8.03 & 0.101 & 19.3 & 2 & S & d & 104 & a \\
CS 0.39-0.05 & 134 & 0.39 & -0.05 & -8.9 & 22.3 & 1429.7 & 1.08 & 0.087 & 10.2 & 2 & CL & e & & \\
CS 0.393-0.087 & 101 & 0.393 & -0.087 & 108.4 & 60.9 & 2180.3 & 3.72 & 0.115 & 20.3 & 2 & L & e & 117 & b \\
CS 0.396-0.023 & 137 & 0.396 & -0.023 & 23.8 & 25.2 & 1194.3 & 0.74 & 0.052 & 17.2 & 1 & S & e & & \\
CS 0.397-0.1 & 148 & 0.397 & -0.1 & 90.6 & 20.8 & 1971.7 & 1.57 & 0.137 & 13.8 & 1 & CU & e & & \\
CS 0.41-0.082 & 225 & 0.41 & -0.082 & 96.5 & 63.8 & 1721.1 & 0.24 & 0.077 & 16.4 & 4 & S & d & & \\
CS 0.411-0.042 & 142 & 0.411 & -0.042 & 52.0 & 20.8 & 1345.2 & 0.99 & 0.081 & 62.3 & 0 & CU & e & & \\
CS 0.411+0.074 & 131 & 0.411 & 0.074 & -19.3 & 25.2 & 1336.1 & 0.69 & 0.079 & 24.1 & 1 & CU & e & & \\
CS 0.416-0.042 & 139 & 0.416 & -0.042 & 52.0 & 19.3 & 2012.0 & 0.83 & 0.088 & 57.9 & 0 & CU & e & & \\
\hline
HVCC Name & ID  & $l$ & $b$ & $V_{\mathrm{LSR}}$ & $\Delta V$ & Verticality & Size & Median &  Fitted & Peaks & PV & Detection &Other & Ref. \\

&&&&&&&&Intensity&FWHM&&Feature&Method&IDs&\\
\hline
 &  & $\mathrm{deg}$ & $\mathrm{deg}$ & $\mathrm{km\,s^{-1}}$  & $\mathrm{km\,s^{-1}}$ & $\mathrm{km\,deg^{-1}\,s^{-1}}$ & $\mathrm{pc}$ & $\mathrm{Jy\,beam^{-1}}$ & $\mathrm{km\,s^{-1}}$ &&&&&\\ \hline
 
\\

CS 0.419-0.065 & 141 & 0.419 & -0.065 & 53.4 & 25.2 & 1312.6 & 2.29 & 0.053 & 62.0 & 0 & S & e & & \\
CS 0.419-0.052 & 128 & 0.419 & -0.052 & -26.7 & 47.5 & 1498.4 & 2.88 & 0.073 & 13.3 & 3 & CU & e & & \\
CS 0.423+0.039 & 189 & 0.423 & 0.039 & 46.0 & 60.9 & 3006.4 & 2.69 & 0.101 & 21.8 & 4 & CL & d & 107 & a \\
CS 0.424-0.078 & 158 & 0.424 & -0.078 & 111.3 & 25.2 & 1864.4 & 1.12 & 0.074 & 19.3 & 2 & S & e & & \\
CS 0.424-0.086 & 160 & 0.424 & -0.086 & 108.4 & 22.3 & 1496.4 & 0.19 & 0.087 & 56.4 & 2 & S & e & & \\
CS 0.427-0.078 & 102 & 0.427 & -0.078 & -26.7 & 68.3 & 1052.8 & 6.52 & 0.047 & 26.4 & 1 & S & e & & \\
CS 0.427-0.072 & 127 & 0.427 & -0.072 & -37.1 & 43.0 & 1772.2 & 2.67 & 0.084 & 38.5 & 1 & CU & e & & \\
CS 0.432+0.019 & 152 & 0.432 & 0.019 & 87.6 & 20.8 & 1608.0 & 1.43 & 0.06 & 13.3 & 1 & S & e & & \\
CS 0.432+0.064 & 156 & 0.432 & 0.064 & 109.9 & 19.3 & 1606.1 & 0.88 & 0.061 & 20.2 & 1 & S & e & & \\
CS 0.438-0.016 & 133 & 0.438 & -0.016 & -13.4 & 20.8 & 1325.0 & 1.87 & 0.086 & 7.1 & 3 & CL & e & & \\
CS 0.446-0.056 & 181 & 0.446 & -0.056 & 5.9 & 65.3 & 5053.9 & 0.5 & 0.084 & 31.7 & 2 & L & d & & \\
CS 0.452-0.021 & 228 & 0.452 & -0.021 & 81.6 & 52.0 & 3562.8 & 0.7 & 0.089 & 27.0 & 2 & CL & d & & \\
CS 0.454+0.049 & 153 & 0.454 & 0.049 & 93.5 & 25.2 & 1206.8 & 0.97 & 0.073 & 14.1 & 1 & VB & e & 121 & b \\
CS 0.454+0.03 & 147 & 0.454 & 0.03 & 83.1 & 31.2 & 1681.3 & 1.7 & 0.069 & 20.5 & 1 & VB & e & & \\
CS 0.473-0.072 & 198 & 0.473 & -0.072 & 41.6 & 50.5 & 2022.1 & 3.42 & 0.079 & 18.2 & 5 & CL & d & & \\
CS 0.493-0.087 & 109 & 0.493 & -0.087 & 3.0 & 54.9 & 1668.7 & 3.82 & 0.048 & 34.2 & 1 & L & e & & \\
CS 0.494-0.032 & 230 & 0.494 & -0.032 & 102.4 & 75.7 & 3694.7 & 3.96 & 0.092 & 23.4 & 2 & S & d & 124 & b \\
CS 0.502+0.022 & 107 & 0.502 & 0.022 & 49.0 & 43.0 & 1890.1 & 1.79 & 0.104 & 12.5 & 1 & CL & e & & \\
CS 0.509-0.02 & 162 & 0.509 & -0.02 & 132.1 & 32.7 & 1325.1 & 1.65 & 0.07 & 22.7 & 1 & S & e & & \\
CS 0.51-0.02 & 103 & 0.51 & -0.02 & 132.1 & 89.1 & 1652.9 & 3.73 & 0.072 & 24.2 & 1 & CL & e & & \\
CS 0.515-0.018 & 232 & 0.515 & -0.018 & 83.1 & 50.5 & 6234.9 & 1.11 & 0.084 & 27.5 & 2 & CL & d & & \\
CS 0.516+0.059 & 138 & 0.516 & 0.059 & 35.6 & 29.7 & 1138.9 & 1.86 & 0.093 & 27.7 & 0 & CU & e & & \\
CS 0.52+0.077 & 157 & 0.52 & 0.077 & 123.2 & 43.0 & 2359.8 & 0.66 & 0.052 & 21.3 & 2 & CL & e & & \\
CS 0.527+0.08 & 108 & 0.527 & 0.08 & 123.2 & 56.4 & 1766.0 & 3.99 & 0.041 & 36.6 & 2 & CU & n & & \\
CS 0.542-0.011 & 161 & 0.542 & -0.011 & 127.7 & 31.2 & 1943.4 & 2.3 & 0.045 & 11.2 & 2 & S & e & & \\
CS 0.547+0.032 & 132 & 0.547 & 0.032 & -17.8 & 25.2 & 1467.1 & 1.82 & 0.077 & 30.6 & 1 & CL & e & & \\
CS 0.554+0.04 & 129 & 0.554 & 0.04 & -17.8 & 22.3 & 1740.0 & 1.52 & 0.068 & 14.8 & 1 & CL & e & & \\
CS 0.559+0.055 & 163 & 0.559 & 0.055 & 123.2 & 22.3 & 2242.2 & 0.75 & 0.065 & 11.0 & 1 & VB & e & & \\
CS 0.56-0.103 & 100 & 0.56 & -0.103 & -8.9 & 43.0 & 1169.7 & 4.6 & 0.031 & 14.7 & 2 & S & n & & \\
CS 0.56-0.082 & 155 & 0.56 & -0.082 & 108.4 & 20.8 & 1389.5 & 1.01 & 0.058 & 14.9 & 1 & S & e & & \\
CS 0.56+0.064 & 135 & 0.56 & 0.064 & -3.0 & 31.2 & 1760.5 & 3.35 & 0.042 & 11.4 & 1 & CL & e & & \\
CS 0.561+0.05 & 144 & 0.561 & 0.05 & 68.3 & 19.3 & 1292.6 & 0.88 & 0.094 & 29.9 & 0 & S & e & & \\
CS 0.565+0.054 & 140 & 0.565 & 0.054 & 50.5 & 23.8 & 1157.4 & 1.51 & 0.083 & 8.3 & 2 & CU & e & & \\
CS 0.574-0.125 & 150 & 0.574 & -0.125 & 90.6 & 19.3 & 1575.2 & 1.28 & 0.085 & 13.9 & 1 & CU & e & & \\
\hline
HVCC Name & ID  & $l$ & $b$ & $V_{\mathrm{LSR}}$ & $\Delta V$ & Verticality & Size & Median &  Fitted & Peaks & PV & Detection &Other & Ref. \\

&&&&&&&&Intensity&FWHM&&Feature&Method&IDs&\\
\hline
 &  & $\mathrm{deg}$ & $\mathrm{deg}$ & $\mathrm{km\,s^{-1}}$  & $\mathrm{km\,s^{-1}}$ & $\mathrm{km\,deg^{-1}\,s^{-1}}$ & $\mathrm{pc}$ & $\mathrm{Jy\,beam^{-1}}$ & $\mathrm{km\,s^{-1}}$ &&&&&\\ \hline
 
\\

CS 0.594-0.124 & 151 & 0.594 & -0.124 & 84.6 & 20.8 & 1312.6 & 2.13 & 0.068 & 19.5 & 1 & CU & e & & \\
CS 0.596-0.005 & 159 & 0.596 & -0.005 & 112.8 & 22.3 & 1687.7 & 0.86 & 0.061 & 66.8 & 1 & CU & e & & \\
CS 0.596-0.028 & 226 & 0.596 & -0.028 & 102.4 & 56.4 & 1187.6 & 1.86 & 0.131 & 26.1 & 2 & VB & d & 115 & a \\
CS 0.602+0.005 & 105 & 0.602 & 0.005 & 108.4 & 60.9 & 2107.0 & 1.72 & 0.093 & 107.1 & 1 & CL & e & 131 & b \\
CS 0.602-0.003 & 104 & 0.602 & -0.003 & 138.1 & 66.8 & 1134.4 & 3.46 & 0.052 & 16.6 & 2 & S & e & & \\
CS 0.602-0.145 & 98 & 0.602 & -0.145 & 93.5 & 19.3 & 2526.5 & 0.72 & 0.114 & 17.2 & 1 & CU & n & & \\
CS 0.61-0.006 & 221 & 0.61 & -0.006 & 75.7 & 52.0 & 962.2 & 3.14 & 0.1 & 15.9 & 4 & S & d & & \\
CS 0.61-0.02 & 110 & 0.61 & -0.02 & 31.2 & 37.1 & 2071.5 & 0.5 & 0.116 & 27.7 & 1 & CU & n & & \\
CS 0.621-0.038 & 217 & 0.621 & -0.038 & 87.6 & 86.1 & 1824.8 & 6.27 & 0.17 & 27.3 & 2 & CU & d & 129 & b \\
CS 0.631-0.032 & 220 & 0.631 & -0.032 & 87.6 & 80.2 & 2101.8 & 2.73 & 0.187 & 23.2 & 2 & CU & d & 118, 133 & a, b \\
CS 0.635-0.062 & 118 & 0.635 & -0.062 & 35.6 & 74.2 & 2873.4 & 1.71 & 0.151 & 32.9 & 3 & CL & e & & \\
CS 0.649-0.081 & 223 & 0.649 & -0.081 & 95.0 & 53.4 & 1247.9 & 4.27 & 0.115 & 18.3 & 3 & CU & d & 136 & b \\
CS 0.652-0.112 & 99 & 0.652 & -0.112 & 114.3 & 40.1 & 1560.0 & 2.08 & 0.027 & 28.4 & 1 & S & n & & \\
CS 0.663-0.04 & 222 & 0.663 & -0.04 & 52.0 & 68.3 & 1725.6 & 6.77 & 0.186 & 16.8 & 3 & S & d & & \\
CS 0.695+0.04 & 205 & 0.695 & 0.04 & 37.1 & 50.5 & 6247.2 & 4.14 & 0.136 & 26.7 & 2 & CU & d & & \\
CS 0.718-0.053 & 116 & 0.718 & -0.053 & 22.3 & 35.6 & 1329.3 & 3.74 & 0.108 & 19.2 & 4 & L & e & & \\
CS 0.72+0.025 & 190 & 0.72 & 0.025 & 11.9 & 66.8 & 3408.3 & 3.1 & 0.102 & 39.5 & 4 & CL & d & & \\
CS 0.726-0.026 & 194 & 0.726 & -0.026 & 34.1 & 53.4 & 1609.6 & 1.92 & 0.086 & 43.0 & 1 & CL & d & & \\
CS 0.727+0.005 & 211 & 0.727 & 0.005 & 77.2 & 65.3 & 1281.0 & 2.6 & 0.116 & 43.6 & 3 & S & d & & \\
CS 0.727-0.085 & 213 & 0.727 & -0.085 & 81.6 & 57.9 & 2758.3 & 1.8 & 0.091 & 24.0 & 2 & CU & d & & \\
CS 0.731-0.215 & 201 & 0.731 & -0.215 & 22.3 & 80.2 & 2340.3 & 3.94 & 0.084 & 20.5 & 5 & CU & d & & \\
CS 0.734-0.143 & 212 & 0.734 & -0.143 & 11.9 & 86.1 & 2760.9 & 9.35 & 0.088 & 27.3 & 4 & CL & d & & \\
CS 0.741-0.223 & 209 & 0.741 & -0.223 & 54.9 & 59.4 & 2613.1 & 3.02 & 0.092 & 29.0 & 2 & S & d & & \\
CS 0.742-0.194 & 202 & 0.742 & -0.194 & 29.7 & 54.9 & 2083.7 & 3.08 & 0.066 & 50.4 & 2 & S & d & & \\
CS 0.743-0.02 & 111 & 0.743 & -0.02 & 72.7 & 25.2 & 435.8 & 1.26 & 0.108 & 22.5 & 1 & S & e & & \\
CS 0.743-0.003 & 112 & 0.743 & -0.003 & 1.5 & 16.3 & 4054.6 & 0.39 & 0.16 & 16.8 & 1 & CL & e & & \\
CS 0.752-0.145 & 120 & 0.752 & -0.145 & 102.4 & 47.5 & 1591.0 & 3.28 & 0.078 & 18.1 & 3 & CU & e & 140 & b \\
CS 0.76-0.07 & 125 & 0.76 & -0.07 & 29.7 & 7.4 & 2111.3 & 0.74 & 0.096 & 16.1 & 1 & CL & e & & \\
CS 0.76-0.07 & 124 & 0.76 & -0.07 & 8.9 & 34.1 & 1272.6 & 4.93 & 0.11 & 102.4 & 1 & S & n & 141 & b \\
CS 0.76-0.078 & 119 & 0.76 & -0.078 & 50.5 & 57.9 & 1820.5 & 2.99 & 0.151 & 21.0 & 3 & S & e & 141 & b \\
CS 0.768-0.062 & 117 & 0.768 & -0.062 & 41.6 & 50.5 & 2202.7 & 1.3 & 0.156 & 20.8 & 2 & S & e & & \\
CS 0.768-0.178 & 121 & 0.768 & -0.178 & 17.8 & 66.8 & 1762.0 & 5.04 & 0.09 & 16.4 & 4 & CL & n & & \\
CS 0.768-0.037 & 113 & 0.768 & -0.037 & 41.6 & 46.0 & 2008.4 & 0.48 & 0.17 & 49.4 & 2 & VB & e & & \\
CS 0.777-0.101 & 195 & 0.777 & -0.101 & 46.0 & 57.9 & 1345.9 & 2.2 & 0.103 & 37.7 & 4 & CL & d & & \\
\hline
HVCC Name & ID  & $l$ & $b$ & $V_{\mathrm{LSR}}$ & $\Delta V$ & Verticality & Size & Median &  Fitted & Peaks & PV & Detection &Other & Ref. \\

&&&&&&&&Intensity&FWHM&&Feature&Method&IDs&\\
\hline
 &  & $\mathrm{deg}$ & $\mathrm{deg}$ & $\mathrm{km\,s^{-1}}$  & $\mathrm{km\,s^{-1}}$ & $\mathrm{km\,deg^{-1}\,s^{-1}}$ & $\mathrm{pc}$ & $\mathrm{Jy\,beam^{-1}}$ & $\mathrm{km\,s^{-1}}$ &&&&&\\ \hline
 
\\

CS 0.777-0.187 & 122 & 0.777 & -0.187 & 19.3 & 56.4 & 1257.6 & 5.89 & 0.082 & 16.8 & 4 & CL & e & & \\
CS 0.785-0.156 & 197 & 0.785 & -0.156 & 38.6 & 54.9 & 1629.7 & 1.78 & 0.095 & 27.1 & 4 & L & d & & \\
CS 0.785-0.203 & 123 & 0.785 & -0.203 & 83.1 & 50.5 & 1808.2 & 3.34 & 0.107 & 18.0 & 1 & VB & e & & \\
CS 0.801-0.103 & 187 & 0.801 & -0.103 & -5.9 & 63.8 & 928.3 & 3.29 & 0.083 & 31.1 & 4 & CL & d & & \\
CS 0.806-0.168 & 204 & 0.806 & -0.168 & 11.9 & 54.9 & 1240.4 & 5.08 & 0.079 & 31.7 & 3 & CL & d & & \\
CS 0.813-0.093 & 191 & 0.813 & -0.093 & -10.4 & 65.3 & 1818.4 & 2.42 & 0.072 & 34.7 & 4 & CL & d & & \\
CS 0.829-0.131 & 200 & 0.829 & -0.131 & 11.9 & 53.4 & 2246.0 & 5.01 & 0.079 & 24.9 & 3 & S & d & & \\
CS 0.842-0.242 & 234 & 0.842 & -0.242 & 133.6 & 53.4 & 7483.0 & 0.96 & 0.147 & 15.6 & 2 & VB & d & & \\
CS 0.846-0.26 & 210 & 0.846 & -0.26 & 49.0 & 49.0 & 1644.5 & 1.96 & 0.08 & 19.2 & 3 & S & d & & \\
CS 0.852-0.22 & 115 & 0.852 & -0.22 & 68.3 & 60.9 & 2077.2 & 2.51 & 0.164 & 22.1 & 3 & VB & e & 145 & b \\
CS 0.852-0.17 & 114 & 0.852 & -0.17 & 4.5 & 40.1 & 1551.8 & 3.83 & 0.096 & 51.0 & 1 & VB & e & 144 & b \\

\hline
\hline

\end{longtable}
\end{longrotatetable}

\begin{figure*}[h!]
\begin{minipage}{\textwidth}
\centering
\includegraphics[width=.95\textwidth]{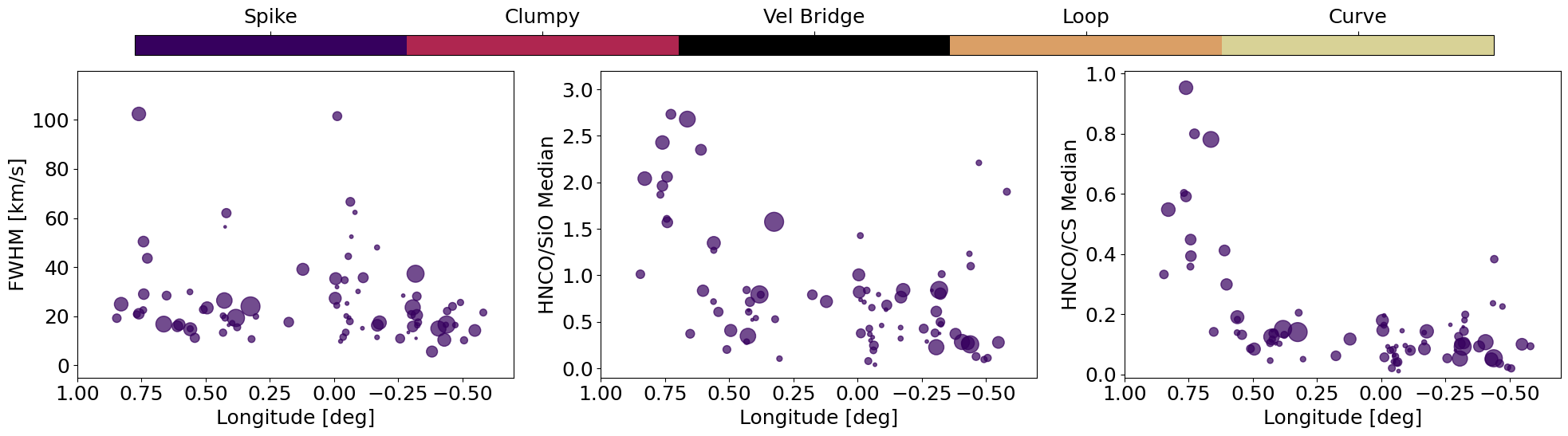}
\end{minipage}
\begin{minipage}{\textwidth}
\centering
\includegraphics[width=.95\textwidth]{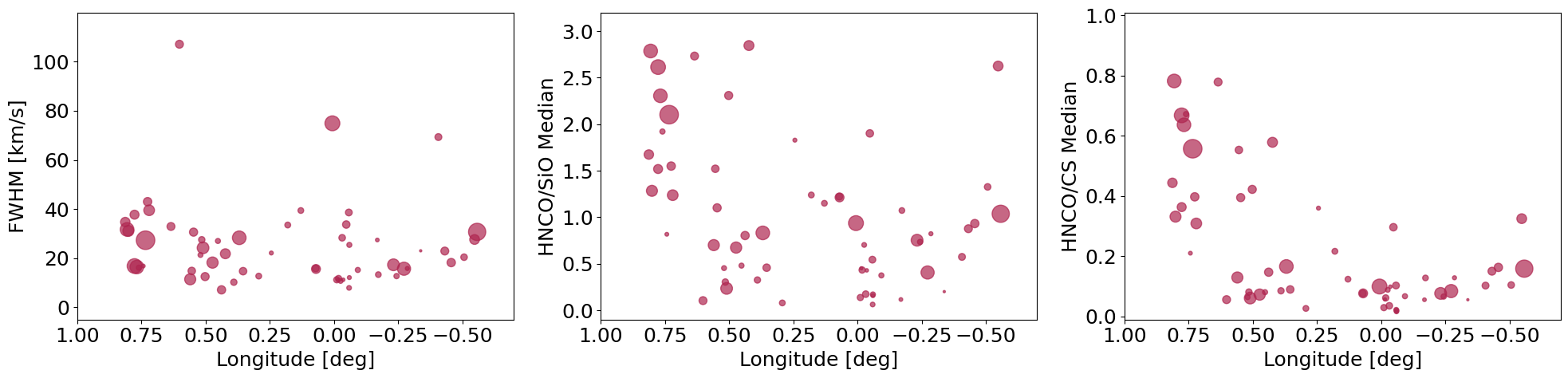}
\end{minipage}
\begin{minipage}{\textwidth}
\centering
\includegraphics[width=.95\textwidth]{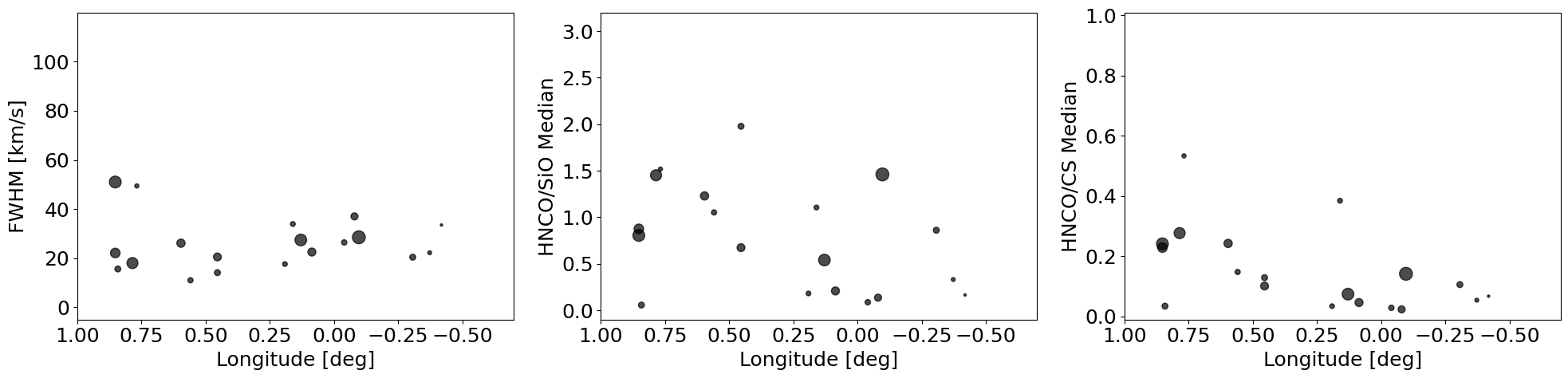}
\end{minipage}
\begin{minipage}{\textwidth}
\centering
\includegraphics[width=.95\textwidth]{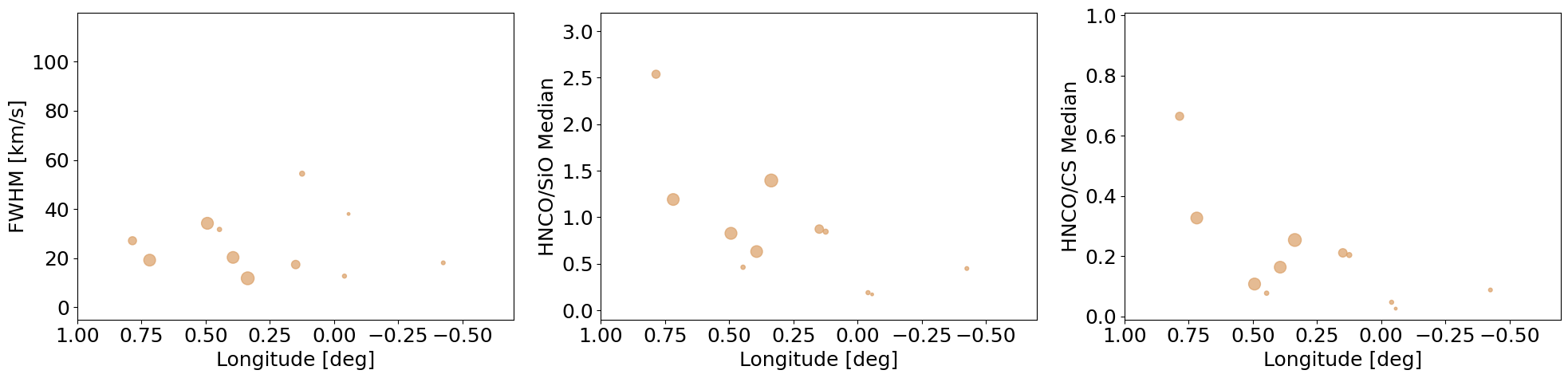}
\end{minipage}
\begin{minipage}{\textwidth}
\centering
\includegraphics[width=.95\textwidth]{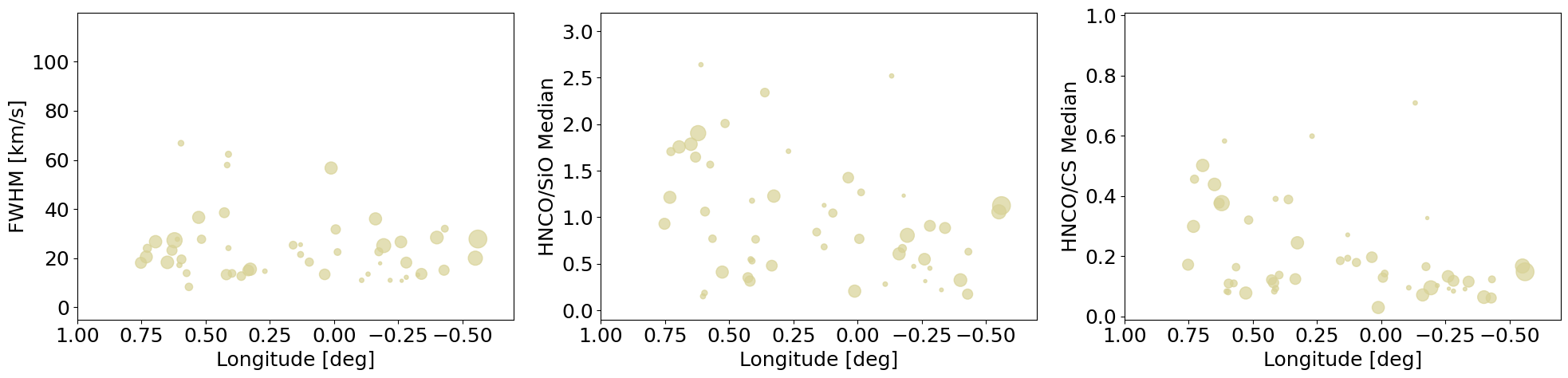}
\end{minipage}
\caption{This figure shows the panels from Figure~\ref{fig: scatter_stats} broken into individual PV classes. Each row shows Galactic longitude compared to the main fitted FWHM peak (left), HNCO/SiO line ratio (middle), and HNCO/CS line ratio (right). Each row shows the HVCCs of a single morphology with colors matching those from Figure~\ref{fig: scatter_stats}, including, from top to bottom: (1) spike, (2) clumpy, (3) velocity bridges, (4) loop, and (5) curve features. Marker sizes are scaled by the reported dendrogram contour size from Table~\ref{tab:physical_properties} in parsecs.}
\label{fig:app_pvmorph_scatters}
\end{figure*}

\clearpage

\bibliography{all_refs}{}

@ARTICLE{Binney1991,
       author = {{Binney}, James and {Gerhard}, Ortwin E. and {Stark}, Antony A. and {Bally}, John and {Uchida}, Keven I.},
        title = "{Understanding the kinematics of Galactic Centre gas.}",
      journal = {\mnras},
         year = 1991,
        month = sep,
       volume = {252},
        pages = {210},
          doi = {10.1093/mnras/252.2.210},
       adsurl = {https://ui.adsabs.harvard.edu/abs/1991MNRAS.252..210B}
}

@ARTICLE{Longmore2026_ACESI,
       author = {{Longmore}, Steven N. and {Bally}, John and {Barnes}, Ashley T. and {Battersby}, Cara and {Colzi}, Laura and {Ginsburg}, Adam and {Henshaw}, Jonathan D. and {Ho}, Paul T.~P. and {Jim{\'e}nez-Serra}, Izaskun and {Kruijssen}, J.~M. Diederik and {Mills}, Elisabeth A.~C. and {Petkova}, Maya A. and {Sormani}, Mattia C. and {Tress}, Robin G. and {Walker}, Daniel L. and {Wallace}, Jennifer and {Alkhuja}, Emad and {Armillotta}, Lucia and {Budaiev}, Nazar and {Buddhacharya}, Rojita and {Bulatek}, Alyssa and {Burton}, Michael and {Butterfield}, Natalie O. and {Busch}, Laura A. and {Caselli}, Paola and {Chevance}, M{\'e}lanie and {Cook}, Claire and {Crowe}, Samuel and {D{\'\i}az-Rodr{\'\i}guez}, Ana Karla and {DiTeodoro}, Enrico and {Dicker}, Simon R. and {Dutkowska}, Katarzyna M. and {Fairley}, Adam and {Federrath}, Christoph and {Fedriani}, Rub{\'e}n and {Feng}, Zi-Xuan and {Fiteni}, Karl and {Fuller}, Gary and {Garc{\'\i}a}, Pablo and {Goicoechea}, Javier and {Girichidis}, Philipp and {Glover}, Simon C.~O. and {Gorski}, Mark and {Gramze}, Savannah R. and {Gu}, Qi-Lao and {Hatchfield}, H. Perry and {Henkel}, Christian and {Houghton}, Rebecca J. and {Hsieh}, Pei-Ying and {Hu}, Yue and {Immer}, Katharina and {Jeff}, Desmond and {Karoly}, Janik and {Kauffmann}, Jens and {Klessen}, Ralf S. and {Krumholz}, Mark R. and {Lazarian}, Alex and {Levesque}, Emily M. and {Liang}, Fu-Heng and {Lipman}, Dani and {Liu}, Xunchuan and {Lu}, Xing and {Luo}, Qiu-yi and {Lupi}, Alessandro and {McCafferty}, Laura and {Mart{\'\i}n}, S. and {Mazoochi}, Farideh and {Morris}, Mark R. and {Nonhebel}, Marie and {Nogueras-Lara}, Francisco and {Oka}, Tomoharu and {Ott}, Juergen and {Padovani}, Marco and {Pan}, Xing and {Pineda}, Jaime E. and {Pillai}, Thushara G.~S. and {Pound}, Marc W. and {Requena Torres}, Miguel and {Riquelme-V{\'a}squez}, Denise and {Rivilla}, V{\'\i}ctor M. and {Salo}, Galaxy and {S{\'a}nchez-Monge}, {\'A}lvaro and {Santa-Maria}, Miriam G. and {Schoedel}, Rainer and {Schmiedeke}, Anika and {Schultheis}, Matthias and {Smith}, Howard A. and {Sofue}, Yoshiaki and {Testi}, Leonardo and {Tremblay}, Grant R. and {Vasini}, Arianna and {Vermari{\"e}n}, Gijs and {Vikhlinin}, Alexey and {Viti}, Serena and {Wang}, Q. Daniel and {Xu}, Fengwei and {Zhang}, Suinan and {Zhang}, Qizhou},
        title = "{ALMA Central Molecular Zone Exploration Survey (ACES) I: Overview}",
      journal = {arXiv e-prints},
         year = 2026,
        month = feb,
          eid = {arXiv:2602.20340},
        pages = {arXiv:2602.20340},
          doi = {10.48550/arXiv.2602.20340},
archivePrefix = {arXiv},
       eprint = {2602.20340},
 primaryClass = {astro-ph.GA},
       adsurl = {https://ui.adsabs.harvard.edu/abs/2026arXiv260220340L}
}

@ARTICLE{Ginsburg2026_ACESII,
       author = {{Ginsburg}, Adam and {Walker}, Daniel L. and {Barnes}, Ashley T. and {Lu}, Xing and {S{\'a}nchez-Monge}, {\'A}lvaro and {Pineda}, Jaime E. and {Pound}, Marc W. and {Hsieh}, Pei-Ying and {Immer}, Katharina and {Zhang}, Qizhou and {Budaiev}, Nazar and {Gramze}, Savannah R. and {Jeff}, Desmond and {Cook}, Claire and {Bulatek}, Alyssa and {Mills}, Elisabeth A.~C. and {Bally}, John and {Colzi}, Laura and {Garc{\'\i}a}, Pablo and {Henshaw}, Jonathan D. and {Jim{\'e}nez-Serra}, Izaskun and {Klessen}, Ralf S. and {Dicker}, Simon R. and {Longmore}, Steven N. and {Nogueras-Lara}, Francisco and {Rivilla}, V{\'\i}ctor M. and {Santa-Maria}, Miriam G. and {Wang}, Q. Daniel and {Xu}, Fengwei and {Battersby}, Cara and {Ho}, Paul T.~P. and {Kruijssen}, J.~M. Diederik and {Petkova}, Maya and {Sormani}, Mattia C. and {Tress}, Robin G. and {Wallace}, Jennifer and {Armijos-Abenda{\~n}o}, J. and {Armillotta}, Lucia and {Bijas}, N. and {Buddhacharya}, Rojita and {Busch}, Laura A. and {Butterfield}, Natalie O. and {Chevance}, M{\'e}lanie and {Crowe}, Samuel and {D{\'\i}az-Rodr{\'\i}guez}, Ana Karla and {Dutkowska}, Katarzyna M. and {Fedriani}, Rub{\'e}n and {Federrath}, Christoph and {Glover}, Simon C.~O. and {Gu}, Qi-Lao and {Houghton}, Rebecca J. and {Hu}, Yue and {Issac}, Namitha and {Karoly}, Janik and {Krumholz}, Mark R. and {Liang}, Fu-Heng and {Mart{\'\i}n}, Sergio and {Mazoochi}, Farideh and {Pan}, Xing and {Par{\'e}}, Dylan and {Pillai}, Thushara G.~S. and {Riquelme-V{\'a}squez}, Denise and {Schmiedeke}, Anika and {Sofue}, Yoshiaki and {Tolls}, Volker and {Williams}, Gwenllian M. and {Zhang}, Suinan and {Moravec}, Emily and {Romero}, Charles E. and {Mason}, Brian S. and {Orlowski-Scherer}, John and {Hatchfield}, H Perry},
        title = "{ALMA Central Molecular Zone Exploration Survey (ACES) II: 3mm continuum images}",
      journal = {arXiv e-prints},
         year = 2026,
        month = feb,
          eid = {arXiv:2602.20240},
        pages = {arXiv:2602.20240},
          doi = {10.48550/arXiv.2602.20240},
archivePrefix = {arXiv},
       eprint = {2602.20240},
 primaryClass = {astro-ph.GA},
       adsurl = {https://ui.adsabs.harvard.edu/abs/2026arXiv260220240G}
}

@ARTICLE{Walker2026_ACESIII,
       author = {{Walker}, Daniel L. and {Ginsburg}, Adam and {Barnes}, Ashley T. and {Lu}, Xing and {Hsieh}, Pei-Ying and {S{\'a}nchez-Monge}, {\'A}lvaro and {Gramze}, Savannah R. and {Budaiev}, Nazar and {Pound}, Marc W. and {Pineda}, Jaime E. and {Bulatek}, Alyssa and {Cook}, Claire and {Henshaw}, Jonathan D. and {Immer}, Katharina and {Issac}, Namitha and {Jeff}, Desmond and {Liang}, Fu-Heng and {Longmore}, Steven N. and {Mills}, Elisabeth A.~C. and {Mart{\'\i}n}, Sergio and {Pan}, Xing and {Pillai}, Thushara G.~S. and {Zhang}, Qizhou and {Bally}, John and {Battersby}, Cara and {Colzi}, Laura and {Ho}, Paul T.~P. and {Jim{\'e}nez-Serra}, Izaskun and {Kruijssen}, J.~M. Diederik and {Petkova}, Maya A. and {Sormani}, Mattia C. and {Tress}, Robin G. and {Wallace}, Jennifer and {Armijos-Abenda{\~n}o}, J. and {Armillotta}, Lucia and {Bijas}, N. and {Buddhacharya}, Rojita and {Busch}, Laura A. and {Butterfield}, Natalie O. and {Chevance}, M{\'e}lanie and {Crowe}, Samuel and {D{\'\i}az-Rodr{\'\i}guez}, Ana Karla and {Dutkowska}, Katarzyna M. and {Federrath}, Christoph and {Fedriani}, Rub{\'e}n and {Garc{\'\i}a}, Pablo and {Glover}, Simon C.~O. and {Gu}, Qi-Lao and {Hatchfield}, H Perry and {Houghton}, Rebecca J. and {Hu}, Yue and {Karoly}, Janik and {Klessen}, Ralf S. and {Krumholz}, Mark R. and {Mazoochi}, Farideh and {Nogueras-Lara}, Francisco and {Par{\'e}}, Dylan and {Riquelme-V{\'a}squez}, Denise and {Rivilla}, V{\'\i}ctor M. and {Santa-Maria}, Miriam G. and {Schmiedeke}, Anika and {Sofue}, Yoshiaki and {Tolls}, Volker and {Wang}, Q. Daniel and {Williams}, Gwenllian M. and {Xu}, Fengwei and {Zhang}, Suinan},
        title = "{ALMA Central molecular zone Exploration Survey (ACES) III: Molecular line data reduction and HNCO and HCO$^{+}$ data}",
      journal = {arXiv e-prints},
         year = 2026,
        month = feb,
          eid = {arXiv:2602.20276},
        pages = {arXiv:2602.20276},
          doi = {10.48550/arXiv.2602.20276},
archivePrefix = {arXiv},
       eprint = {2602.20276},
 primaryClass = {astro-ph.GA},
       adsurl = {https://ui.adsabs.harvard.edu/abs/2026arXiv260220276W}
}

@ARTICLE{Lu2026_ACESIV,
       author = {{Lu}, Xing and {Walker}, Daniel L. and {Ginsburg}, Adam and {Barnes}, Ashley T. and {Hsieh}, Pei-Ying and {Sanchez-Monge}, Alvaro and {Gramze}, Savannah R. and {Budaiev}, Nazar and {Pound}, Marc W. and {Pineda}, Jaime E. and {Bulatek}, Alyssa and {Cook}, Claire and {Henshaw}, Jonathan D. and {Immer}, Katharina and {Issac}, Namitha and {Jeff}, Desmond and {Liang}, Fu-Heng and {Longmore}, Steven N. and {Mills}, Elisabeth A.~C. and {Martin}, Sergio and {Pan}, Xing and {Zhang}, Qizhou and {Bally}, John and {Battersby}, Cara and {Colzi}, Laura and {Ho}, Paul T.~P. and {Jimenez-Serra}, Izaskun and {Kruijssen}, J.~M. Diederik and {Petkova}, Maya A. and {Sormani}, Mattia C. and {Tress}, Robin G. and {Wallace}, Jennifer and {Armijos-Abendano}, J. and {Armillotta}, Lucia and {Bijas}, N. and {Buddhacharya}, Rojita and {Busch}, Laura A. and {Butterfield}, Natalie O. and {Chevance}, Melanie and {Diaz-Rodriguez}, Ana Karla and {Federrath}, Christoph and {Fedriani}, Ruben and {Garcia}, Pablo and {Gu}, Qi-Lao and {Hatchfield}, H Perry and {Houghton}, Rebecca J. and {Hu}, Yue and {Karoly}, Janik and {Klessen}, Ralf S. and {Krumholz}, Mark R. and {Liu}, Xunchuan and {Mazoochi}, Farideh and {Nogueras-Lara}, Francisco and {Pare}, Dylan and {Riquelme-Vasquez}, Denise and {Rivilla}, Victor M. and {Santa-Maria}, Miriam G. and {Schmiedeke}, Anika and {Sofue}, Yoshiaki and {Tolls}, Volker and {Wang}, Q. Daniel and {Williams}, Gwenllian M. and {Xu}, Fengwei and {Zhang}, Suinan},
        title = "{ALMA Central Molecular Zone Exploration Survey (ACES)-IV. Data of the two intermediate-width spectral windows}",
      journal = {arXiv e-prints},
         year = 2026,
        month = feb,
          eid = {arXiv:2602.20445},
        pages = {arXiv:2602.20445},
          doi = {10.48550/arXiv.2602.20445},
archivePrefix = {arXiv},
       eprint = {2602.20445},
 primaryClass = {astro-ph.GA},
       adsurl = {https://ui.adsabs.harvard.edu/abs/2026arXiv260220445L}
}

@ARTICLE{Hsieh2026_ACESV,
       author = {{Hsieh}, Pei-Ying and {Walker}, Daniel L. and {Ginsburg}, Adam and {Barnes}, Ashley T. and {Lu}, Xing and {S{\'a}nchez-Monge}, {\'A}lvaro and {Gramze}, Savannah R. and {Budaiev}, Nazar and {Pound}, Marc W. and {Pineda}, Jaime E. and {Cook}, Claire and {Henshaw}, Jonathan D. and {Immer}, Katharina and {Issac}, Namitha and {Jeff}, Desmond and {Liang}, Fu-Heng and {Longmore}, Steven N. and {Mills}, Elisabeth A.~C. and {Mart{\'\i}n}, Sergio and {Pan}, Xing and {Pillai}, Thushara G.~S. and {Zhang}, Qizhou and {Bally}, John and {Battersby}, Cara and {Colzi}, Laura and {Ho}, Paul T.~P. and {Jim{\'e}nez-Serra}, Izaskun and {Kruijssen}, J.~M. Diederik and {Petkova}, Maya and {Sormani}, Mattia C. and {Tress}, Robin G. and {Wallace}, Jennifer and {Armijos-Abenda{\~n}o}, J. and {Armillotta}, Lucia and {Bijas}, N. and {Budhathoki-Chhetrya}, Rojita and {Busch}, Laura A. and {Butterfield}, Natalie O. and {Chevance}, M{\'e}lanie and {D{\'\i}az-Rodr{\'\i}guez}, Ana Karla and {Federrath}, Christoph and {Fedriani}, Rub{\'e}n and {Garc{\'\i}a}, Pablo and {Gu}, Qi-Lao and {Houghton}, Rebecca J. and {Hu}, Yue and {Karoly}, Janik and {Klessen}, Ralf S. and {Krumholz}, Mark R. and {Mazoochi}, Farideh and {Nogueras-Lara}, Francisco and {Par{\'e}}, Dylan and {Riquelme-V{\'a}squez}, Denise and {Rivilla}, V{\'\i}ctor M. and {Santa-Maria}, Miriam G. and {Schmiedeke}, Anika and {Sofue}, Yoshiaki and {Tolls}, Volker and {Wang}, Q. Daniel and {Williams}, Gwenllian M. and {Xu}, Fengwei and {Zhang}, Suinan},
        title = "{ALMA Central molecular zone Exploration Survey (ACES) V: CS(2-1), SO(2\_3-1\_2), CH3CHO(5\_1,4-4\_1,3), HC3N(11-10), and H40a lines data}",
      journal = {arXiv e-prints},
         year = 2026,
        month = mar,
          eid = {arXiv:2603.00863},
        pages = {arXiv:2603.00863},
          doi = {10.48550/arXiv.2603.00863},
archivePrefix = {arXiv},
       eprint = {2603.00863},
 primaryClass = {astro-ph.GA},
       adsurl = {https://ui.adsabs.harvard.edu/abs/2026arXiv260300863H}
}

@ARTICLE{Ginsburg2024_MUBLO,
       author = {{Ginsburg}, Adam and {Bally}, John and {Barnes}, Ashley T. and {Battersby}, Cara and {Budaiev}, Nazar and {Butterfield}, Natalie O. and {Caselli}, Paola and {Colzi}, Laura and {Dutkowska}, Katarzyna M. and {Garc{\'\i}a}, Pablo and {Gramze}, Savannah and {Henshaw}, Jonathan D. and {Hu}, Yue and {Jeff}, Desmond and {Jim{\'e}nez-Serra}, Izaskun and {Kauffmann}, Jens and {Klessen}, Ralf S. and {Levesque}, Emily M. and {Longmore}, Steven N. and {Lu}, Xing and {Mills}, Elisabeth A.~C. and {Morris}, Mark R. and {Nogueras-Lara}, Francisco and {Oka}, Tomoharu and {Pineda}, Jaime E. and {Pillai}, Thushara G.~S. and {Rivilla}, V{\'\i}ctor M. and {S{\'a}nchez-Monge}, {\'A}lvaro and {Santa-Maria}, Miriam G. and {Smith}, Howard A. and {Sofue}, Yoshiaki and {Sormani}, Mattia C. and {Tremblay}, Grant R. and {Vermari{\"e}n}, Gijs and {Vikhlinin}, Alexey and {Viti}, Serena and {Walker}, Dan and {Wang}, Q. Daniel and {Xu}, Fengwei and {Zhang}, Qizhou},
        title = "{A Broad Line-width, Compact, Millimeter-bright Molecular Emission Line Source near the Galactic Center}",
      journal = {\apjl},
         year = 2024,
        month = jun,
       volume = {968},
       number = {1},
          eid = {L11},
        pages = {L11},
          doi = {10.3847/2041-8213/ad47fa},
archivePrefix = {arXiv},
       eprint = {2404.07808},
 primaryClass = {astro-ph.GA},
       adsurl = {https://ui.adsabs.harvard.edu/abs/2024ApJ...968L..11G}
}

@ARTICLE{Uehara2019,
       author = {{Uehara}, Kenta and {Tsuboi}, Masato and {Kitamura}, Yoshimi and {Miyawaki}, Ryosuke and {Miyazaki}, Atsushi},
        title = "{Molecular Cloud Cores in the Galactic Center 50 km s$^{-1}$ Molecular Cloud}",
      journal = {\apj},
         year = 2019,
        month = feb,
       volume = {872},
       number = {2},
          eid = {121},
        pages = {121},
          doi = {10.3847/1538-4357/aafee7},
archivePrefix = {arXiv},
       eprint = {1903.01759},
 primaryClass = {astro-ph.GA},
       adsurl = {https://ui.adsabs.harvard.edu/abs/2019ApJ...872..121U}
}

@ARTICLE{Clark2021,
       author = {{Clark}, J.~S. and {Patrick}, L.~R. and {Najarro}, F. and {Evans}, C.~J. and {Lohr}, M.},
        title = "{Constraining the population of isolated massive stars within the Central Molecular Zone}",
      journal = {\aap},
         year = 2021,
        month = may,
       volume = {649},
          eid = {A43},
        pages = {A43},
          doi = {10.1051/0004-6361/202039205},
archivePrefix = {arXiv},
       eprint = {2102.08126},
 primaryClass = {astro-ph.GA},
       adsurl = {https://ui.adsabs.harvard.edu/abs/2021A&A...649A..43C}
}

@ARTICLE{Figer1999_mar,
       author = {{Figer}, Donald F. and {McLean}, Ian S. and {Morris}, Mark},
        title = "{Massive Stars in the Quintuplet Cluster}",
      journal = {\apj},
         year = 1999,
        month = mar,
       volume = {514},
       number = {1},
        pages = {202-220},
          doi = {10.1086/306931},
archivePrefix = {arXiv},
       eprint = {astro-ph/9903281},
 primaryClass = {astro-ph},
       adsurl = {https://ui.adsabs.harvard.edu/abs/1999ApJ...514..202F}
}

@ARTICLE{Figer1999_nov,
       author = {{Figer}, Donald F. and {Kim}, Sungsoo S. and {Morris}, Mark and {Serabyn}, Eugene and {Rich}, R. Michael and {McLean}, Ian S.},
        title = "{Hubble Space Telescope/NICMOS Observations of Massive Stellar Clusters near the Galactic Center}",
      journal = {\apj},
         year = 1999,
        month = nov,
       volume = {525},
       number = {2},
        pages = {750-758},
          doi = {10.1086/307937},
archivePrefix = {arXiv},
       eprint = {astro-ph/9906299},
 primaryClass = {astro-ph},
       adsurl = {https://ui.adsabs.harvard.edu/abs/1999ApJ...525..750F}
}

@ARTICLE{Clarkson2012,
       author = {{Clarkson}, W.~I. and {Ghez}, A.~M. and {Morris}, M.~R. and {Lu}, J.~R. and {Stolte}, A. and {McCrady}, N. and {Do}, T. and {Yelda}, S.},
        title = "{Proper Motions of the Arches Cluster with Keck Laser Guide Star Adaptive Optics: The First Kinematic Mass Measurement of the Arches}",
      journal = {\apj},
         year = 2012,
        month = jun,
       volume = {751},
       number = {2},
          eid = {132},
        pages = {132},
          doi = {10.1088/0004-637X/751/2/132},
archivePrefix = {arXiv},
       eprint = {1112.5458},
 primaryClass = {astro-ph.GA},
       adsurl = {https://ui.adsabs.harvard.edu/abs/2012ApJ...751..132C}
}

@ARTICLE{Kaneko2023,
       author = {{Kaneko}, Miyuki and {Oka}, Tomoharu and {Yokozuka}, Hiroki and {Enokiya}, Rei and {Takekawa}, Shunya and {Iwata}, Yuhei and {Tsujimoto}, Shiho},
        title = "{Discovery of the Tadpole Molecular Cloud near the Galactic Nucleus}",
      journal = {\apj},
         year = 2023,
        month = jan,
       volume = {942},
       number = {1},
          eid = {46},
        pages = {46},
          doi = {10.3847/1538-4357/aca66a},
archivePrefix = {arXiv},
       eprint = {2301.04831},
 primaryClass = {astro-ph.GA},
       adsurl = {https://ui.adsabs.harvard.edu/abs/2023ApJ...942...46K}
}

@ARTICLE{Tideswell2010,
       author = {{Tideswell}, D.~M. and {Fuller}, G.~A. and {Millar}, T.~J. and {Markwick}, A.~J.},
        title = "{The abundance of HNCO and its use as a diagnostic of environment}",
      journal = {\aap},
         year = 2010,
        month = feb,
       volume = {510},
          eid = {A85},
        pages = {A85},
          doi = {10.1051/0004-6361/200810820},
archivePrefix = {arXiv},
       eprint = {0909.4337},
 primaryClass = {astro-ph.GA},
       adsurl = {https://ui.adsabs.harvard.edu/abs/2010A&A...510A..85T}
}

@ARTICLE{Rico-Villas2020,
       author = {{Rico-Villas}, F. and {Mart{\'\i}n-Pintado}, J. and {Gonz{\'a}lez-Alfonso}, E. and {Mart{\'\i}n}, S. and {Rivilla}, V.~M.},
        title = "{Super Hot Cores in NGC 253: witnessing the formation and early evolution of super star clusters}",
      journal = {\mnras},
         year = 2020,
        month = jan,
       volume = {491},
       number = {3},
        pages = {4573-4589},
          doi = {10.1093/mnras/stz3347},
archivePrefix = {arXiv},
       eprint = {1909.11385},
 primaryClass = {astro-ph.GA},
       adsurl = {https://ui.adsabs.harvard.edu/abs/2020MNRAS.491.4573R}
}

@ARTICLE{Schilke1997,
       author = {{Schilke}, P. and {Walmsley}, C.~M. and {Pineau des Forets}, G. and {Flower}, D.~R.},
        title = "{SiO production in interstellar shocks.}",
      journal = {\aap},
         year = 1997,
        month = may,
       volume = {321},
        pages = {293-304},
       adsurl = {https://ui.adsabs.harvard.edu/abs/1997A&A...321..293S}
}

@ARTICLE{Sormani2019b,
       author = {{Sormani}, Mattia C. and {Tre{\ss}}, Robin G. and {Glover}, Simon C.~O. and {Klessen}, Ralf S. and {Barnes}, Ashley T. and {Battersby}, Cara D. and {Clark}, Paul C. and {Hatchfield}, H. Perry and {Smith}, Rowan J.},
        title = "{The geometry of the gas surrounding the Central Molecular Zone: on the origin of localized molecular clouds with extreme velocity dispersions}",
      journal = {\mnras},
         year = 2019,
        month = oct,
       volume = {488},
       number = {4},
        pages = {4663-4673},
          doi = {10.1093/mnras/stz2054},
archivePrefix = {arXiv},
       eprint = {1906.10129},
 primaryClass = {astro-ph.GA},
       adsurl = {https://ui.adsabs.harvard.edu/abs/2019MNRAS.488.4663S}
}

@ARTICLE{Barnes_Priestley2024,
       author = {{Barnes}, Rees A. and {Priestley}, Felix D.},
        title = "{Cloud Collision Signatures in the Central Molecular Zone}",
      journal = {The Open Journal of Astrophysics},
         year = 2024,
        month = oct,
       volume = {7},
          eid = {95},
        pages = {95},
          doi = {10.33232/001c.124112},
archivePrefix = {arXiv},
       eprint = {2407.21575},
 primaryClass = {astro-ph.GA},
       adsurl = {https://ui.adsabs.harvard.edu/abs/2024OJAp....7E..95B}
}

@ARTICLE{Battersby2025a,
       author = {{Battersby}, Cara and {Walker}, Daniel L. and {Barnes}, Ashley and {Ginsburg}, Adam and {Lipman}, Dani and {Alboslani}, Danya and {Hatchfield}, H. Perry and {Bally}, John and {Glover}, Simon C.~O. and {Henshaw}, Jonathan D. and {Immer}, Katharina and {Klessen}, Ralf S. and {Longmore}, Steven N. and {Mills}, Elisabeth A.~C. and {Molinari}, Sergio and {Smith}, Rowan and {Sormani}, Mattia C. and {Tress}, Robin G. and {Zhang}, Qizhou},
        title = "{3D CMZ. I. Central Molecular Zone Overview}",
      journal = {\apj},
         year = 2025,
        month = may,
       volume = {984},
       number = {2},
          eid = {156},
        pages = {156},
          doi = {10.3847/1538-4357/adb5f0},
archivePrefix = {arXiv},
       eprint = {2410.17334},
 primaryClass = {astro-ph.GA},
       adsurl = {https://ui.adsabs.harvard.edu/abs/2025ApJ...984..156B}
}

@ARTICLE{Battersby2025b,
       author = {{Battersby}, Cara and {Walker}, Daniel L. and {Barnes}, Ashley and {Ginsburg}, Adam and {Lipman}, Dani and {Alboslani}, Danya and {Hatchfield}, H. Perry and {Bally}, John and {Glover}, Simon C.~O. and {Henshaw}, Jonathan D. and {Immer}, Katharina and {Klessen}, Ralf S. and {Longmore}, Steven N. and {Mills}, Elisabeth A.~C. and {Molinari}, Sergio and {Smith}, Rowan and {Sormani}, Mattia C. and {Tress}, Robin G. and {Zhang}, Qizhou},
        title = "{3D CMZ. II. Hierarchical Structure Analysis of the Central Molecular Zone}",
      journal = {\apj},
         year = 2025,
        month = may,
       volume = {984},
       number = {2},
          eid = {157},
        pages = {157},
          doi = {10.3847/1538-4357/adb844},
archivePrefix = {arXiv},
       eprint = {2410.17332},
 primaryClass = {astro-ph.GA},
       adsurl = {https://ui.adsabs.harvard.edu/abs/2025ApJ...984..157B}
}

@ARTICLE{Walker2025,
       author = {{Walker}, Daniel L. and {Battersby}, Cara and {Lipman}, Dani and {Sormani}, Mattia C. and {Ginsburg}, Adam and {Glover}, Simon C.~O. and {Henshaw}, Jonathan D. and {Longmore}, Steven N. and {Klessen}, Ralf S. and {Immer}, Katharina and {Alboslani}, Danya and {Bally}, John and {Barnes}, Ashley and {Hatchfield}, H. Perry and {Mills}, Elisabeth A.~C. and {Smith}, Rowan and {Tress}, Robin G. and {Zhang}, Qizhou},
        title = "{3D CMZ. III. Constraining the 3D Structure of the Central Molecular Zone via Molecular Line Emission and Absorption}",
      journal = {\apj},
         year = 2025,
        month = may,
       volume = {984},
       number = {2},
          eid = {158},
        pages = {158},
          doi = {10.3847/1538-4357/adb5ef},
archivePrefix = {arXiv},
       eprint = {2410.17320},
 primaryClass = {astro-ph.GA},
       adsurl = {https://ui.adsabs.harvard.edu/abs/2025ApJ...984..158W}
}

@ARTICLE{Pare2025,
       author = {{Par{\'e}}, Dylan M. and {Chuss}, David T. and {Karpovich}, Kaitlyn and {Butterfield}, Natalie O. and {Iuliano}, Jeffrey Inara and {Pan}, Xing and {Wollack}, Edward J. and {Zhang}, Qizhou and {Morris}, Mark R. and {Nilsson}, Mathilda and {Zhao}, Roy J.},
        title = "{SOFIA/HAWC+ Far-infrared Polarimetric Large-area CMZ Exploration Survey. IV. Relative Magnetic Field Orientation throughout the CMZ}",
      journal = {\apj},
         year = 2025,
        month = jan,
       volume = {978},
       number = {1},
          eid = {28},
        pages = {28},
          doi = {10.3847/1538-4357/ad9586},
archivePrefix = {arXiv},
       eprint = {2410.10597},
 primaryClass = {astro-ph.GA},
       adsurl = {https://ui.adsabs.harvard.edu/abs/2025ApJ...978...28P}
}

@ARTICLE{GRAVITYCollaboration2021,
       author = {{GRAVITY Collaboration} and {Abuter}, R. and {Amorim}, A. and {Baub{\"o}ck}, M. and {Berger}, J.~P. and {Bonnet}, H. and {Brandner}, W. and {Cl{\'e}net}, Y. and {Davies}, R. and {de Zeeuw}, P.~T. and {Dexter}, J. and {Dallilar}, Y. and {Drescher}, A. and {Eckart}, A. and {Eisenhauer}, F. and {F{\"o}rster Schreiber}, N.~M. and {Garcia}, P. and {Gao}, F. and {Gendron}, E. and {Genzel}, R. and {Gillessen}, S. and {Habibi}, M. and {Haubois}, X. and {Hei{\ss}el}, G. and {Henning}, T. and {Hippler}, S. and {Horrobin}, M. and {Jim{\'e}nez-Rosales}, A. and {Jochum}, L. and {Jocou}, L. and {Kaufer}, A. and {Kervella}, P. and {Lacour}, S. and {Lapeyr{\`e}re}, V. and {Le Bouquin}, J. -B. and {L{\'e}na}, P. and {Lutz}, D. and {Nowak}, M. and {Ott}, T. and {Paumard}, T. and {Perraut}, K. and {Perrin}, G. and {Pfuhl}, O. and {Rabien}, S. and {Rodr{\'\i}guez-Coira}, G. and {Shangguan}, J. and {Shimizu}, T. and {Scheithauer}, S. and {Stadler}, J. and {Straub}, O. and {Straubmeier}, C. and {Sturm}, E. and {Tacconi}, L.~J. and {Vincent}, F. and {von Fellenberg}, S. and {Waisberg}, I. and {Widmann}, F. and {Wieprecht}, E. and {Wiezorrek}, E. and {Woillez}, J. and {Yazici}, S. and {Young}, A. and {Zins}, G.},
        title = "{Improved GRAVITY astrometric accuracy from modeling optical aberrations}",
      journal = {\aap},
         year = 2021,
        month = mar,
       volume = {647},
          eid = {A59},
        pages = {A59},
          doi = {10.1051/0004-6361/202040208},
archivePrefix = {arXiv},
       eprint = {2101.12098},
 primaryClass = {astro-ph.GA},
       adsurl = {https://ui.adsabs.harvard.edu/abs/2021A&A...647A..59G}
}

@ARTICLE{Jimenez2008,
       author = {{Jim{\'e}nez-Serra}, I. and {Caselli}, P. and {Mart{\'\i}n-Pintado}, J. and {Hartquist}, T.~W.},
        title = "{Parametrization of C-shocks. Evolution of the sputtering of grains}",
      journal = {\aap},
         year = 2008,
        month = may,
       volume = {482},
       number = {2},
        pages = {549-559},
          doi = {10.1051/0004-6361:20078054},
archivePrefix = {arXiv},
       eprint = {0802.0594},
 primaryClass = {astro-ph},
       adsurl = {https://ui.adsabs.harvard.edu/abs/2008A&A...482..549J}
}

@ARTICLE{Zeng2020,
       author = {{Zeng}, S. and {Zhang}, Q. and {Jim{\'e}nez-Serra}, I. and {Tercero}, B. and {Lu}, X. and {Mart{\'\i}n-Pintado}, J. and {de Vicente}, P. and {Rivilla}, V.~M. and {Li}, S.},
        title = "{Cloud-cloud collision as drivers of the chemical complexity in Galactic Centre molecular clouds}",
      journal = {\mnras},
         year = 2020,
        month = oct,
       volume = {497},
       number = {4},
        pages = {4896-4909},
          doi = {10.1093/mnras/staa2187},
archivePrefix = {arXiv},
       eprint = {2007.14362},
 primaryClass = {astro-ph.GA},
       adsurl = {https://ui.adsabs.harvard.edu/abs/2020MNRAS.497.4896Z}
}

@ARTICLE{Armijos-Abendano2020,
       author = {{Armijos-Abenda{\~n}o}, J. and {Banda-Barrag{\'a}n}, W.~E. and {Mart{\'\i}n-Pintado}, J. and {D{\'e}nes}, H. and {Federrath}, C. and {Requena-Torres}, M.~A.},
        title = "{Structure and kinematics of shocked gas in Sgr B2: further evidence of a cloud-cloud collision from SiO emission maps}",
      journal = {\mnras},
         year = 2020,
        month = dec,
       volume = {499},
       number = {4},
        pages = {4918-4939},
          doi = {10.1093/mnras/staa3119},
archivePrefix = {arXiv},
       eprint = {2010.02757},
 primaryClass = {astro-ph.GA},
       adsurl = {https://ui.adsabs.harvard.edu/abs/2020MNRAS.499.4918A}
}

@ARTICLE{Yokozuka2021,
       author = {{Yokozuka}, Hiroki and {Oka}, Tomoharu and {Takekawa}, Shunya and {Iwata}, Yuhei and {Tsujimoto}, Shiho},
        title = "{Broad-velocity-width Molecular Features in the Galactic Plane}",
      journal = {\apj},
         year = 2021,
        month = feb,
       volume = {908},
       number = {2},
          eid = {246},
        pages = {246},
          doi = {10.3847/1538-4357/abd556},
archivePrefix = {arXiv},
       eprint = {2103.13728},
 primaryClass = {astro-ph.GA},
       adsurl = {https://ui.adsabs.harvard.edu/abs/2021ApJ...908..246Y}
}

@ARTICLE{Yusef-Zadeh1987,
       author = {{Yusef-Zadeh}, Farhad and {Morris}, Mark},
        title = "{G0.18-0.04: Interaction of Thermal and Nonthermal Radio Structures in the Arc Near the Galactic Center}",
      journal = {\aj},
         year = 1987,
        month = nov,
       volume = {94},
        pages = {1178},
          doi = {10.1086/114555},
       adsurl = {https://ui.adsabs.harvard.edu/abs/1987AJ.....94.1178Y}
}

@ARTICLE{Kauffmann2017c,
       author = {{Kauffmann}, Jens and {Pillai}, Thushara and {Zhang}, Qizhou and {Menten}, Karl M. and {Goldsmith}, Paul F. and {Lu}, Xing and {Guzm{\'a}n}, Andr{\'e}s E.},
        title = "{The Galactic Center Molecular Cloud Survey. I. A steep linewidth-size relation and suppression of star formation}",
      journal = {\aap},
         year = 2017,
        month = jul,
       volume = {603},
          eid = {A89},
        pages = {A89},
          doi = {10.1051/0004-6361/201628088},
archivePrefix = {arXiv},
       eprint = {1610.03499},
 primaryClass = {astro-ph.GA},
       adsurl = {https://ui.adsabs.harvard.edu/abs/2017A&A...603A..89K}
}

@ARTICLE{Lu2021,
       author = {{Lu}, Xing and {Li}, Shanghuo and {Ginsburg}, Adam and {Longmore}, Steven N. and {Kruijssen}, J.~M. Diederik and {Walker}, Daniel L. and {Feng}, Siyi and {Zhang}, Qizhou and {Battersby}, Cara and {Pillai}, Thushara and {Mills}, Elisabeth A.~C. and {Kauffmann}, Jens and {Cheng}, Yu and {Inutsuka}, Shu-ichiro},
        title = "{ALMA Observations of Massive Clouds in the Central Molecular Zone: Ubiquitous Protostellar Outflows}",
      journal = {\apj},
         year = 2021,
        month = mar,
       volume = {909},
       number = {2},
          eid = {177},
        pages = {177},
          doi = {10.3847/1538-4357/abde3c},
archivePrefix = {arXiv},
       eprint = {2101.07925},
 primaryClass = {astro-ph.GA},
       adsurl = {https://ui.adsabs.harvard.edu/abs/2021ApJ...909..177L}
}

@ARTICLE{Gramze2025,
       author = {{Gramze}, Savannah and {Ginsburg}, Adam and {Budaiev}, Nazar and {Bulatek}, Alyssa and {Richardson}, Theo and {Barnes}, A.~T. and {Santa-Maria}, Miriam G. and {Sormani}, Mattia C. and {Lu}, Xing and {Nogueras-Lara}, Francisco and {Gaches}, Brandt A.~L. and {Battersby}, Cara D. and {Wallace}, Jennifer and {Walker}, Daniel L. and {Mills}, Elisabeth A.~C. and {Mattern}, Michael},
        title = "{Mapping CO Ice in a Star-Forming Filament in the 3 kpc Arm with JWST}",
      journal = {arXiv e-prints},
         year = 2025,
        month = sep,
          eid = {arXiv:2509.21763},
        pages = {arXiv:2509.21763},
          doi = {10.48550/arXiv.2509.21763},
archivePrefix = {arXiv},
       eprint = {2509.21763},
 primaryClass = {astro-ph.GA},
       adsurl = {https://ui.adsabs.harvard.edu/abs/2025arXiv250921763G}
}

@ARTICLE{Sormani_and_Barnes_2019,
       author = {{Sormani}, Mattia C. and {Barnes}, Ashley T.},
        title = "{Mass inflow rate into the Central Molecular Zone: observational determination and evidence of episodic accretion}",
      journal = {\mnras},
         year = 2019,
        month = mar,
       volume = {484},
       number = {1},
        pages = {1213-1219},
          doi = {10.1093/mnras/stz046},
archivePrefix = {arXiv},
       eprint = {1901.00867},
 primaryClass = {astro-ph.GA},
       adsurl = {https://ui.adsabs.harvard.edu/abs/2019MNRAS.484.1213S}
}

@ARTICLE{Heyer2015,
       author = {{Heyer}, Mark and {Dame}, T.~M.},
        title = "{Molecular Clouds in the Milky Way}",
      journal = {\araa},
         year = 2015,
        month = aug,
       volume = {53},
        pages = {583-629},
          doi = {10.1146/annurev-astro-082214-122324},
       adsurl = {https://ui.adsabs.harvard.edu/abs/2015ARA&A..53..583H}
}

@ARTICLE{RomanDuval2016,
       author = {{Roman-Duval}, Julia and {Heyer}, Mark and {Brunt}, Christopher M. and {Clark}, Paul and {Klessen}, Ralf and {Shetty}, Rahul},
        title = "{Distribution and Mass of Diffuse and Dense CO Gas in the Milky Way}",
      journal = {\apj},
         year = 2016,
        month = feb,
       volume = {818},
       number = {2},
          eid = {144},
        pages = {144},
          doi = {10.3847/0004-637X/818/2/144},
archivePrefix = {arXiv},
       eprint = {1601.00937},
 primaryClass = {astro-ph.GA},
       adsurl = {https://ui.adsabs.harvard.edu/abs/2016ApJ...818..144R}
}

@ARTICLE{Haworth2015a,
       author = {{Haworth}, T.~J. and {Tasker}, E.~J. and {Fukui}, Y. and {Torii}, K. and {Dale}, J.~E. and {Shima}, K. and {Takahira}, K. and {Habe}, A. and {Hasegawa}, K.},
        title = "{Isolating signatures of major cloud-cloud collisions using position-velocity diagrams}",
      journal = {\mnras},
         year = 2015,
        month = jun,
       volume = {450},
       number = {1},
        pages = {10-20},
          doi = {10.1093/mnras/stv639},
archivePrefix = {arXiv},
       eprint = {1503.06795},
 primaryClass = {astro-ph.GA},
       adsurl = {https://ui.adsabs.harvard.edu/abs/2015MNRAS.450...10H}
}

@ARTICLE{Haworth2015b,
       author = {{Haworth}, T.~J. and {Shima}, K. and {Tasker}, E.~J. and {Fukui}, Y. and {Torii}, K. and {Dale}, J.~E. and {Takahira}, K. and {Habe}, A.},
        title = "{Isolating signatures of major cloud-cloud collisions - II. The lifetimes of broad bridge features}",
      journal = {\mnras},
         year = 2015,
        month = dec,
       volume = {454},
       number = {2},
        pages = {1634-1643},
          doi = {10.1093/mnras/stv2068},
archivePrefix = {arXiv},
       eprint = {1509.00859},
 primaryClass = {astro-ph.GA},
       adsurl = {https://ui.adsabs.harvard.edu/abs/2015MNRAS.454.1634H}
}

@ARTICLE{Butterfield2024,
       author = {{Butterfield}, Natalie O. and {Chuss}, David T. and {Guerra}, Jordan A. and {Morris}, Mark R. and {Par{\'e}}, Dylan and {Wollack}, Edward J. and {Dowell}, C. Darren and {Hankins}, Matthew J. and {Karpovich}, Kaitlyn and {Siah}, Javad and {Staguhn}, Johannes and {Zweibel}, Ellen},
        title = "{SOFIA/HAWC+ Far-Infrared Polarimetric Large Area CMZ Exploration Survey. I. General Results from the Pilot Program}",
      journal = {\apj},
         year = 2024,
        month = mar,
       volume = {963},
       number = {2},
          eid = {130},
        pages = {130},
          doi = {10.3847/1538-4357/ad12b9},
archivePrefix = {arXiv},
       eprint = {2306.01681},
 primaryClass = {astro-ph.GA},
       adsurl = {https://ui.adsabs.harvard.edu/abs/2024ApJ...963..130B}
}

@phdthesis{Nagai2008,
    author = {{Nagai}, Makoto},
    title = "PhD Thesis",
    school = "Univ of Tokyo" ,
    year = 2008 
}

@ARTICLE{Chapman2009,
       author = {{Chapman}, J.~F. and {Millar}, T.~J. and {Wardle}, M. and {Burton}, M.~G. and {Walsh}, A.~J.},
        title = "{Cyanopolyynes in hot cores: modelling G305.2+0.2}",
      journal = {\mnras},
         year = 2009,
        month = mar,
       volume = {394},
       number = {1},
        pages = {221-230},
          doi = {10.1111/j.1365-2966.2008.14144.x},
       adsurl = {https://ui.adsabs.harvard.edu/abs/2009MNRAS.394..221C}
}

@ARTICLE{Miettinen2014,
       author = {{Miettinen}, O.},
        title = "{A MALT90 study of the chemical properties of massive clumps and filaments of infrared dark clouds}",
      journal = {\aap},
         year = 2014,
        month = feb,
       volume = {562},
          eid = {A3},
        pages = {A3},
          doi = {10.1051/0004-6361/201322596},
archivePrefix = {arXiv},
       eprint = {1311.4300},
 primaryClass = {astro-ph.GA},
       adsurl = {https://ui.adsabs.harvard.edu/abs/2014A&A...562A...3M}
}

@ARTICLE{Higuchi2014,
       author = {{Higuchi}, Aya E. and {Chibueze}, James O. and {Habe}, Asao and {Takahira}, Ken and {Takano}, Shuro},
        title = "{ALMA View of G0.253+0.016: Can Cloud-Cloud Collision form the Cloud?}",
      journal = {\aj},
         year = 2014,
        month = jun,
       volume = {147},
       number = {6},
          eid = {141},
        pages = {141},
          doi = {10.1088/0004-6256/147/6/141},
archivePrefix = {arXiv},
       eprint = {1403.4734},
 primaryClass = {astro-ph.GA},
       adsurl = {https://ui.adsabs.harvard.edu/abs/2014AJ....147..141H}
}

@ARTICLE{coil1999,
       author = {{Coil}, Alison L. and {Ho}, Paul T.~P.},
        title = "{Infalling Gas toward the Galactic Center}",
      journal = {\apj},
         year = 1999,
        month = mar,
       volume = {513},
       number = {2},
        pages = {752-766},
          doi = {10.1086/306875},
archivePrefix = {arXiv},
       eprint = {astro-ph/9809068},
 primaryClass = {astro-ph},
       adsurl = {https://ui.adsabs.harvard.edu/abs/1999ApJ...513..752C}
}

@ARTICLE{Guesten1987,
       author = {{Guesten}, R. and {Genzel}, R. and {Wright}, M.~C.~H. and {Jaffe}, D.~T. and {Stutzki}, J. and {Harris}, A.~I.},
        title = "{Aperture Synthesis Observations of the Circumnuclear Ring in the Galactic Center}",
      journal = {\apj},
         year = 1987,
        month = jul,
       volume = {318},
        pages = {124},
          doi = {10.1086/165355},
       adsurl = {https://ui.adsabs.harvard.edu/abs/1987ApJ...318..124G}
}

@INPROCEEDINGS{Hsieh2021,
       author = {{Hsieh}, P.-Y.},
        title = "{Dynamics of Streamers Towards the Circumnuclear Disk: On-going Mass Accretion}",
    booktitle = {New Horizons in Galactic Center Astronomy and Beyond},
         year = 2021,
       editor = {{Tsuboi}, M. and {Oka}, T.},
       series = {Astronomical Society of the Pacific Conference Series},
       volume = {528},
        month = jul,
        pages = {197},
       adsurl = {https://ui.adsabs.harvard.edu/abs/2021ASPC..528..197H}
}

@ARTICLE{Tanaka2026,
       author = {{Tanaka}, Kunihiko and {Nagai}, Makoto and {Kamegai}, Kazuhisa},
        title = "{ALMA [C I] Image of the Circumnuclear Disk of the Milky Way: Inflowing Low-density Molecular Gas}",
      journal = {\apj},
         year = 2026,
        month = mar,
       volume = {999},
       number = {2},
          eid = {185},
        pages = {185},
          doi = {10.3847/1538-4357/ae3aa5},
archivePrefix = {arXiv},
       eprint = {2602.09251},
 primaryClass = {astro-ph.GA},
       adsurl = {https://ui.adsabs.harvard.edu/abs/2026ApJ...999..185T}
}

@ARTICLE{Goicoechea2018,
       author = {{Goicoechea}, Javier R. and {Pety}, Jerome and {Chapillon}, Edwige and {Cernicharo}, Jos{\'e} and {Gerin}, Maryvonne and {Herrera}, Cinthya and {Requena-Torres}, Miguel A. and {Santa-Maria}, Miriam G.},
        title = "{High-speed molecular cloudlets around the Galactic center's supermassive black hole}",
      journal = {\aap},
         year = 2018,
        month = oct,
       volume = {618},
          eid = {A35},
        pages = {A35},
          doi = {10.1051/0004-6361/201833558},
archivePrefix = {arXiv},
       eprint = {1806.01748},
 primaryClass = {astro-ph.GA},
       adsurl = {https://ui.adsabs.harvard.edu/abs/2018A&A...618A..35G}
}

@ARTICLE{Tsuboi2018,
       author = {{Tsuboi}, Masato and {Kitamura}, Yoshimi and {Uehara}, Kenta and {Tsutsumi}, Takahiro and {Miyawaki}, Ryosuke and {Miyoshi}, Makoto and {Miyazaki}, Atsushi},
        title = "{ALMA view of the circumnuclear disk of the Galactic Center: tidally disrupted molecular clouds falling to the Galactic Center}",
      journal = {\pasj},
         year = 2018,
        month = oct,
       volume = {70},
       number = {5},
          eid = {85},
        pages = {85},
          doi = {10.1093/pasj/psy080},
archivePrefix = {arXiv},
       eprint = {1806.10246},
 primaryClass = {astro-ph.GA},
       adsurl = {https://ui.adsabs.harvard.edu/abs/2018PASJ...70...85T}
}

@ARTICLE{Tsuboi2015,
       author = {{Tsuboi}, Masato and {Miyazaki}, Atsushi and {Uehara}, Kenta},
        title = "{Cloud-cloud collision in the Galactic center 50 km s$^{-1}$ molecular cloud}",
      journal = {\pasj},
         year = 2015,
        month = dec,
       volume = {67},
       number = {6},
          eid = {109},
        pages = {109},
          doi = {10.1093/pasj/psv076},
archivePrefix = {arXiv},
       eprint = {1507.08351},
 primaryClass = {astro-ph.GA},
       adsurl = {https://ui.adsabs.harvard.edu/abs/2015PASJ...67..109T}
}

@ARTICLE{Mills2018a,
       author = {{Mills}, E.~A.~C. and {Ginsburg}, A. and {Immer}, K. and {Barnes}, J.~M. and {Wiesenfeld}, L. and {Faure}, A. and {Morris}, M.~R. and {Requena-Torres}, M.~A.},
        title = "{The Dense Gas Fraction in Galactic Center Clouds}",
      journal = {\apj},
         year = 2018,
        month = nov,
       volume = {868},
       number = {1},
          eid = {7},
        pages = {7},
          doi = {10.3847/1538-4357/aae581},
archivePrefix = {arXiv},
       eprint = {1810.00266},
 primaryClass = {astro-ph.GA},
       adsurl = {https://ui.adsabs.harvard.edu/abs/2018ApJ...868....7M}
}

@ARTICLE{Hatchfield2021,
       author = {{Hatchfield}, H. Perry and {Sormani}, Mattia C. and {Tress}, Robin G. and {Battersby}, Cara and {Smith}, Rowan J. and {Glover}, Simon C.~O. and {Klessen}, Ralf S.},
        title = "{Dynamically Driven Inflow onto the Galactic Center and its Effect upon Molecular Clouds}",
      journal = {\apj},
         year = 2021,
        month = nov,
       volume = {922},
       number = {1},
          eid = {79},
        pages = {79},
          doi = {10.3847/1538-4357/ac1e89},
archivePrefix = {arXiv},
       eprint = {2106.08461},
 primaryClass = {astro-ph.GA},
       adsurl = {https://ui.adsabs.harvard.edu/abs/2021ApJ...922...79H}
}

@ARTICLE{Eden2020,
       author = {{Eden}, D.~J. and {Moore}, T.~J.~T. and {Currie}, M.~J. and {Rigby}, A.~J. and {Rosolowsky}, E. and {Su}, Y. and {Kim}, Kee-Tae and {Parsons}, H. and {Morata}, O. and {Chen}, H.-R. and et al.},
        title = "{CHIMPS2: survey description and $^{12}$CO emission in the Galactic Centre}",
      journal = {\mnras},
         year = 2020,
        month = nov,
       volume = {498},
       number = {4},
        pages = {5936-5951},
          doi = {10.1093/mnras/staa2734},
archivePrefix = {arXiv},
       eprint = {2009.05073},
 primaryClass = {astro-ph.GA},
       adsurl = {https://ui.adsabs.harvard.edu/abs/2020MNRAS.498.5936E}
}

@ARTICLE{Gramze2023,
       author = {{Gramze}, Savannah R. and {Ginsburg}, Adam and {Meier}, David S. and {Ott}, Juergen and {Shirley}, Yancy and {Sormani}, Mattia C. and {Svoboda}, Brian E.},
        title = "{Evidence of a Cloud{\textendash}Cloud Collision from Overshooting Gas in the Galactic Center}",
      journal = {\apj},
         year = 2023,
        month = dec,
       volume = {959},
       number = {2},
          eid = {93},
        pages = {93},
          doi = {10.3847/1538-4357/ad01be},
archivePrefix = {arXiv},
       eprint = {2309.16403},
 primaryClass = {astro-ph.GA},
       adsurl = {https://ui.adsabs.harvard.edu/abs/2023ApJ...959...93G}
}

@ARTICLE{Schmiedeke2016,
       author = {{Schmiedeke}, A. and {Schilke}, P. and {M{\"o}ller}, Th. and {S{\'a}nchez-Monge}, {\'A}. and {Bergin}, E. and {Comito}, C. and {Csengeri}, T. and {Lis}, D.~C. and {Molinari}, S. and {Qin}, S. -L. and {Rolffs}, R.},
        title = "{The physical and chemical structure of Sagittarius B2. I. Three-dimensional thermal dust and free-free continuum modeling on 100 au to 45 pc scales}",
      journal = {\aap},
         year = 2016,
        month = apr,
       volume = {588},
          eid = {A143},
        pages = {A143},
          doi = {10.1051/0004-6361/201527311},
archivePrefix = {arXiv},
       eprint = {1602.02274},
 primaryClass = {astro-ph.GA},
       adsurl = {https://ui.adsabs.harvard.edu/abs/2016A&A...588A.143S}
}

@ARTICLE{Sormani2020,
       author = {{Sormani}, Mattia C. and {Tress}, Robin G. and {Glover}, Simon C.~O. and {Klessen}, Ralf S. and {Battersby}, Cara D. and {Clark}, Paul C. and {Hatchfield}, H. Perry and {Smith}, Rowan J.},
        title = "{Simulations of the Milky Way's Central Molecular Zone - II. Star formation}",
      journal = {\mnras},
         year = 2020,
        month = oct,
       volume = {497},
       number = {4},
        pages = {5024-5040},
          doi = {10.1093/mnras/staa1999},
archivePrefix = {arXiv},
       eprint = {2004.06731},
 primaryClass = {astro-ph.GA},
       adsurl = {https://ui.adsabs.harvard.edu/abs/2020MNRAS.497.5024S}
}

@ARTICLE{Tress2020,
       author = {{Tress}, Robin G. and {Sormani}, Mattia C. and {Glover}, Simon C.~O. and {Klessen}, Ralf S. and {Battersby}, Cara D. and {Clark}, Paul C. and {Hatchfield}, H. Perry and {Smith}, Rowan J.},
        title = "{Simulations of the Milky Way's central molecular zone - I. Gas dynamics}",
      journal = {\mnras},
         year = 2020,
        month = dec,
       volume = {499},
       number = {3},
        pages = {4455-4478},
          doi = {10.1093/mnras/staa3120},
archivePrefix = {arXiv},
       eprint = {2004.06724},
 primaryClass = {astro-ph.GA},
       adsurl = {https://ui.adsabs.harvard.edu/abs/2020MNRAS.499.4455T}
}

@ARTICLE{Sormani2018b,
       author = {{Sormani}, Mattia C. and {Sobacchi}, Emanuele and {Fragkoudi}, Francesca and {Ridley}, Matthew and {Tre{\ss}}, Robin G. and {Glover}, Simon C.~O. and {Klessen}, Ralf S.},
        title = "{A dynamical mechanism for the origin of nuclear rings}",
      journal = {\mnras},
         year = 2018,
        month = nov,
       volume = {481},
       number = {1},
        pages = {2-19},
          doi = {10.1093/mnras/sty2246},
archivePrefix = {arXiv},
       eprint = {1805.07969},
 primaryClass = {astro-ph.GA},
       adsurl = {https://ui.adsabs.harvard.edu/abs/2018MNRAS.481....2S}
}

@ARTICLE{Longmore2013a,
       author = {{Longmore}, S.~N. and {Bally}, J. and {Testi}, L. and {Purcell}, C.~R. and {Walsh}, A.~J. and {Bressert}, E. and {Pestalozzi}, M. and {Molinari}, S. and {Ott}, J. and {Cortese}, L. and {Battersby}, C. and {Murray}, N. and {Lee}, E. and {Kruijssen}, J.~M.~D. and {Schisano}, E. and {Elia}, D.},
        title = "{Variations in the Galactic star formation rate and density thresholds for star formation}",
      journal = {\mnras},
         year = 2013,
        month = feb,
       volume = {429},
       number = {2},
        pages = {987-1000},
          doi = {10.1093/mnras/sts376},
archivePrefix = {arXiv},
       eprint = {1208.4256},
 primaryClass = {astro-ph.GA},
       adsurl = {https://ui.adsabs.harvard.edu/abs/2013MNRAS.429..987L}
}

@ARTICLE{Gusdorf2008,
       author = {{Gusdorf}, A. and {Cabrit}, S. and {Flower}, D.~R. and {Pineau Des For{\^e}ts}, G.},
        title = "{SiO line emission from C-type shock waves: interstellar jets and outflows}",
      journal = {\aap},
         year = 2008,
        month = may,
       volume = {482},
       number = {3},
        pages = {809-829},
          doi = {10.1051/0004-6361:20078900},
archivePrefix = {arXiv},
       eprint = {0803.2791},
 primaryClass = {astro-ph},
       adsurl = {https://ui.adsabs.harvard.edu/abs/2008A&A...482..809G}
}

@ARTICLE{Federrath2016,
       author = {{Federrath}, C. and {Rathborne}, J.~M. and {Longmore}, S.~N. and
         {Kruijssen}, J.~M.~D. and {Bally}, J. and {Contreras}, Y. and
         {Crocker}, R.~M. and {Garay}, G. and {Jackson}, J.~M. and {Testi}, L. and
         {Walsh}, A.~J.},
        title = "{The Link between Turbulence, Magnetic Fields, Filaments, and Star Formation in the Central Molecular Zone Cloud G0.253+0.016}",
      journal = {\apj},
         year = 2016,
        month = dec,
       volume = {832},
       number = {2},
          eid = {143},
        pages = {143},
          doi = {10.3847/0004-637X/832/2/143},
archivePrefix = {arXiv},
       eprint = {1609.05911},
 primaryClass = {astro-ph.GA},
       adsurl = {https://ui.adsabs.harvard.edu/abs/2016ApJ...832..143F}
}

@article{Pillai2015,
	Adsurl = {http://adsabs.harvard.edu/abs/2015ApJ...799...74P},
	Archiveprefix = {arXiv},
	Author = {{Pillai}, T. and {Kauffmann}, J. and {Tan}, J.~C. and {Goldsmith}, P.~F. and {Carey}, S.~J. and {Menten}, K.~M.},
	Doi = {10.1088/0004-637X/799/1/74},
	Eid = {74},
	Eprint = {1410.7390},
	Journal = {\apj},
	Month = jan,
	Pages = {74},
	Title = {{Magnetic Fields in High-mass Infrared Dark Clouds}},
	Volume = 799,
	Year = 2015}

@ARTICLE{Kruijssen2015,
       author = {{Kruijssen}, J.~M. Diederik and {Dale}, James E. and {Longmore}, Steven N.},
        title = "{The dynamical evolution of molecular clouds near the Galactic Centre - I. Orbital structure and evolutionary timeline}",
      journal = {\mnras},
         year = 2015,
        month = feb,
       volume = {447},
       number = {2},
        pages = {1059-1079},
          doi = {10.1093/mnras/stu2526},
archivePrefix = {arXiv},
       eprint = {1412.0664},
 primaryClass = {astro-ph.GA},
       adsurl = {https://ui.adsabs.harvard.edu/abs/2015MNRAS.447.1059K}
}

@article{astropy:2013,
Adsurl = {http://adsabs.harvard.edu/abs/2013A%26A...558A..33A},
Archiveprefix = {arXiv},
Author = {{Astropy Collaboration} and {Robitaille}, T.~P. and {Tollerud}, E.~J. and {Greenfield}, P. and {Droettboom}, M. and {Bray}, E. and {Aldcroft}, T. and {Davis}, M. and {Ginsburg}, A. and {Price-Whelan}, A.~M. and {Kerzendorf}, W.~E. and {Conley}, A. and {Crighton}, N. and {Barbary}, K. and {Muna}, D. and {Ferguson}, H. and {Grollier}, F. and {Parikh}, M.~M. and {Nair}, P.~H. and {Unther}, H.~M. and {Deil}, C. and {Woillez}, J. and {Conseil}, S. and {Kramer}, R. and {Turner}, J.~E.~H. and {Singer}, L. and {Fox}, R. and {Weaver}, B.~A. and {Zabalza}, V. and {Edwards}, Z.~I. and {Azalee Bostroem}, K. and {Burke}, D.~J. and {Casey}, A.~R. and {Crawford}, S.~M. and {Dencheva}, N. and {Ely}, J. and {Jenness}, T. and {Labrie}, K. and {Lim}, P.~L. and {Pierfederici}, F. and {Pontzen}, A. and {Ptak}, A. and {Refsdal}, B. and {Servillat}, M. and {Streicher}, O.},
Doi = {10.1051/0004-6361/201322068},
Eid = {A33},
Eprint = {1307.6212},
Journal = {\aap},
Month = oct,
Pages = {A33},
Primaryclass = {astro-ph.IM},
Title = {{Astropy: A community Python package for astronomy}},
Volume = 558,
Year = 2013}

@ARTICLE{astropy:2018,
       author = {{Astropy Collaboration} and {Price-Whelan}, A.~M. and
         {Sip{\H{o}}cz}, B.~M. and {G{\"u}nther}, H.~M. and {Lim}, P.~L. and
         {Crawford}, S.~M. and {Conseil}, S. and {Shupe}, D.~L. and
         {Craig}, M.~W. and {Dencheva}, N. and {Ginsburg}, A. and {Vand
        erPlas}, J.~T. and {Bradley}, L.~D. and {P{\'e}rez-Su{\'a}rez}, D. and
         {de Val-Borro}, M. and {Aldcroft}, T.~L. and {Cruz}, K.~L. and
         {Robitaille}, T.~P. and {Tollerud}, E.~J. and {Ardelean}, C. and
         {Babej}, T. and {Bach}, Y.~P. and {Bachetti}, M. and {Bakanov}, A.~V. and
         {Bamford}, S.~P. and {Barentsen}, G. and {Barmby}, P. and
         {Baumbach}, A. and {Berry}, K.~L. and {Biscani}, F. and {Boquien}, M. and
         {Bostroem}, K.~A. and {Bouma}, L.~G. and {Brammer}, G.~B. and
         {Bray}, E.~M. and {Breytenbach}, H. and {Buddelmeijer}, H. and
         {Burke}, D.~J. and {Calderone}, G. and {Cano Rodr{\'\i}guez}, J.~L. and
         {Cara}, M. and {Cardoso}, J.~V.~M. and {Cheedella}, S. and {Copin}, Y. and
         {Corrales}, L. and {Crichton}, D. and {D'Avella}, D. and {Deil}, C. and
         {Depagne}, {\'E}. and {Dietrich}, J.~P. and {Donath}, A. and
         {Droettboom}, M. and {Earl}, N. and {Erben}, T. and {Fabbro}, S. and
         {Ferreira}, L.~A. and {Finethy}, T. and {Fox}, R.~T. and
         {Garrison}, L.~H. and {Gibbons}, S.~L.~J. and {Goldstein}, D.~A. and
         {Gommers}, R. and {Greco}, J.~P. and {Greenfield}, P. and
         {Groener}, A.~M. and {Grollier}, F. and {Hagen}, A. and {Hirst}, P. and
         {Homeier}, D. and {Horton}, A.~J. and {Hosseinzadeh}, G. and {Hu}, L. and
         {Hunkeler}, J.~S. and {Ivezi{\'c}}, {\v{Z}}. and {Jain}, A. and
         {Jenness}, T. and {Kanarek}, G. and {Kendrew}, S. and {Kern}, N.~S. and
         {Kerzendorf}, W.~E. and {Khvalko}, A. and {King}, J. and {Kirkby}, D. and
         {Kulkarni}, A.~M. and {Kumar}, A. and {Lee}, A. and {Lenz}, D. and
         {Littlefair}, S.~P. and {Ma}, Z. and {Macleod}, D.~M. and
         {Mastropietro}, M. and {McCully}, C. and {Montagnac}, S. and
         {Morris}, B.~M. and {Mueller}, M. and {Mumford}, S.~J. and {Muna}, D. and
         {Murphy}, N.~A. and {Nelson}, S. and {Nguyen}, G.~H. and
         {Ninan}, J.~P. and {N{\"o}the}, M. and {Ogaz}, S. and {Oh}, S. and
         {Parejko}, J.~K. and {Parley}, N. and {Pascual}, S. and {Patil}, R. and
         {Patil}, A.~A. and {Plunkett}, A.~L. and {Prochaska}, J.~X. and
         {Rastogi}, T. and {Reddy Janga}, V. and {Sabater}, J. and
         {Sakurikar}, P. and {Seifert}, M. and {Sherbert}, L.~E. and
         {Sherwood-Taylor}, H. and {Shih}, A.~Y. and {Sick}, J. and
         {Silbiger}, M.~T. and {Singanamalla}, S. and {Singer}, L.~P. and
         {Sladen}, P.~H. and {Sooley}, K.~A. and {Sornarajah}, S. and
         {Streicher}, O. and {Teuben}, P. and {Thomas}, S.~W. and
         {Tremblay}, G.~R. and {Turner}, J.~E.~H. and {Terr{\'o}n}, V. and
         {van Kerkwijk}, M.~H. and {de la Vega}, A. and {Watkins}, L.~L. and
         {Weaver}, B.~A. and {Whitmore}, J.~B. and {Woillez}, J. and
         {Zabalza}, V. and {Astropy Contributors}},
        title = "{The Astropy Project: Building an Open-science Project and Status of the v2.0 Core Package}",
      journal = {\aj},
         year = 2018,
        month = sep,
       volume = {156},
       number = {3},
          eid = {123},
        pages = {123},
          doi = {10.3847/1538-3881/aabc4f},
archivePrefix = {arXiv},
       eprint = {1801.02634},
 primaryClass = {astro-ph.IM},
       adsurl = {https://ui.adsabs.harvard.edu/abs/2018AJ....156..123A}
}

@ARTICLE{astropy:2022,
       author = {{Astropy Collaboration} and {Price-Whelan}, Adrian M. and {Lim}, Pey Lian and {Earl}, Nicholas and {Starkman}, Nathaniel and {Bradley}, Larry and {Shupe}, David L. and {Patil}, Aarya A. and {Corrales}, Lia and {Brasseur}, C.~E. and {N{"o}the}, Maximilian and {Donath}, Axel and {Tollerud}, Erik and {Morris}, Brett M. and {Ginsburg}, Adam and {Vaher}, Eero and {Weaver}, Benjamin A. and {Tocknell}, James and {Jamieson}, William and {van Kerkwijk}, Marten H. and {Robitaille}, Thomas P. and {Merry}, Bruce and {Bachetti}, Matteo and {G{"u}nther}, H. Moritz and {Aldcroft}, Thomas L. and {Alvarado-Montes}, Jaime A. and {Archibald}, Anne M. and {B{'o}di}, Attila and {Bapat}, Shreyas and {Barentsen}, Geert and {Baz{'a}n}, Juanjo and {Biswas}, Manish and {Boquien}, M{'e}d{'e}ric and {Burke}, D.~J. and {Cara}, Daria and {Cara}, Mihai and {Conroy}, Kyle E. and {Conseil}, Simon and {Craig}, Matthew W. and {Cross}, Robert M. and {Cruz}, Kelle L. and {D'Eugenio}, Francesco and {Dencheva}, Nadia and {Devillepoix}, Hadrien A.~R. and {Dietrich}, J{"o}rg P. and {Eigenbrot}, Arthur Davis and {Erben}, Thomas and {Ferreira}, Leonardo and {Foreman-Mackey}, Daniel and {Fox}, Ryan and {Freij}, Nabil and {Garg}, Suyog and {Geda}, Robel and {Glattly}, Lauren and {Gondhalekar}, Yash and {Gordon}, Karl D. and {Grant}, David and {Greenfield}, Perry and {Groener}, Austen M. and {Guest}, Steve and {Gurovich}, Sebastian and {Handberg}, Rasmus and {Hart}, Akeem and {Hatfield-Dodds}, Zac and {Homeier}, Derek and {Hosseinzadeh}, Griffin and {Jenness}, Tim and {Jones}, Craig K. and {Joseph}, Prajwel and {Kalmbach}, J. Bryce and {Karamehmetoglu}, Emir and {Ka{l}uszy{'n}ski}, Miko{l}aj and {Kelley}, Michael S.~P. and {Kern}, Nicholas and {Kerzendorf}, Wolfgang E. and {Koch}, Eric W. and {Kulumani}, Shankar and {Lee}, Antony and {Ly}, Chun and {Ma}, Zhiyuan and {MacBride}, Conor and {Maljaars}, Jakob M. and {Muna}, Demitri and {Murphy}, N.~A. and {Norman}, Henrik and {O'Steen}, Richard and {Oman}, Kyle A. and {Pacifici}, Camilla and {Pascual}, Sergio and {Pascual-Granado}, J. and {Patil}, Rohit R. and {Perren}, Gabriel I. and {Pickering}, Timothy E. and {Rastogi}, Tanuj and {Roulston}, Benjamin R. and {Ryan}, Daniel F. and {Rykoff}, Eli S. and {Sabater}, Jose and {Sakurikar}, Parikshit and {Salgado}, Jes{'u}s and {Sanghi}, Aniket and {Saunders}, Nicholas and {Savchenko}, Volodymyr and {Schwardt}, Ludwig and {Seifert-Eckert}, Michael and {Shih}, Albert Y. and {Jain}, Anany Shrey and {Shukla}, Gyanendra and {Sick}, Jonathan and {Simpson}, Chris and {Singanamalla}, Sudheesh and {Singer}, Leo P. and {Singhal}, Jaladh and {Sinha}, Manodeep and {Sip{H{o}}cz}, Brigitta M. and {Spitler}, Lee R. and {Stansby}, David and {Streicher}, Ole and {{{S}}umak}, Jani and {Swinbank}, John D. and {Taranu}, Dan S. and {Tewary}, Nikita and {Tremblay}, Grant R. and {Val-Borro}, Miguel de and {Van Kooten}, Samuel J. and {Vasovi{'c}}, Zlatan and {Verma}, Shresth and {de Miranda Cardoso}, Jos{'e} Vin{'i}cius and {Williams}, Peter K.~G. and {Wilson}, Tom J. and {Winkel}, Benjamin and {Wood-Vasey}, W.~M. and {Xue}, Rui and {Yoachim}, Peter and {Zhang}, Chen and {Zonca}, Andrea and {Astropy Project Contributors}},
        title = "{The Astropy Project: Sustaining and Growing a Community-oriented Open-source Project and the Latest Major Release (v5.0) of the Core Package}",
      journal = {apj},
         year = 2022,
        month = aug,
       volume = {935},
       number = {2},
          eid = {167},
        pages = {167},
          doi = {10.3847/1538-4357/ac7c74},
archivePrefix = {arXiv},
       eprint = {2206.14220},
 primaryClass = {astro-ph.IM},
       adsurl = {https://ui.adsabs.harvard.edu/abs/2022ApJ...935..167A}
}

@misc{astrodendro:2019,
       author = {{Robitaille}, Thomas and {Rice}, Tom and {Beaumont}, Chris and {Ginsburg}, Adam and {MacDonald}, Braden and {Rosolowsky}, Erik},
        title = "{astrodendro: Astronomical data dendrogram creator}",
 howpublished = {Astrophysics Source Code Library, record ascl:1907.016},
         year = 2019,
        month = jul,
          eid = {ascl:1907.016},
archivePrefix = {ascl},
       eprint = {1907.016},
       adsurl = {https://ui.adsabs.harvard.edu/abs/2019ascl.soft07016R}
}

@ARTICLE{Longmore2013,
       author = {{Longmore}, S.~N. and {Bally}, J. and {Testi}, L. and {Purcell}, C.~R. and {Walsh}, A.~J. and {Bressert}, E. and {Pestalozzi}, M. and {Molinari}, S. and {Ott}, J. and {Cortese}, L. and {Battersby}, C. and {Murray}, N. and {Lee}, E. and {Kruijssen}, J.~M.~D. and {Schisano}, E. and {Elia}, D.},
        title = "{Variations in the Galactic star formation rate and density thresholds for star formation}",
      journal = {\mnras},
         year = 2013,
        month = feb,
       volume = {429},
       number = {2},
        pages = {987-1000},
          doi = {10.1093/mnras/sts376},
archivePrefix = {arXiv},
       eprint = {1208.4256},
 primaryClass = {astro-ph.GA},
       adsurl = {https://ui.adsabs.harvard.edu/abs/2013MNRAS.429..987L}
}

@misc{LMFIT_newville_2025_16175987,
  author       = {Newville, Matthew and
                  Otten, Renee and
                  Nelson, Andrew and
                  Stensitzki, Till and
                  Ingargiola, Antonino and
                  Allan, Daniel and
                  Fox, Austin and
                  Carter, Faustin and
                  Rawlik, Michal},
  title        = {LMFIT: Non-Linear Least-Squares Minimization and
                   Curve-Fitting for Python
                  },
  month        = jul,
  year         = 2025,
  publisher    = {Zenodo},
  version      = {1.3.4},
  doi          = {10.5281/zenodo.16175987},
  url          = {https://doi.org/10.5281/zenodo.16175987},
  swhid        = {swh:1:dir:76742b0e41b1d2bff5a3716dd2376531f3a21db8
                   ;origin=https://doi.org/10.5281/zenodo.598352;visi
                   t=swh:1:snp:89f98ce93be53a573d85de0145f9174c827b7e
                   92;anchor=swh:1:rel:9528e5133d2d78c4036335f023b212
                   487cf1b632;path=lmfit-lmfit-py-0566445
                  },
}

@ARTICLE{Kruijssen2013,
       author = {{Kruijssen}, J.~M. Diederik and {Longmore}, Steven N.},
        title = "{Comparing molecular gas across cosmic time-scales: the Milky Way as both a typical spiral galaxy and a high-redshift galaxy analogue}",
      journal = {\mnras},
         year = 2013,
        month = nov,
       volume = {435},
       number = {3},
        pages = {2598-2603},
          doi = {10.1093/mnras/stt1634},
archivePrefix = {arXiv},
       eprint = {1309.0505},
 primaryClass = {astro-ph.CO},
       adsurl = {https://ui.adsabs.harvard.edu/abs/2013MNRAS.435.2598K}
}

@ARTICLE{Federrath_Klessen_2012,
       author = {{Federrath}, Christoph and {Klessen}, Ralf S.},
        title = "{The Star Formation Rate of Turbulent Magnetized Clouds: Comparing Theory, Simulations, and Observations}",
      journal = {\apj},
         year = 2012,
        month = dec,
       volume = {761},
       number = {2},
          eid = {156},
        pages = {156},
          doi = {10.1088/0004-637X/761/2/156},
archivePrefix = {arXiv},
       eprint = {1209.2856},
 primaryClass = {astro-ph.SR},
       adsurl = {https://ui.adsabs.harvard.edu/abs/2012ApJ...761..156F}
}

@ARTICLE{Padoan2011,
       author = {{Padoan}, Paolo and {Nordlund}, {\r{A}}ke},
        title = "{The Star Formation Rate of Supersonic Magnetohydrodynamic Turbulence}",
      journal = {\apj},
         year = 2011,
        month = mar,
       volume = {730},
       number = {1},
          eid = {40},
        pages = {40},
          doi = {10.1088/0004-637X/730/1/40},
archivePrefix = {arXiv},
       eprint = {0907.0248},
 primaryClass = {astro-ph.GA},
       adsurl = {https://ui.adsabs.harvard.edu/abs/2011ApJ...730...40P}
}

@ARTICLE{Hennebelle2011,
       author = {{Hennebelle}, Patrick and {Chabrier}, Gilles},
        title = "{Analytical Star Formation Rate from Gravoturbulent Fragmentation}",
      journal = {\apjl},
         year = 2011,
        month = dec,
       volume = {743},
       number = {2},
          eid = {L29},
        pages = {L29},
          doi = {10.1088/2041-8205/743/2/L29},
archivePrefix = {arXiv},
       eprint = {1110.0033},
 primaryClass = {astro-ph.GA},
       adsurl = {https://ui.adsabs.harvard.edu/abs/2011ApJ...743L..29H}
}

@ARTICLE{Humire2020,
       author = {{Humire}, P.~K. and {Thiel}, V. and {Henkel}, C. and {Belloche}, A. and {Loison}, J.-C. and {Pillai}, T. and {Riquelme}, D. and {Wakelam}, V. and {Langer}, N. and {Hern{\'a}ndez-G{\'o}mez}, A. and {Mauersberger}, R. and {Menten}, K.~M.},
        title = "{Sulphur and carbon isotopes towards Galactic centre clouds}",
      journal = {\aap},
         year = 2020,
        month = oct,
       volume = {642},
          eid = {A222},
        pages = {A222},
          doi = {10.1051/0004-6361/202038216},
archivePrefix = {arXiv},
       eprint = {2009.07306},
 primaryClass = {astro-ph.GA},
       adsurl = {https://ui.adsabs.harvard.edu/abs/2020A&A...642A.222H}
}

@ARTICLE{Rod-Baras2021,
       author = {{Rodr{\'\i}guez-Baras}, M. and {Fuente}, A. and {Rivi{\'e}re-Marichalar}, P. and {Navarro-Almaida}, D. and {Caselli}, P. and {Gerin}, M. and {Kramer}, C. and {Roueff}, E. and {Wakelam}, V. and {Esplugues}, G. and {Garc{\'\i}a-Burillo}, S. and {Le Gal}, R. and {Spezzano}, S. and {Alonso-Albi}, T. and {Bachiller}, R. and {Cazaux}, S. and {Commercon}, B. and {Goicoechea}, J.~R. and {Loison}, J.~C. and {Trevi{\~n}o-Morales}, S.~P. and {Roncero}, O. and {Jim{\'e}nez-Serra}, I. and {Laas}, J. and {Hacar}, A. and {Kirk}, J. and {Lattanzi}, V. and {Mart{\'\i}n-Dom{\'e}nech}, R. and {Mu{\~n}oz-Caro}, G. and {Pineda}, J.~E. and {Tercero}, B. and {Ward-Thompson}, D. and {Tafalla}, M. and {Marcelino}, N. and {Malinen}, J. and {Friesen}, R. and {Giuliano}, B.~M.},
        title = "{Gas phase Elemental abundances in Molecular cloudS (GEMS). IV. Observational results and statistical trends}",
      journal = {\aap},
         year = 2021,
        month = apr,
       volume = {648},
          eid = {A120},
        pages = {A120},
          doi = {10.1051/0004-6361/202040112},
archivePrefix = {arXiv},
       eprint = {2102.13153},
 primaryClass = {astro-ph.GA},
       adsurl = {https://ui.adsabs.harvard.edu/abs/2021A&A...648A.120R}
}

@ARTICLE{Burkhart2018,
       author = {{Burkhart}, Blakesley},
        title = "{The Star Formation Rate in the Gravoturbulent Interstellar Medium}",
      journal = {\apj},
         year = 2018,
        month = aug,
       volume = {863},
       number = {2},
          eid = {118},
        pages = {118},
          doi = {10.3847/1538-4357/aad002},
archivePrefix = {arXiv},
       eprint = {1801.05428},
 primaryClass = {astro-ph.GA},
       adsurl = {https://ui.adsabs.harvard.edu/abs/2018ApJ...863..118B}
}

@ARTICLE{Meier_Turner_2012,
       author = {{Meier}, David S. and {Turner}, Jean L.},
        title = "{Spatially Resolved Chemistry in nearby Galaxies. II. The Nuclear Bar in Maffei 2}",
      journal = {\apj},
         year = 2012,
        month = aug,
       volume = {755},
       number = {2},
          eid = {104},
        pages = {104},
          doi = {10.1088/0004-637X/755/2/104},
archivePrefix = {arXiv},
       eprint = {1206.4098},
 primaryClass = {astro-ph.CO},
       adsurl = {https://ui.adsabs.harvard.edu/abs/2012ApJ...755..104M}
}

@ARTICLE{Meier2015,
       author = {{Meier}, David S. and {Walter}, Fabian and {Bolatto}, Alberto D. and {Leroy}, Adam K. and {Ott}, J{\"u}rgen and {Rosolowsky}, Erik and {Veilleux}, Sylvain and {Warren}, Steven R. and {Wei{\ss}}, Axel and {Zwaan}, Martin A. and {Zschaechner}, Laura K.},
        title = "{ALMA Multi-line Imaging of the Nearby Starburst NGC 253}",
      journal = {\apj},
         year = 2015,
        month = mar,
       volume = {801},
       number = {1},
          eid = {63},
        pages = {63},
          doi = {10.1088/0004-637X/801/1/63},
archivePrefix = {arXiv},
       eprint = {1501.05694},
 primaryClass = {astro-ph.GA},
       adsurl = {https://ui.adsabs.harvard.edu/abs/2015ApJ...801...63M}
}

@ARTICLE{Gusten1981,
       author = {{G{\"u}sten}, R. and {Walmsley}, C.~M. and {Pauls}, T.},
        title = "{Ammonia in the neighbourhood of the Galactic Center.}",
      journal = {\aap},
         year = 1981,
        month = nov,
       volume = {103},
        pages = {197-206},
       adsurl = {https://ui.adsabs.harvard.edu/abs/1981A&A...103..197G}
}

@ARTICLE{Walker2018,
       author = {{Walker}, D.~L. and {Longmore}, S.~N. and {Zhang}, Q. and {Battersby}, C. and {Keto}, E. and {Kruijssen}, J.~M.~D. and {Ginsburg}, A. and {Lu}, X. and {Henshaw}, J.~D. and {Kauffmann}, J. and {Pillai}, T. and {Mills}, E.~A.~C. and {Walsh}, A.~J. and {Bally}, J. and {Ho}, L.~C. and {Immer}, K. and {Johnston}, K.~G.},
        title = "{Star formation in a high-pressure environment: an SMA view of the Galactic Centre dust ridge}",
      journal = {\mnras},
         year = 2018,
        month = feb,
       volume = {474},
       number = {2},
        pages = {2373-2388},
          doi = {10.1093/mnras/stx2898},
archivePrefix = {arXiv},
       eprint = {1711.00781},
 primaryClass = {astro-ph.GA},
       adsurl = {https://ui.adsabs.harvard.edu/abs/2018MNRAS.474.2373W}
}

@ARTICLE{Ginsburg2016,
       author = {{Ginsburg}, Adam and {Henkel}, Christian and {Ao}, Yiping and {Riquelme}, Denise and {Kauffmann}, Jens and {Pillai}, Thushara and {Mills}, Elisabeth A.~C. and {Requena-Torres}, Miguel A. and {Immer}, Katharina and {Testi}, Leonardo and {Ott}, Juergen and {Bally}, John and {Battersby}, Cara and {Darling}, Jeremy and {Aalto}, Susanne and {Stanke}, Thomas and {Kendrew}, Sarah and {Kruijssen}, J.~M. Diederik and {Longmore}, Steven and {Dale}, James and {Guesten}, Rolf and {Menten}, Karl M.},
        title = "{Dense gas in the Galactic central molecular zone is warm and heated by turbulence}",
      journal = {\aap},
         year = 2016,
        month = feb,
       volume = {586},
          eid = {A50},
        pages = {A50},
          doi = {10.1051/0004-6361/201526100},
archivePrefix = {arXiv},
       eprint = {1509.01583},
 primaryClass = {astro-ph.GA},
       adsurl = {https://ui.adsabs.harvard.edu/abs/2016A&A...586A..50G}
}

@ARTICLE{Mills2013,
       author = {{Mills}, E.~A.~C. and {Morris}, M.~R.},
        title = "{Detection of Widespread Hot Ammonia in the Galactic Center}",
      journal = {\apj},
         year = 2013,
        month = aug,
       volume = {772},
       number = {2},
          eid = {105},
        pages = {105},
          doi = {10.1088/0004-637X/772/2/105},
archivePrefix = {arXiv},
       eprint = {1306.0953},
 primaryClass = {astro-ph.GA},
       adsurl = {https://ui.adsabs.harvard.edu/abs/2013ApJ...772..105M}
}

@INPROCEEDINGS{Henshaw2023,
       author = {{Henshaw}, J.~D. and {Barnes}, A.~T. and {Battersby}, C. and {Ginsburg}, A. and {Sormani}, M.~C. and {Walker}, D.~L.},
        title = "{Star Formation in the Central Molecular Zone of the Milky Way}",
    booktitle = {Protostars and Planets VII},
         year = 2023,
       editor = {{Inutsuka}, S. and {Aikawa}, Y. and {Muto}, T. and {Tomida}, K. and {Tamura}, M.},
       series = {Astronomical Society of the Pacific Conference Series},
       volume = {534},
        month = jul,
        pages = {83},
          doi = {10.48550/arXiv.2203.11223},
archivePrefix = {arXiv},
       eprint = {2203.11223},
 primaryClass = {astro-ph.GA},
       adsurl = {https://ui.adsabs.harvard.edu/abs/2023ASPC..534...83H}
}

@ARTICLE{Feng2026,
       author = {{Feng}, Zi-Xuan and {Sormani}, Mattia C. and {Tress}, Robin G. and {Glover}, Simon C.~O. and {Klessen}, Ralf S. and {Petersson}, Jonathan and {Hirschmann}, Michaela and {Barnes}, Ashley T. and {Battersby}, Cara and {Donati}, Marco and et al.},
        title = "{Simulations of gas inflow in the Milky Way I. Stellar-Feedback-Regulated Transport from the Central Molecular Zone to the Circumnuclear disk}",
      journal = {arXiv e-prints},
         year = 2026,
        month = may,
          eid = {arXiv:2605.16192},
        pages = {arXiv:2605.16192},
          doi = {10.48550/arXiv.2605.16192},
archivePrefix = {arXiv},
       eprint = {2605.16192},
 primaryClass = {astro-ph.GA},
       adsurl = {https://ui.adsabs.harvard.edu/abs/2026arXiv260516192F}
}

@ARTICLE{Dutkowska2025,
       author = {{Dutkowska}, Katarzyna M. and {Vermari{\"e}n}, Gijs and {Viti}, Serena and {Jim{\'e}nez-Serra}, Izaskun and {Colzi}, Laura and {Busch}, Laura A. and {Rivilla}, V{\'\i}ctor M. and {Mills}, Elisabeth A.~C. and {Mart{\'\i}n}, Sergio and {Henkel}, Christian and et al.},
        title = "{Chemical templates of the Central Molecular Zone: Shock and protostellar object signatures under Galactic Center conditions}",
      journal = {\aap},
         year = 2025,
        month = nov,
       volume = {703},
          eid = {A46},
        pages = {A46},
          doi = {10.1051/0004-6361/202556188},
archivePrefix = {arXiv},
       eprint = {2508.10759},
 primaryClass = {astro-ph.GA},
       adsurl = {https://ui.adsabs.harvard.edu/abs/2025A&A...703A..46D}
}

@ARTICLE{Bally1987,
       author = {{Bally}, John and {Stark}, Antony A. and {Wilson}, Robert W. and {Henkel}, Christian},
        title = "{Galactic Center Molecular Clouds. I. Spatial and Spatial Velocity Maps}",
      journal = {\apjs},
         year = 1987,
        month = sep,
       volume = {65},
        pages = {13},
          doi = {10.1086/191217},
       adsurl = {https://ui.adsabs.harvard.edu/abs/1987ApJS...65...13B}
}

@ARTICLE{Bally2010,
       author = {{Bally}, John and {Aguirre}, James and {Battersby}, Cara and {Bradley}, Eric Todd and {Cyganowski}, Claudia and {Dowell}, Darren and {Drosback}, Meredith and {Dunham}, Miranda K. and {Evans}, Neal J., II and {Ginsburg}, Adam and {Glenn}, Jason and {Harvey}, Paul and {Mills}, Elisabeth and {Merello}, Manuel and {Rosolowsky}, Erik and {Schlingman}, Wayne and {Shirley}, Yancy L. and {Stringfellow}, Guy S. and {Walawender}, Josh and {Williams}, Jonathan},
        title = "{The Bolocam Galactic Plane Survey: {\ensuremath{\lambda}} = 1.1 and 0.35 mm Dust Continuum Emission in the Galactic Center Region}",
      journal = {\apj},
         year = 2010,
        month = sep,
       volume = {721},
       number = {1},
        pages = {137-163},
          doi = {10.1088/0004-637X/721/1/137},
archivePrefix = {arXiv},
       eprint = {1011.0932},
 primaryClass = {astro-ph.GA},
       adsurl = {https://ui.adsabs.harvard.edu/abs/2010ApJ...721..137B}
}

@ARTICLE{Longmore2012,
       author = {{Longmore}, Steven N. and {Rathborne}, Jill and {Bastian}, Nate and {Alves}, Joao and {Ascenso}, Joana and {Bally}, John and {Testi}, Leonardo and {Longmore}, Andy and {Battersby}, Cara and {Bressert}, Eli and {Purcell}, Cormac and {Walsh}, Andrew and {Jackson}, James and {Foster}, Jonathan and {Molinari}, Sergio and {Meingast}, Stefan and {Amorim}, A. and {Lima}, J. and {Marques}, R. and {Moitinho}, A. and {Pinhao}, J. and {Rebordao}, J. and {Santos}, F.~D.},
        title = "{G0.253 + 0.016: A Molecular Cloud Progenitor of an Arches-like Cluster}",
      journal = {\apj},
         year = 2012,
        month = feb,
       volume = {746},
       number = {2},
          eid = {117},
        pages = {117},
          doi = {10.1088/0004-637X/746/2/117},
archivePrefix = {arXiv},
       eprint = {1111.3199},
 primaryClass = {astro-ph.GA},
       adsurl = {https://ui.adsabs.harvard.edu/abs/2012ApJ...746..117L}
}

@ARTICLE{Colzi2022,
       author = {{Colzi}, Laura and {Mart{\'\i}n-Pintado}, Jes{\'u}s and {Rivilla}, V{\'\i}ctor M. and {Jim{\'e}nez-Serra}, Izaskun and {Zeng}, Shaoshan and {Rodr{\'\i}guez-Almeida}, Lucas F. and {Rico-Villas}, Fernando and {Mart{\'\i}n}, Sergio and {Requena-Torres}, Miguel A.},
        title = "{Deuterium Fractionation as a Multiphase Component Tracer in the Galactic Center}",
      journal = {\apjl},
         year = 2022,
        month = feb,
       volume = {926},
       number = {2},
          eid = {L22},
        pages = {L22},
          doi = {10.3847/2041-8213/ac52ac},
archivePrefix = {arXiv},
       eprint = {2202.04111},
 primaryClass = {astro-ph.GA},
       adsurl = {https://ui.adsabs.harvard.edu/abs/2022ApJ...926L..22C}
}

@ARTICLE{Colzi2024,
       author = {{Colzi}, L. and {Mart{\'\i}n-Pintado}, J. and {Zeng}, S. and {Jim{\'e}nez-Serra}, I. and {Rivilla}, V.~M. and {Sanz-Novo}, M. and {Mart{\'\i}n}, S. and {Zhang}, Q. and {Lu}, X.},
        title = "{Excitation and spatial study of a prestellar cluster towards G+0.693-0.027 in the Galactic centre}",
      journal = {arXiv e-prints},
         year = 2024,
        month = aug,
          eid = {arXiv:2408.17141},
        pages = {arXiv:2408.17141},
          doi = {10.48550/arXiv.2408.17141},
archivePrefix = {arXiv},
       eprint = {2408.17141},
 primaryClass = {astro-ph.GA},
       adsurl = {https://ui.adsabs.harvard.edu/abs/2024arXiv240817141C}
}

@ARTICLE{Cosentino2018,
       author = {{Cosentino}, G. and {Jim{\'e}nez-Serra}, I. and {Henshaw}, J.~D. and {Caselli}, P. and {Viti}, S. and {Barnes}, A.~T. and {Fontani}, F. and {Tan}, J.~C. and {Pon}, A.},
        title = "{Widespread SiO and CH$_{3}$OH emission in filamentary infrared dark clouds}",
      journal = {\mnras},
         year = 2018,
        month = mar,
       volume = {474},
       number = {3},
        pages = {3760-3781},
          doi = {10.1093/mnras/stx3013},
archivePrefix = {arXiv},
       eprint = {1711.09679},
 primaryClass = {astro-ph.GA},
       adsurl = {https://ui.adsabs.harvard.edu/abs/2018MNRAS.474.3760C}
}

@ARTICLE{Cosentino2019,
       author = {{Cosentino}, Giuliana and {Jim{\'e}nez-Serra}, Izaskun and {Caselli}, Paola and {Henshaw}, Jonathan D. and {Barnes}, Ashley T. and {Tan}, Jonathan C. and {Viti}, Serena and {Fontani}, Francesco and {Wu}, Benjamin},
        title = "{Interstellar Plunging Waves: ALMA Resolves the Physical Structure of Nonstationary MHD Shocks}",
      journal = {\apjl},
         year = 2019,
        month = aug,
       volume = {881},
       number = {2},
          eid = {L42},
        pages = {L42},
          doi = {10.3847/2041-8213/ab38c5},
archivePrefix = {arXiv},
       eprint = {1908.09743},
 primaryClass = {astro-ph.GA},
       adsurl = {https://ui.adsabs.harvard.edu/abs/2019ApJ...881L..42C}
}

@ARTICLE{Henshaw2016_gas_kinematics,
       author = {{Henshaw}, J.~D. and {Longmore}, S.~N. and {Kruijssen}, J.~M.~D. and {Davies}, B. and {Bally}, J. and {Barnes}, A. and {Battersby}, C. and {Burton}, M. and {Cunningham}, M.~R. and {Dale}, J.~E. and {Ginsburg}, A. and {Immer}, K. and {Jones}, P.~A. and {Kendrew}, S. and {Mills}, E.~A.~C. and {Molinari}, S. and {Moore}, T.~J.~T. and {Ott}, J. and {Pillai}, T. and {Rathborne}, J. and {Schilke}, P. and {Schmiedeke}, A. and {Testi}, L. and {Walker}, D. and {Walsh}, A. and {Zhang}, Q.},
        title = "{Molecular gas kinematics within the central 250 pc of the Milky Way}",
      journal = {\mnras},
         year = 2016,
        month = apr,
       volume = {457},
       number = {3},
        pages = {2675-2702},
          doi = {10.1093/mnras/stw121},
archivePrefix = {arXiv},
       eprint = {1601.03732},
 primaryClass = {astro-ph.GA},
       adsurl = {https://ui.adsabs.harvard.edu/abs/2016MNRAS.457.2675H}
}

@ARTICLE{Bouvier2024,
       author = {{Bouvier}, M. and {Viti}, S. and {Behrens}, E. and {Butterworth}, J. and {Huang}, K. -Y. and {Mangum}, J.~G. and {Harada}, N. and {Mart{\'\i}n}, S. and {Rivilla}, V.~M. and {Muller}, S. and {Sakamoto}, K. and {Yoshimura}, Y. and {Tanaka}, K. and {Nakanishi}, K. and {Herrero-Illana}, R. and {Colzi}, L. and {Gorski}, M.~D. and {Henkel}, C. and {Humire}, P.~K. and {Meier}, D.~S. and {van der Werf}, P.~P. and {Yan}, Y.~T.},
        title = "{An ALCHEMI inspection of sulphur-bearing species towards the central molecular zone of NGC 253}",
      journal = {\aap},
         year = 2024,
        month = sep,
       volume = {689},
          eid = {A64},
        pages = {A64},
          doi = {10.1051/0004-6361/202449186},
archivePrefix = {arXiv},
       eprint = {2405.08408},
 primaryClass = {astro-ph.GA},
       adsurl = {https://ui.adsabs.harvard.edu/abs/2024A&A...689A..64B}
}

@ARTICLE{Huang2023,
       author = {{Huang}, K. -Y. and {Viti}, S. and {Holdship}, J. and {Mangum}, J.~G. and {Mart{\'\i}n}, S. and {Harada}, N. and {Muller}, S. and {Sakamoto}, K. and {Tanaka}, K. and {Yoshimura}, Y. and {Herrero-Illana}, R. and {Meier}, D.~S. and {Behrens}, E. and {van der Werf}, P.~P. and {Henkel}, C. and {Garc{\'\i}a-Burillo}, S. and {Rivilla}, V.~M. and {Emig}, K.~L. and {Colzi}, L. and {Humire}, P.~K. and {Aladro}, R. and {Bouvier}, M.},
        title = "{Reconstructing the shock history in the CMZ of NGC 253 with ALCHEMI}",
      journal = {\aap},
         year = 2023,
        month = jul,
       volume = {675},
          eid = {A151},
        pages = {A151},
          doi = {10.1051/0004-6361/202245659},
archivePrefix = {arXiv},
       eprint = {2303.12685},
 primaryClass = {astro-ph.GA},
       adsurl = {https://ui.adsabs.harvard.edu/abs/2023A&A...675A.151H}
}

@ARTICLE{Beaumont2013,
       author = {{Beaumont}, Christopher N. and {Offner}, Stella S.~R. and {Shetty}, Rahul and {Glover}, Simon C.~O. and {Goodman}, Alyssa A.},
        title = "{Quantifying Observational Projection Effects Using Molecular Cloud Simulations}",
      journal = {\apj},
         year = 2013,
        month = nov,
       volume = {777},
       number = {2},
          eid = {173},
        pages = {173},
          doi = {10.1088/0004-637X/777/2/173},
archivePrefix = {arXiv},
       eprint = {1310.1929},
 primaryClass = {astro-ph.GA},
       adsurl = {https://ui.adsabs.harvard.edu/abs/2013ApJ...777..173B}
}

@ARTICLE{Caselli1998,
       author = {{Caselli}, P. and {Walmsley}, C.~M. and {Terzieva}, R. and {Herbst}, Eric},
        title = "{The Ionization Fraction in Dense Cloud Cores}",
      journal = {\apj},
         year = 1998,
        month = may,
       volume = {499},
       number = {1},
        pages = {234-249},
          doi = {10.1086/305624},
       adsurl = {https://ui.adsabs.harvard.edu/abs/1998ApJ...499..234C}
}

@ARTICLE{Barnes2017,
   author = {{Barnes}, A.~T. and {Longmore}, S.~N. and {Battersby}, C. and 
	{Bally}, J. and {Kruijssen}, J.~M.~D. and {Henshaw}, J.~D. and 
	{Walker}, D.~L.},
    title = "{Star formation rates and efficiencies in the Galactic Centre}",
  journal = {\mnras},
archivePrefix = "arXiv",
   eprint = {1704.03572},
     year = 2017,
    month = apr,
   adsurl = {http://adsabs.harvard.edu/abs/2017arXiv170403572B}
}

@ARTICLE{Jones2012,
   author = {{Jones}, P.~A. and {Burton}, M.~G. and {Cunningham}, M.~R. and 
	{Requena-Torres}, M.~A. and {Menten}, K.~M. and {Schilke}, P. and 
	{Belloche}, A. and {Leurini}, S. and {Mart{\'{\i}}n-Pintado}, J. and 
	{Ott}, J. and {Walsh}, A.~J.},
    title = "{Spectral imaging of the Central Molecular Zone in multiple 3-mm molecular lines}",
  journal = {\mnras},
archivePrefix = "arXiv",
   eprint = {1110.1421},
     year = 2012,
    month = feb,
   volume = 419,
    pages = {2961-2986},
      doi = {10.1111/j.1365-2966.2011.19941.x},
   adsurl = {http://adsabs.harvard.edu/abs/2012MNRAS.419.2961J}
}

@ARTICLE{Reid2016,
       author = {{Reid}, M.~J. and {Dame}, T.~M. and {Menten}, K.~M. and {Brunthaler}, A.},
        title = "{A Parallax-based Distance Estimator for Spiral Arm Sources}",
      journal = {\apj},
         year = 2016,
        month = jun,
       volume = {823},
       number = {2},
          eid = {77},
        pages = {77},
          doi = {10.3847/0004-637X/823/2/77},
archivePrefix = {arXiv},
       eprint = {1604.02433},
 primaryClass = {astro-ph.GA},
       adsurl = {https://ui.adsabs.harvard.edu/abs/2016ApJ...823...77R}
}

@ARTICLE{Reid2019,
       author = {{Reid}, M.~J. and {Menten}, K.~M. and {Brunthaler}, A. and
         {Zheng}, X.~W. and {Dame}, T.~M. and {Xu}, Y. and {Li}, J. and
         {Sakai}, N. and {Wu}, Y. and {Immer}, K. and {Zhang}, B. and
         {Sanna}, A. and {Moscadelli}, L. and {Rygl}, K.~L.~J. and
         {Bartkiewicz}, A. and {Hu}, B. and {Quiroga-Nu{\~n}ez}, L.~H. and
         {van Langevelde}, H.~J.},
        title = "{Trigonometric Parallaxes of High-mass Star-forming Regions: Our View of the Milky Way}",
      journal = {\apj},
         year = 2019,
        month = nov,
       volume = {885},
       number = {2},
          eid = {131},
        pages = {131},
          doi = {10.3847/1538-4357/ab4a11},
archivePrefix = {arXiv},
       eprint = {1910.03357},
 primaryClass = {astro-ph.GA},
       adsurl = {https://ui.adsabs.harvard.edu/abs/2019ApJ...885..131R}
}

@ARTICLE{Requena-Torres2006,
       author = {{Requena-Torres}, M.~A. and {Mart{\'\i}n-Pintado}, J. and {Rodr{\'\i}guez-Franco}, A. and {Mart{\'\i}n}, S. and {Rodr{\'\i}guez-Fern{\'a}ndez}, N.~J. and {de Vicente}, P.},
        title = "{Organic molecules in the Galactic center. Hot core chemistry without hot cores}",
      journal = {\aap},
         year = 2006,
        month = sep,
       volume = {455},
       number = {3},
        pages = {971-985},
          doi = {10.1051/0004-6361:20065190},
archivePrefix = {arXiv},
       eprint = {astro-ph/0605031},
 primaryClass = {astro-ph},
       adsurl = {https://ui.adsabs.harvard.edu/abs/2006A&A...455..971R}
}

@ARTICLE{Gravity19,
       author = {{Gravity Collaboration} and {Abuter}, R. and {Amorim}, A. and
         {Baub{\"o}ck}, M. and {Berger}, J.~P. and {Bonnet}, H. and {Brand
        ner}, W. and {Cl{\'e}net}, Y. and {Coud{\'e} Du Foresto}, V. and
         {de Zeeuw}, P.~T. and {Dexter}, J. and {Duvert}, G. and {Eckart}, A. and
         {Eisenhauer}, F. and {F{\"o}rster Schreiber}, N.~M. and {Garcia}, P. and
         {Gao}, F. and {Gendron}, E. and {Genzel}, R. and {Gerhard}, O. and
         {Gillessen}, S. and {Habibi}, M. and {Haubois}, X. and {Henning}, T. and
         {Hippler}, S. and {Horrobin}, M. and {Jim{\'e}nez-Rosales}, A. and
         {Jocou}, L. and {Kervella}, P. and {Lacour}, S. and
         {Lapeyr{\`e}re}, V. and {Le Bouquin}, J. -B. and {L{\'e}na}, P. and
         {Ott}, T. and {Paumard}, T. and {Perraut}, K. and {Perrin}, G. and
         {Pfuhl}, O. and {Rabien}, S. and {Rodriguez Coira}, G. and
         {Rousset}, G. and {Scheithauer}, S. and {Sternberg}, A. and
         {Straub}, O. and {Straubmeier}, C. and {Sturm}, E. and
         {Tacconi}, L.~J. and {Vincent}, F. and {von Fellenberg}, S. and
         {Waisberg}, I. and {Widmann}, F. and {Wieprecht}, E. and
         {Wiezorrek}, E. and {Woillez}, J. and {Yazici}, S.},
        title = "{A geometric distance measurement to the Galactic center black hole with 0.3\% uncertainty}",
      journal = {\aap},
         year = "2019",
        month = "May",
       volume = {625},
          eid = {L10},
        pages = {L10},
          doi = {10.1051/0004-6361/201935656},
archivePrefix = {arXiv},
       eprint = {1904.05721},
 primaryClass = {astro-ph.GA},
       adsurl = {https://ui.adsabs.harvard.edu/abs/2019A&A...625L..10G}
}

@ARTICLE{Rosolowsky2008b,
   author = {{Rosolowsky}, E.~W. and {Pineda}, J.~E. and {Kauffmann}, J. and 
	{Goodman}, A.~A.},
    title = "{Structural Analysis of Molecular Clouds: Dendrograms}",
  journal = {\apj},
archivePrefix = "arXiv",
   eprint = {0802.2944},
     year = 2008,
    month = jun,
   volume = 679,
    pages = {1338-1351},
      doi = {10.1086/587685},
   adsurl = {http://adsabs.harvard.edu/abs/2008ApJ...679.1338R}
}

@ARTICLE{Shetty2012,
   author = {{Shetty}, R. and {Beaumont}, C.~N. and {Burton}, M.~G. and {Kelly}, B.~C. and 
	{Klessen}, R.~S.},
    title = "{The linewidth-size relationship in the dense interstellar medium of the Central Molecular Zone}",
  journal = {\mnras},
archivePrefix = "arXiv",
   eprint = {1206.5803},
     year = 2012,
    month = sep,
   volume = 425,
    pages = {720-729},
      doi = {10.1111/j.1365-2966.2012.21588.x},
   adsurl = {http://adsabs.harvard.edu/abs/2012MNRAS.425..720S}
}

@ARTICLE{Goicoechea2006,
       author = {{Goicoechea}, J.~R. and {Pety}, J. and {Gerin}, M. and {Teyssier}, D. and {Roueff}, E. and {Hily-Blant}, P. and {Baek}, S.},
        title = "{Low sulfur depletion in the Horsehead PDR}",
      journal = {\aap},
         year = 2006,
        month = sep,
       volume = {456},
       number = {2},
        pages = {565-580},
          doi = {10.1051/0004-6361:20065260},
archivePrefix = {arXiv},
       eprint = {astro-ph/0605716},
 primaryClass = {astro-ph},
       adsurl = {https://ui.adsabs.harvard.edu/abs/2006A&A...456..565G}
}

@ARTICLE{Martin2006,
       author = {{Mart{\'\i}n}, S. and {Mart{\'\i}n-Pintado}, J. and {Mauersberger}, R.},
        title = "{Methanol detection in <ASTROBJ>M 82</ASTROBJ>}",
      journal = {\aap},
         year = 2006,
        month = apr,
       volume = {450},
       number = {1},
        pages = {L13-L16},
          doi = {10.1051/0004-6361:200600021},
archivePrefix = {arXiv},
       eprint = {astro-ph/0603173},
 primaryClass = {astro-ph},
       adsurl = {https://ui.adsabs.harvard.edu/abs/2006A&A...450L..13M}
}

@ARTICLE{Woodall2007,
       author = {{Woodall}, J. and {Ag{\'u}ndez}, M. and {Markwick-Kemper}, A.~J. and {Millar}, T.~J.},
        title = "{The UMIST database for astrochemistry 2006}",
      journal = {\aap},
         year = 2007,
        month = may,
       volume = {466},
       number = {3},
        pages = {1197-1204},
          doi = {10.1051/0004-6361:20064981},
       adsurl = {https://ui.adsabs.harvard.edu/abs/2007A&A...466.1197W}
}

@ARTICLE{Nogueras-Lara2026,
       author = {{Nogueras-Lara}, Francisco and {Barnes}, Ashley T. and {Henshaw}, Jonathan D. and {Fiteni}, Karl and {Sofue}, Yoshiaki and {Sch{\"o}del}, Rainer and {Mart{\'\i}nez-Arranz}, {\'A}lvaro and {Sormani}, Mattia C. and {Armijos-Abenda{\~n}o}, Jairo and {Colzi}, Laura and et al.},
        title = "{Unveiling the 3D structure of the central molecular zone from stellar kinematics and photometry: The 50 and 20 km/s clouds}",
      journal = {\aap},
         year = 2026,
        month = jan,
       volume = {706},
          eid = {A18},
        pages = {A18},
          doi = {10.1051/0004-6361/202556047},
archivePrefix = {arXiv},
       eprint = {2601.05252},
 primaryClass = {astro-ph.GA},
       adsurl = {https://ui.adsabs.harvard.edu/abs/2026A&A...706A..18N}
}

@ARTICLE{Nogueras-Lara2024,
       author = {{Nogueras-Lara}, F.},
        title = "{Hunting young stars in the Galactic centre. Hundreds of thousands of solar masses of young stars in the Sagittarius C region}",
      journal = {\aap},
         year = 2024,
        month = jan,
       volume = {681},
          eid = {L21},
        pages = {L21},
          doi = {10.1051/0004-6361/202348712},
archivePrefix = {arXiv},
       eprint = {2401.07900},
 primaryClass = {astro-ph.GA},
       adsurl = {https://ui.adsabs.harvard.edu/abs/2024A&A...681L..21N}
}

@ARTICLE{Nogueras-Lara2022,
       author = {{Nogueras-Lara}, Francisco and {Sch{\"o}del}, Rainer and {Neumayer}, Nadine},
        title = "{Detection of an excess of young stars in the Galactic Centre Sagittarius B1 region}",
      journal = {Nature Astronomy},
         year = 2022,
        month = aug,
       volume = {6},
        pages = {1178-1184},
          doi = {10.1038/s41550-022-01755-3},
archivePrefix = {arXiv},
       eprint = {2207.02227},
 primaryClass = {astro-ph.GA},
       adsurl = {https://ui.adsabs.harvard.edu/abs/2022NatAs...6.1178N}
}

@ARTICLE{Lipman2025,
       author = {{Lipman}, Dani and {Battersby}, Cara and {Walker}, Daniel L. and {Sormani}, Mattia C. and {Bally}, John and {Barnes}, Ashley and {Ginsburg}, Adam and {Glover}, Simon C.~O. and {Henshaw}, Jonathan D. and {Hatchfield}, H. Perry and {Immer}, Katharina and {Klessen}, Ralf S. and {Longmore}, Steven N. and {Mills}, Elisabeth A.~C. and {Smith}, Rowan and {Tress}, R.~G. and {Alboslani}, Danya and {Zhang}, Qizhou},
        title = "{3D CMZ. IV. Distinguishing Near versus Far Distances in the Galactic Center Using Spitzer and Herschel}",
      journal = {\apj},
         year = 2025,
        month = may,
       volume = {984},
       number = {2},
          eid = {159},
        pages = {159},
          doi = {10.3847/1538-4357/adb5ee},
archivePrefix = {arXiv},
       eprint = {2410.17321},
 primaryClass = {astro-ph.GA},
       adsurl = {https://ui.adsabs.harvard.edu/abs/2025ApJ...984..159L}
}

@ARTICLE{Lipman2026,
       author = {{Lipman}, Dani R. and {Battersby}, Cara and {Walker}, Daniel and {Clavel}, Ma{\"\i}ca and {DuBois}, B.~L. and {Ginsburg}, Adam and {Henshaw}, Jonathan D. and {Klessen}, Ralf S. and {Mills}, Elisabeth A.~C. and {Nogueras-Lara}, Francisco and et al.},
        title = "{3D CMZ. V. A New Orbital Model of Our Galaxy's Center, Informed by Data Across the Electromagnetic Spectrum}",
      journal = {\apj},
         year = 2026,
        month = may,
       volume = {1002},
       number = {1},
          eid = {24},
        pages = {24},
          doi = {10.3847/1538-4357/ae561f},
       adsurl = {https://ui.adsabs.harvard.edu/abs/2026ApJ..1002...24L}
}

@ARTICLE{Martin-Pintado1992,
       author = {{Martin-Pintado}, J. and {Bachiller}, R. and {Fuente}, A.},
        title = "{SiO emission as a tracer of shocked gas in molecular outflows.}",
      journal = {\aap},
         year = 1992,
        month = feb,
       volume = {254},
        pages = {315-326},
       adsurl = {https://ui.adsabs.harvard.edu/abs/1992A&A...254..315M}
}

@ARTICLE{Jimenez-Serra2010,
       author = {{Jim{\'e}nez-Serra}, I. and {Caselli}, P. and {Tan}, J.~C. and {Hernandez}, A.~K. and {Fontani}, F. and {Butler}, M.~J. and {van Loo}, S.},
        title = "{Parsec-scale SiO emission in an infrared dark cloud}",
      journal = {\mnras},
         year = 2010,
        month = jul,
       volume = {406},
       number = {1},
        pages = {187-196},
          doi = {10.1111/j.1365-2966.2010.16698.x},
archivePrefix = {arXiv},
       eprint = {1003.3463},
 primaryClass = {astro-ph.SR},
       adsurl = {https://ui.adsabs.harvard.edu/abs/2010MNRAS.406..187J}
}

@ARTICLE{Martin-Pintado1997,
       author = {{Mart{\'\i}n-Pintado}, J. and {de Vicente}, P. and {Fuente}, A. and {Planesas}, P.},
        title = "{SiO Emission from the Galactic Center Molecular Clouds}",
      journal = {\apjl},
         year = 1997,
        month = jun,
       volume = {482},
       number = {1},
        pages = {L45-L48},
          doi = {10.1086/310691},
archivePrefix = {arXiv},
       eprint = {astro-ph/9704006},
 primaryClass = {astro-ph},
       adsurl = {https://ui.adsabs.harvard.edu/abs/1997ApJ...482L..45M}
}

@ARTICLE{He2021,
       author = {{He}, Yu-Xin and {Henkel}, Christian and {Zhou}, Jian-Jun and {Esimbek}, Jarken and {Stutz}, Amelia M. and {Liu}, Hong-Li and {Ji}, Wei-Guang and {Li}, Da-Lei and {Wu}, Gang and {Tang}, Xin-Di and {Komesh}, Toktarkhan and {Sailanbek}, Serikbek},
        title = "{Extended HNCO, SiO, and HC$_{3}$N Emission in 43 Southern Star-forming Regions}",
      journal = {\apjs},
         year = 2021,
        month = mar,
       volume = {253},
       number = {1},
          eid = {2},
        pages = {2},
          doi = {10.3847/1538-4365/abd0fb},
archivePrefix = {arXiv},
       eprint = {2012.04354},
 primaryClass = {astro-ph.GA},
       adsurl = {https://ui.adsabs.harvard.edu/abs/2021ApJS..253....2H}
}

@ARTICLE{Kelly2017,
       author = {{Kelly}, G. and {Viti}, S. and {Garc{\'\i}a-Burillo}, S. and {Fuente}, A. and {Usero}, A. and {Krips}, M. and {Neri}, R.},
        title = "{Molecular shock tracers in NGC 1068: SiO and HNCO}",
      journal = {\aap},
         year = 2017,
        month = jan,
       volume = {597},
          eid = {A11},
        pages = {A11},
          doi = {10.1051/0004-6361/201628946},
archivePrefix = {arXiv},
       eprint = {1609.02023},
 primaryClass = {astro-ph.GA},
       adsurl = {https://ui.adsabs.harvard.edu/abs/2017A&A...597A..11K}
}

@ARTICLE{Tanaka2018,
       author = {{Tanaka}, Kunihiko and {Nagai}, Makoto and {Kamegai}, Kazuhisa and {Iino}, Takahiro and {Sakai}, Takeshi},
        title = "{HCN J = 4-3, HNC J = 1-0, H$^{13}$CN J = 1-0, and HC$_{3}$N J = 10-9 Maps of the Galactic Center Region. I. Spatially Resolved Measurements of Physical Conditions and Chemical Composition}",
      journal = {\apjs},
         year = 2018,
        month = jun,
       volume = {236},
       number = {2},
          eid = {40},
        pages = {40},
          doi = {10.3847/1538-4365/aab9a5},
archivePrefix = {arXiv},
       eprint = {1804.00666},
 primaryClass = {astro-ph.GA},
       adsurl = {https://ui.adsabs.harvard.edu/abs/2018ApJS..236...40T}
}

@ARTICLE{Taniguchi2019,
       author = {{Taniguchi}, Kotomi and {Saito}, Masao and {Sridharan}, T.~K. and {Minamidani}, Tetsuhiro},
        title = "{Survey Observations to Study Chemical Evolution from High-mass Starless Cores to High-mass Protostellar Objects. II. HC$_{3}$N and N$_{2}$H$^{+}$}",
      journal = {\apj},
         year = 2019,
        month = feb,
       volume = {872},
       number = {2},
          eid = {154},
        pages = {154},
          doi = {10.3847/1538-4357/ab001e},
archivePrefix = {arXiv},
       eprint = {1901.06446},
 primaryClass = {astro-ph.GA},
       adsurl = {https://ui.adsabs.harvard.edu/abs/2019ApJ...872..154T}
}

@ARTICLE{Goldsmith2017,
       author = {{Goldsmith}, Paul F. and {Kauffmann}, Jens},
        title = "{Electron Excitation of High Dipole Moment Molecules Re-examined}",
      journal = {\apj},
         year = 2017,
        month = may,
       volume = {841},
       number = {1},
          eid = {25},
        pages = {25},
          doi = {10.3847/1538-4357/aa6f12},
archivePrefix = {arXiv},
       eprint = {1708.07553},
 primaryClass = {astro-ph.GA},
       adsurl = {https://ui.adsabs.harvard.edu/abs/2017ApJ...841...25G}
}

@article{Martin2012,
	Adsurl = {http://adsabs.harvard.edu/abs/2012A%26A...539A..29M},
	Archiveprefix = {arXiv},
	Author = {{Mart{\'{\i}}n}, S. and {Mart{\'{\i}}n-Pintado}, J. and {Montero-Casta{\~n}o}, M. and {Ho}, P.~T.~P. and {Blundell}, R.},
	Doi = {10.1051/0004-6361/201117268},
	Eid = {A29},
	Eprint = {1112.0566},
	Journal = {\aap},
	Month = mar,
	Pages = {A29},
	Primaryclass = {astro-ph.GA},
	Title = {{Surviving the hole. I. Spatially resolved chemistry around Sagittarius A$^{∗}$}},
	Volume = 539,
	Year = 2012}

@ARTICLE{Martin2008,
       author = {{Mart{\'\i}n}, Sergio and {Requena-Torres}, M.~A. and {Mart{\'\i}n-Pintado}, J. and {Mauersberger}, R.},
        title = "{Tracing Shocks and Photodissociation in the Galactic Center Region}",
      journal = {\apj},
         year = 2008,
        month = may,
       volume = {678},
       number = {1},
        pages = {245-254},
          doi = {10.1086/533409},
archivePrefix = {arXiv},
       eprint = {0801.3614},
 primaryClass = {astro-ph},
       adsurl = {https://ui.adsabs.harvard.edu/abs/2008ApJ...678..245M}
}

@ARTICLE{Oka1998,
       author = {{Oka}, Tomoharu and {Hasegawa}, Tetsuo and {Sato}, Fumio and {Tsuboi}, Masato and {Miyazaki}, Atsushi},
        title = "{A Large-Scale CO Survey of the Galactic Center}",
      journal = {\apjs},
         year = 1998,
        month = oct,
       volume = {118},
       number = {2},
        pages = {455-515},
          doi = {10.1086/313138},
       adsurl = {https://ui.adsabs.harvard.edu/abs/1998ApJS..118..455O}
}

@ARTICLE{Oka1999,
       author = {{Oka}, Tomoharu and {White}, Glenn J. and {Hasegawa}, Tetsuo and {Sato}, Fumio and {Tsuboi}, Masato and {Miyazaki}, Atsushi},
        title = "{A High-Velocity Molecular Cloud near the Center of the Galaxy}",
      journal = {\apj},
         year = 1999,
        month = apr,
       volume = {515},
       number = {1},
        pages = {249-255},
          doi = {10.1086/307029},
archivePrefix = {arXiv},
       eprint = {astro-ph/9810434},
 primaryClass = {astro-ph},
       adsurl = {https://ui.adsabs.harvard.edu/abs/1999ApJ...515..249O}
}

@ARTICLE{Mills2026,
       author = {{Mills}, Elisabeth A.~C. and {Butterfield}, Natalie O. and {Liu}, Hauyu Baobab and {Lipman}, Dani and {Ginsburg}, Adam and {Sormani}, Mattia C. and {Henshaw}, Jonathan D. and {Battersby}, Cara D. and {Barnes}, Ashley T. and {Glover}, Simon C.~O. and {Nogueras-Lara}, Francisco and {Morris}, Mark R. and {Ott}, Juergen and {Lang}, Cornelia and {Cook}, Claire and {Mai}, Xinyu},
        title = "{Reconciling 3D Models for the Central 10 pc of the Milky Way}",
      journal = {\apj},
         year = 2026,
        month = apr,
       volume = {1001},
       number = {2},
          eid = {184},
        pages = {184},
          doi = {10.3847/1538-4357/ae4d3a},
archivePrefix = {arXiv},
       eprint = {2603.02211},
 primaryClass = {astro-ph.GA},
       adsurl = {https://ui.adsabs.harvard.edu/abs/2026ApJ..1001..184M}
}

@ARTICLE{Makita2026,
       author = {{Makita}, Momoko and {Oka}, Tomoharu and {Tsujimoto}, Shiho and {Kotani}, Tatsuya},
        title = "{Discovery of Multiple Ultra-broad-velocity Molecular Features Associated with the W44 Molecular Cloud}",
      journal = {\apjl},
         year = 2026,
        month = mar,
       volume = {999},
       number = {1},
          eid = {L3},
        pages = {L3},
          doi = {10.3847/2041-8213/ae36aa},
archivePrefix = {arXiv},
       eprint = {2601.12728},
 primaryClass = {astro-ph.GA},
       adsurl = {https://ui.adsabs.harvard.edu/abs/2026ApJ...999L...3M}
}

@ARTICLE{Sofue2025_GCarms,
       author = {{Sofue}, Yoshiaki and {Oka}, Tomoharu and {Longmore}, Steven N. and {Walker}, Daniel and {Ginsburg}, Adam and {Henshaw}, Jonathan D. and {Bally}, John and {Barnes}, Ashley T. and {Battersby}, Cara and {Colzi}, Laura and {Ho}, Paul and {Jimenez-Serra}, Izaskun and {Kruijssen}, J.~M. Diederik and {Mills}, Elizabeth and {Petkova}, Maya A. and {Sormani}, Mattia C. and {Wallace}, Jen and {Armijos-Abenda{\~n}o}, Jairo and {Dutkowska}, Katarzyna M. and {Enokiya}, Rei and {Fukui}, Yasuo and {Garc{\'\i}a}, Pablo and {Guzman}, Andres and {Henkel}, Christian and {Hsieh}, Pei-Ying and {Hu}, Yue and {Immer}, Katharina and {Jeff}, Desmond and {Klessen}, Ralf S. and {Kohno}, Kotaro and {Krumholz}, Mark R. and {Lipman}, Dani and {Morris}, Mark R. and {Nogueras-Lara}, Francisco and {Nonhebel}, M. and {Ott}, J{\"u}rgen and {Pineda}, Jaime E. and {Mart{\'\i}n}, Sergio and {Requena-Torres}, Miguel Angel and {Rivilla}, V{\'\i}ctor M. and {Riquelme-V{\'a}squez}, Denise and {S{\'a}nchez-Monge}, {\'A}lvaro and {Santa-Maria}, Miriam G. and {Smith}, Howard A. and {Tanvir}, Tabassum S. and {Tolls}, Volker and {Wang}, Q. Daniel},
        title = "{The Galactic Center arms inferred from the ALMA CMZ Exploration Survey (ACES)}",
      journal = {\pasj},
         year = 2025,
        month = aug,
       volume = {77},
       number = {4},
        pages = {687-706},
          doi = {10.1093/pasj/psaf034},
archivePrefix = {arXiv},
       eprint = {2504.03331},
 primaryClass = {astro-ph.GA},
       adsurl = {https://ui.adsabs.harvard.edu/abs/2025PASJ...77..687S}
}

@ARTICLE{Sofue2025,
       author = {{Sofue}, Yoshiaki and {Oka}, Tomoharu and {Longmore}, Steven N. and {Walker}, Daniel and {Ginsburg}, Adam and {Henshaw}, Jonathan D. and {Bally}, John and {Barnes}, Ashley T. and {Battersby}, Cara and {Colzi}, Laura and {Ho}, Paul and {Jimenez-Serra}, Izaskun and {Kruijssen}, J.~M. Diederik and {Mills}, Elizabeth and {Petkova}, Maya A. and {Sormani}, Mattia C. and {Wallace}, Jennifer and {Armijos-Abenda{\~n}o}, Jairo and {Dutkowska}, Katarzyna M. and {Enokiya}, Rei and {Garc{\'\i}a}, Pablo and {Gramze}, Savannah and {Henkel}, Christian and {Hsieh}, Pei-Ying and {Hu}, Yue and {Immer}, Katharina and {Iwata}, Yuhei and {Karoly}, Janik and {Klessen}, Ralf S. and {Kohno}, Kotaro and {Krumholz}, Mark R. and {Lipman}, Dani and {Morris}, Mark R. and {Nogueras-Lara}, Francisco and {Pineda}, Jaime E. and {Mart{\'\i}n}, Sergio and {Requena-Torres}, Miguel Angel and {Rivilla}, V{\'\i}ctor M. and {Riquelme-V{\'a}squez}, Denise and {S{\'a}nchez-Monge}, {\'A}lvaro and {Santa-Maria}, Miriam G. and {Smith}, Howard A. and {Tolls}, Volker and {Wang}, Q. Daniel},
        title = "{Circumnuclear eccentric gas flow in the Galactic Center revealed by ALMA CMZ Exploration Survey (ACES)}",
      journal = {\pasj},
         year = 2025,
        month = jul,
          doi = {10.1093/pasj/psaf072},
       adsurl = {https://ui.adsabs.harvard.edu/abs/2025PASJ..tmp...79S}
}

@ARTICLE{Kumar1997,
       author = {{Kumar}, Pawan and {Riffert}, Harald},
        title = "{Possible explanations for some unusually large velocity dispersion molecular clouds near the Galactic Centre}",
      journal = {\mnras},
         year = 1997,
        month = dec,
       volume = {292},
       number = {4},
        pages = {871-878},
          doi = {10.1093/mnras/292.4.871},
       adsurl = {https://ui.adsabs.harvard.edu/abs/1997MNRAS.292..871K}
}

@ARTICLE{stark1986,
       author = {{Stark}, A.~A. and {Bania}, T.~M.},
        title = "{Clump 2: an Inner Spiral Arm?}",
      journal = {\apjl},
         year = 1986,
        month = jul,
       volume = {306},
        pages = {L17},
          doi = {10.1086/184695},
       adsurl = {https://ui.adsabs.harvard.edu/abs/1986ApJ...306L..17S}
}

@ARTICLE{Riffert1997,
       author = {{Riffert}, H. and {Kumar}, P. and {Huchtmeier}, W.~K.},
        title = "{H i observations of two molecular clouds with extremely large velocity dispersions}",
      journal = {\mnras},
         year = 1997,
        month = jan,
       volume = {284},
       number = {3},
        pages = {749-753},
          doi = {10.1093/mnras/284.3.749},
archivePrefix = {arXiv},
       eprint = {astro-ph/9611124},
 primaryClass = {astro-ph},
       adsurl = {https://ui.adsabs.harvard.edu/abs/1997MNRAS.284..749R}
}

@ARTICLE{Kolcu2025,
       author = {{Kolcu}, Tutku and {Sormani}, Mattia C. and {Maciejewski}, Witold and {Stuber}, Sophia K. and {Schinnerer}, Eva and {Fragkoudi}, Francesca and {Barnes}, Ashley T. and {Bigiel}, Frank and {Chevance}, M{\'e}lanie and {Colombo}, Dario and {Emsellem}, {\'E}ric and {Glover}, Simon C.~O. and {Henshaw}, Jonathan D. and {Klessen}, Ralf S. and {Meidt}, Sharon E. and {Neumann}, Justus and {Pinna}, Francesca and {Querejeta}, Miguel and {Williams}, Thomas G.},
        title = "{Extreme cloud collisions in nearby barred galaxies}",
      journal = {\mnras},
         year = 2025,
        month = jul,
          doi = {10.1093/mnras/staf1080},
archivePrefix = {arXiv},
       eprint = {2507.04530},
 primaryClass = {astro-ph.GA},
       adsurl = {https://ui.adsabs.harvard.edu/abs/2025MNRAS.tmp.1035K}
}

@ARTICLE{Busch2022,
       author = {{Busch}, Laura A. and {Riquelme}, Denise and {G{\"u}sten}, Rolf and {Menten}, Karl M. and {Pillai}, Thushara G.~S. and {Kauffmann}, Jens},
        title = "{Living on the edge of the Milky Way's central molecular zone. G1.3 is the more likely candidate for gas accretion into the CMZ}",
      journal = {\aap},
         year = 2022,
        month = dec,
       volume = {668},
          eid = {A183},
        pages = {A183},
          doi = {10.1051/0004-6361/202244870},
archivePrefix = {arXiv},
       eprint = {2210.12980},
 primaryClass = {astro-ph.GA},
       adsurl = {https://ui.adsabs.harvard.edu/abs/2022A&A...668A.183B}
}

@ARTICLE{Liszt2006,
       author = {{Liszt}, H.~S.},
        title = "{Molecular cloud shredding in the Galactic Bar}",
      journal = {\aap},
         year = 2006,
        month = feb,
       volume = {447},
       number = {2},
        pages = {533-544},
          doi = {10.1051/0004-6361:20054070},
       adsurl = {https://ui.adsabs.harvard.edu/abs/2006A&A...447..533L}
}

@ARTICLE{Oka2022,
       author = {{Oka}, Tomoharu and {Uruno}, Asaka and {Enokiya}, Rei and {Nakamura}, Taichi and {Yamasaki}, Yuto and {Watanabe}, Yuto and {Tokuyama}, Sekito and {Iwata}, Yuhei},
        title = "{Catalog of High-velocity Dispersion Compact Clouds in the Central Molecular Zone of Our Galaxy}",
      journal = {\apjs},
         year = 2022,
        month = aug,
       volume = {261},
       number = {2},
          eid = {13},
        pages = {13},
          doi = {10.3847/1538-4365/ac6bfc},
archivePrefix = {arXiv},
       eprint = {2209.12395},
 primaryClass = {astro-ph.GA},
       adsurl = {https://ui.adsabs.harvard.edu/abs/2022ApJS..261...13O}
}

@ARTICLE{Oka2012,
       author = {{Oka}, Tomoharu and {Onodera}, Yui and {Nagai}, Makoto and {Tanaka}, Kunihiko and {Matsumura}, Shinji and {Kamegai}, Kazuhisa},
        title = "{ASTE CO J = 3-2 Survey of the Galactic Center}",
      journal = {\apjs},
         year = 2012,
        month = aug,
       volume = {201},
       number = {2},
          eid = {14},
        pages = {14},
          doi = {10.1088/0067-0049/201/2/14},
       adsurl = {https://ui.adsabs.harvard.edu/abs/2012ApJS..201...14O}
}

@ARTICLE{Oka2017,
       author = {{Oka}, Tomoharu and {Tsujimoto}, Shiho and {Iwata}, Yuhei and {Nomura}, Mariko and {Takekawa}, Shunya},
        title = "{Millimetre-wave emission from an intermediate-mass black hole candidate in the Milky Way}",
      journal = {Nature Astronomy},
         year = 2017,
        month = sep,
       volume = {1},
        pages = {709-712},
          doi = {10.1038/s41550-017-0224-z},
archivePrefix = {arXiv},
       eprint = {1707.07603},
 primaryClass = {astro-ph.GA},
       adsurl = {https://ui.adsabs.harvard.edu/abs/2017NatAs...1..709O}
}

@ARTICLE{Takekawa2020,
       author = {{Takekawa}, Shunya and {Oka}, Tomoharu and {Iwata}, Yuhei and {Tsujimoto}, Shiho and {Nomura}, Mariko},
        title = "{The Fifth Candidate for an Intermediate-mass Black Hole in the Galactic Center}",
      journal = {\apj},
         year = 2020,
        month = feb,
       volume = {890},
       number = {2},
          eid = {167},
        pages = {167},
          doi = {10.3847/1538-4357/ab6f6f},
archivePrefix = {arXiv},
       eprint = {2002.05173},
 primaryClass = {astro-ph.GA},
       adsurl = {https://ui.adsabs.harvard.edu/abs/2020ApJ...890..167T}
}

@ARTICLE{Takekawa2024,
       author = {{Takekawa}, Shunya and {Oka}, Tomoharu and {Tsujimoto}, Shiho and {Yokozuka}, Hiroki and {Harada}, Nanase and {Kaneko}, Miyuki and {Enokiya}, Rei and {Iwata}, Yuhei},
        title = "{Parabolic-like Trend in SiO Ratios throughout the Central Molecular Zone: Possible Signature of a Past Nuclear Activity in the Galactic Center}",
      journal = {\apjl},
         year = 2024,
        month = sep,
       volume = {972},
       number = {1},
          eid = {L3},
        pages = {L3},
          doi = {10.3847/2041-8213/ad6c51},
archivePrefix = {arXiv},
       eprint = {2408.09252},
 primaryClass = {astro-ph.GA},
       adsurl = {https://ui.adsabs.harvard.edu/abs/2024ApJ...972L...3T}
}

@ARTICLE{Takekawa2017,
       author = {{Takekawa}, Shunya and {Oka}, Tomoharu and {Iwata}, Yuhei and {Tokuyama}, Sekito and {Nomura}, Mariko},
        title = "{Discovery of Two Small High-velocity Compact Clouds in the Central 10 pc of Our Galaxy}",
      journal = {\apjl},
         year = 2017,
        month = jul,
       volume = {843},
       number = {1},
          eid = {L11},
        pages = {L11},
          doi = {10.3847/2041-8213/aa79ee},
archivePrefix = {arXiv},
       eprint = {1706.04810},
 primaryClass = {astro-ph.GA},
       adsurl = {https://ui.adsabs.harvard.edu/abs/2017ApJ...843L..11T}
}

@ARTICLE{Takekawa2019,
       author = {{Takekawa}, Shunya and {Oka}, Tomoharu and {Tokuyama}, Sekito and {Tanabe}, Kyosuke and {Iwata}, Yuhei and {Tsujimoto}, Shiho and {Nomura}, Mariko and {Shibuya}, Yukihiro},
        title = "{An energetic high-velocity compact cloud: CO-0.31+0.11}",
      journal = {\pasj},
         year = 2019,
        month = dec,
       volume = {71},
          eid = {S21},
        pages = {S21},
          doi = {10.1093/pasj/psz027},
archivePrefix = {arXiv},
       eprint = {1903.08896},
 primaryClass = {astro-ph.GA},
       adsurl = {https://ui.adsabs.harvard.edu/abs/2019PASJ...71S..21T}
}

@ARTICLE{Luisi2021,
       author = {{Luisi}, Matteo and {Anderson}, Loren D. and {Schneider}, Nicola and {Simon}, Robert and {Kabanovic}, Slawa and {G{\"u}sten}, Rolf and {Zavagno}, Annie and {Broos}, Patrick S. and {Buchbender}, Christof and {Guevara}, Cristian and {Jacobs}, Karl and {Justen}, Matthias and {Klein}, Bernd and {Linville}, Dylan and {R{\"o}llig}, Markus and {Russeil}, Delphine and {Stutzki}, J{\"u}rgen and {Tiwari}, Maitraiyee and {Townsley}, Leisa K. and {Tielens}, Alexander G.~G.~M.},
        title = "{Stellar feedback and triggered star formation in the prototypical bubble RCW 120}",
      journal = {Science Advances},
         year = 2021,
        month = apr,
       volume = {7},
       number = {15},
        pages = {eabe9511},
          doi = {10.1126/sciadv.abe9511},
archivePrefix = {arXiv},
       eprint = {2104.04568},
 primaryClass = {astro-ph.GA},
       adsurl = {https://ui.adsabs.harvard.edu/abs/2021SciA....7.9511L}
}

@ARTICLE{Keilmann2025,
       author = {{Keilmann}, E. and {Dannhauer}, S. and {Kabanovic}, S. and {Schneider}, N. and {Ossenkopf-Okada}, V. and {Simon}, R. and {Bonne}, L. and {Goldsmith}, P.~F. and {G{\"u}sten}, R. and {Zavagno}, A. and {Stutzki}, J. and {Riechers}, D. and {R{\"o}llig}, M. and {Verbena}, J.~L. and {Tielens}, A.~G.~G.~M.},
        title = "{[C II]-deficit caused by self-absorption in an ionized carbon-filled bubble in RCW79}",
      journal = {\aap},
         year = 2025,
        month = may,
       volume = {697},
          eid = {L2},
        pages = {L2},
          doi = {10.1051/0004-6361/202453445},
archivePrefix = {arXiv},
       eprint = {2504.08976},
 primaryClass = {astro-ph.GA},
       adsurl = {https://ui.adsabs.harvard.edu/abs/2025A&A...697L...2K}
}

@ARTICLE{Bonne2022,
       author = {{Bonne}, L. and {Schneider}, N. and {Garc{\'\i}a}, P. and {Bij}, A. and {Broos}, P. and {Fissel}, L. and {Guesten}, R. and {Jackson}, J. and {Simon}, R. and {Townsley}, L. and {Zavagno}, A. and {Aladro}, R. and {Buchbender}, C. and {Guevara}, C. and {Higgins}, R. and {Jacob}, A.~M. and {Kabanovic}, S. and {Karim}, R. and {Soam}, A. and {Stutzki}, J. and {Tiwari}, M. and {Wyrowski}, F. and {Tielens}, A.~G.~G.~M.},
        title = "{The SOFIA FEEDBACK Legacy Survey Dynamics and Mass Ejection in the Bipolar H II Region RCW 36}",
      journal = {\apj},
         year = 2022,
        month = aug,
       volume = {935},
       number = {2},
          eid = {171},
        pages = {171},
          doi = {10.3847/1538-4357/ac8052},
archivePrefix = {arXiv},
       eprint = {2207.06479},
 primaryClass = {astro-ph.GA},
       adsurl = {https://ui.adsabs.harvard.edu/abs/2022ApJ...935..171B}
}

@ARTICLE{Feddersen2018,
       author = {{Feddersen}, Jesse R. and {Arce}, H{\'e}ctor G. and {Kong}, Shuo and {Shimajiri}, Yoshito and {Nakamura}, Fumitaka and {Hara}, Chihomi and {Ishii}, Shun and {Sasaki}, Kazushige and {Kawabe}, Ryohei},
        title = "{Expanding CO Shells in the Orion A Molecular Cloud}",
      journal = {\apj},
         year = 2018,
        month = aug,
       volume = {862},
       number = {2},
          eid = {121},
        pages = {121},
          doi = {10.3847/1538-4357/aacaf2},
archivePrefix = {arXiv},
       eprint = {1806.01893},
 primaryClass = {astro-ph.GA},
       adsurl = {https://ui.adsabs.harvard.edu/abs/2018ApJ...862..121F}
}

@article{Hsieh2017,
	Author = {Hsieh, Pei-Ying and Koch, Patrick M. and Ho, Paul T. P. and Kim, Woong-Tae and Tang, Ya-Wen and Wang, Hsiang-Hsu and Yen, Hsi-Wei and Hwang, Chorng-Yuan},
	Doi = {10.3847/1538-4357/aa8329},
	Journal = {The Astrophysical Journal},
	Month = {sep},
	Number = {1},
	Pages = {3},
	Publisher = {The American Astronomical Society},
	Title = {Molecular Gas Feeding the Circumnuclear Disk of the Galactic Center},
	Url = {https://dx.doi.org/10.3847/1538-4357/aa8329},
	Volume = {847},
	Year = {2017}}

@ARTICLE{Dong2015,
       author = {{Dong}, Hui and {Mauerhan}, Jon and {Morris}, Mark R. and {Wang}, Q. Daniel and {Cotera}, Angela},
        title = "{Origins of massive field stars in the Galactic Centre: a spectroscopic study}",
      journal = {\mnras},
         year = 2015,
        month = jan,
       volume = {446},
       number = {1},
        pages = {842-856},
          doi = {10.1093/mnras/stu2116},
       adsurl = {https://ui.adsabs.harvard.edu/abs/2015MNRAS.446..842D}
}

@ARTICLE{Garcia2016,
       author = {{Garc{\'\i}a}, P. and {Simon}, R. and {Stutzki}, J. and {G{\"u}sten}, R. and {Requena-Torres}, M.~A. and {Higgins}, R.},
        title = "{Warm ISM in the Sagittarius A Complex. I. Mid-J CO, atomic carbon, ionized atomic carbon, and ionized nitrogen sub-mm/FIR line observations with the Herschel-HIFI and NANTEN2/SMART telescopes}",
      journal = {\aap},
         year = 2016,
        month = apr,
       volume = {588},
          eid = {A131},
        pages = {A131},
          doi = {10.1051/0004-6361/201526600},
archivePrefix = {arXiv},
       eprint = {1606.04402},
 primaryClass = {astro-ph.GA},
       adsurl = {https://ui.adsabs.harvard.edu/abs/2016A&A...588A.131G}
}

@INCOLLECTION{Figer2009,
       author = {{Figer}, Don. F.},
        title = "{Massive-star formation in the Galactic center:}",
    booktitle = {Massive Stars: From Pop III and GRBs to the Milky Way. Space Telescope Science Institute Symposium Series No. 20. Edited by Mario Livio and Eva Villaver. Cambridge University Press},
         year = 2009,
       editor = {{Livio}, Mario and {Villaver}, Eva},
        pages = {40-59},
        publisher = {Cambridge University Press},
          doi = {10.1017/CBO9780511770593.004},
       adsurl = {https://ui.adsabs.harvard.edu/abs/2009msfp.book...40F}
}

@ARTICLE{Ohashi2014,
       author = {{Ohashi}, Nagayoshi and {Saigo}, Kazuya and {Aso}, Yusuke and {Aikawa}, Yuri and {Koyamatsu}, Shin and {Machida}, Masahiro N. and {Saito}, Masao and {Takahashi}, Sanemichi Z. and {Takakuwa}, Shigehisa and {Tomida}, Kengo and {Tomisaka}, Kohji and {Yen}, Hsi-Wei},
        title = "{Formation of a Keplerian Disk in the Infalling Envelope around L1527 IRS: Transformation from Infalling Motions to Kepler Motions}",
      journal = {\apj},
         year = 2014,
        month = dec,
       volume = {796},
       number = {2},
          eid = {131},
        pages = {131},
          doi = {10.1088/0004-637X/796/2/131},
archivePrefix = {arXiv},
       eprint = {1410.0172},
 primaryClass = {astro-ph.GA},
       adsurl = {https://ui.adsabs.harvard.edu/abs/2014ApJ...796..131O}
}

@ARTICLE{Lopez2025,
       author = {{L{\'o}pez-V{\'a}zquez}, J.~A. and {Fern{\'a}ndez-L{\'o}pez}, M. and {Girart}, J.~M. and {Curiel}, S. and {Estalella}, R. and {Busquet}, G. and {Zapata}, L.~A. and {Lee}, C. -F. and {Galv{\'a}n-Madrid}, R.},
        title = "{Erosion of a dense molecular core by a strong outflow from a massive protostar}",
      journal = {\aap},
         year = 2025,
        month = mar,
       volume = {695},
          eid = {A236},
        pages = {A236},
          doi = {10.1051/0004-6361/202453196},
archivePrefix = {arXiv},
       eprint = {2502.17786},
 primaryClass = {astro-ph.SR},
       adsurl = {https://ui.adsabs.harvard.edu/abs/2025A&A...695A.236L}
}

@ARTICLE{Tress2024,
       author = {{Tress}, R.~G. and {Sormani}, M.~C. and {Girichidis}, P. and {Glover}, S.~C.~O. and {Klessen}, R.~S. and {Smith}, R.~J. and {Sobacchi}, E. and {Armillotta}, L. and {Barnes}, A.~T. and {Battersby}, C. and {Bogue}, K.~R.~J. and {Brucy}, N. and {Colzi}, L. and {Federrath}, C. and {Garc{\'\i}a}, P. and {Ginsburg}, A. and {G{\"o}ller}, J. and {Hatchfield}, H.~P. and {Henkel}, C. and {Hennebelle}, P. and {Henshaw}, J.~D. and {Hirschmann}, M. and {Hu}, Y. and {Kauffmann}, J. and {Kruijssen}, J.~M.~D. and {Lazarian}, A. and {Lipman}, D. and {Longmore}, S.~N. and {Morris}, M.~R. and {Nogueras-Lara}, F. and {Petkova}, M.~A. and {Pillai}, T.~G.~S. and {Rivilla}, V.~M. and {S{\'a}nchez-Monge}, {\'A}. and {Soler}, J.~D. and {Whitworth}, D. and {Zhang}, Q.},
        title = "{Magnetic field morphology and evolution in the Central Molecular Zone and its effect on gas dynamics}",
      journal = {\aap},
         year = 2024,
        month = nov,
       volume = {691},
          eid = {A303},
        pages = {A303},
          doi = {10.1051/0004-6361/202450035},
archivePrefix = {arXiv},
       eprint = {2403.13048},
 primaryClass = {astro-ph.GA},
       adsurl = {https://ui.adsabs.harvard.edu/abs/2024A&A...691A.303T}
}
\bibliographystyle{aasjournalv7.1}

\end{document}